\documentclass[useAMS, usenatbib, fleqn]{mnras}

\usepackage{graphicx}
\usepackage{color}
\usepackage{times}
\usepackage{natbib}
\usepackage{setspace}
\usepackage{amsmath}
\usepackage{amsbsy}
\usepackage{gensymb}
\usepackage{hyperref}

\newif\ifAMStwofonts
\AMStwofontstrue
\definecolor{red}{rgb}{1,0.,0.}

\newcommand{\arepo}{{\sc arepo}}

\newcommand{\kms} {{\rm km~s}^{-1}}

\newcommand{\gsim}{\lower.7ex\hbox{$\;\stackrel{\textstyle>}{\sim}\;$}}
\newcommand{\lsim}{\lower.7ex\hbox{$\;\stackrel{\textstyle<}{\sim}\;$}}
\newcommand{\sgsim}{\lower.6ex\hbox{$\;\stackrel{\scriptstyle>}{\scriptstyle\sim}\;$}}
\newcommand{\slsim}{\lower.6ex\hbox{$\;\stackrel{\scriptstyle<}{\scriptstyle\sim}\;$}}

\newcommand{\muG}{{\rm \mu G}}

\newcommand{\FM}[1]{#1}

\ifpdf
  \DeclareGraphicsExtensions{.pdf}
\else
  \DeclareGraphicsExtensions{.eps}
\fi

\date{Accepted 2018 February 12. Received 2018 February 10; in original form 2017 October 27}

\begin{document}
\title[Non-ideal magnetohydrodynamics on a moving mesh]
{Non-ideal magnetohydrodynamics on a moving mesh} 
\author[F. Marinacci et al.]
{Federico Marinacci$^1$\thanks{E-mail:
fmarinac@mit.edu}, Mark Vogelsberger$^1$\thanks{Alfred P. Sloan Fellow}, Rahul Kannan$^{1,2}$\thanks{Einstein Fellow},
Philip Mocz$^3${\color{blue}{\footnotemark[3]}}, \newauthor R\"udiger Pakmor$^4$, and Volker Springel$^{4,5,6}$
\vspace*{0.2cm}\\
$^1$Kavli Institute for Astrophysics and Space Research, 
  Massachusetts Institute of Technology, 77 Massachusetts Ave, Cambridge, MA 02139, USA\\
$^2$Harvard-Smithsonian Center for Astrophysics, 60 Garden Street, 
  Cambridge, MA 02138, USA\\
$^3$Department of Astrophysical Sciences, Princeton University, 4 Ivy Lane, Princeton, NJ, 08544, USA\\
$^4$Heidelberger Institut f\"{u}r Theoretische Studien,
  Schloss-Wolfsbrunnenweg 35, 69118 Heidelberg, Germany\\
$^5$Zentrum f\"ur Astronomie der Universit\"at Heidelberg,
  Astronomisches Recheninstitut, M\"{o}nchhofstr. 12-14, 69120
  Heidelberg, Germany\\
$^6$Max-Planck-Institut f\"ur Astrophysik,
 Karl-Schwarzschild-Str. 1, D-85748, Garching, Germany
}

\pagerange{\pageref{firstpage}--\pageref{lastpage}}
\pubyear{2018}

\maketitle

\label{firstpage}

\begin{abstract}
In certain astrophysical systems the commonly employed ideal 
magnetohydrodynamics (MHD) approximation breaks down. Here, we introduce novel 
explicit and implicit numerical schemes of ohmic resistivity terms in the 
moving-mesh code \arepo. We include these non-ideal terms for two MHD 
techniques: the Powell 8-wave formalism and a constrained transport scheme, 
which evolves the cell-centred magnetic vector potential. We test our 
implementation against problems of increasing complexity, such as one- and 
two-dimensional diffusion problems, and the evolution of progressive and 
stationary Alfv\'en waves. \FM{On these test problems, our} implementation 
recovers the analytic solutions \FM{to second-order accuracy}.  As first 
applications, we investigate the tearing instability in magnetized plasmas and 
the gravitational collapse of a rotating magnetized gas cloud. In both systems, 
resistivity plays a key role. In the former case, it allows for the development 
of the tearing instability through reconnection of the magnetic field lines. In 
the latter, \FM{the adopted (constant) value of ohmic resistivity has an impact 
on} both the gas distribution around the emerging protostar and the mass 
loading of magnetically driven outflows. Our new non-ideal MHD implementation 
opens up the possibility to study magneto-hydrodynamical systems on a moving 
mesh beyond the ideal MHD approximation. 
\end{abstract}


\section{Introduction}\label{sec:intro}

Magnetic fields are an essential component of the Universe. They are present at 
all spatial scales~\citep{Vallee1998, Feretti2012, Beck2013c}, and directly 
influence a large amount of processes that play a key role in shaping the 
properties of the objects populating the cosmos. Therefore, a complete 
understanding of many astrophysical phenomena requires taking into account the 
effects of magnetic fields on the dynamics of conducting 
gases~\citep{Ferriere2001, Cox2005} and charged relativistic 
particles~\citep{Fermi1949, Kotera2011}. 

Numerical simulations represent the most comprehensive approach to describe the 
evolution of complex physical systems. The inclusion of magnetic fields in 
numerical astrophysical magneto-hydrodynamical simulations often makes use of 
the so-called ideal magnetohydrodynamics (MHD) approximation~\citep[e.g.][]{Fromang2006, Mignone2007, 
Stone2008, Dolag2009, Pakmor2011, Pakmor2013, Hopkins2016b}. Under many 
circumstances, this approximation is an excellent description for the behaviour 
of partially ionized gases in the presence of magnetic fields. Indeed, 
simulations using this approach have become quite sophisticated, and are 
modelling systems of increasing complexity. These range from small-scale 
calculations studying the development of turbulence and the structure of the 
interstellar medium of galaxies~\citep[e.g][]{deAvillez2005, Iffrig2017} to 
larger scale simulations studying the origin and the evolution of magnetic 
fields in galaxies~\citep[e.g.][]{Pakmor2014, Pakmor2017} and galaxy 
clusters~\citep[e.g.][]{Dolag1999, Dolag2002}, and to large-scale cosmological 
simulations~\citep{Marinacci2015, dolag2016, Marinacci2016, Marinacci2017}.

However, there are situations, especially at small spatial scales, e.g. below
those of giant molecular clouds, where the ideal MHD approximation is not an
accurate description of the underlying physics any more. Here the assumptions of
ideal MHD break down, and non-ideal MHD terms, such as ambipolar diffusion and
ohmic resistivity, must be taken into account for a correct description of the
physical system.

For example, in studies of galactic molecular clouds, it is well established 
that ambipolar diffusion, which arises in partially ionized plasmas, is a key 
physical process for the mechanism of star formation~\citep[e.g.][]{Mestel1956, 
Mouschovias1976a, Mouschovias1976b, Shu1987} because it allows for the 
decoupling of neutral gas from magnetic fields~\citep{Basu2004}, which would 
otherwise hinder gravitational collapse and star formation. Ambipolar diffusion 
is also advocated to solve the so-called fragmentation crisis, i.e. the 
stabilizing effect that comparatively weak magnetic fields have on the 
fragmentation of a collapsing star-forming cloud~\citep[e.g.][]{Hennebelle2008}. 
Moreover, ambipolar diffusion can have a non-negligible effect on MHD 
turbulence, by steepening the velocity and magnetic field power 
spectrum~\citep{Li2008} and changing the morphology of the velocity and density 
structures of the gas~\citep{Ntormousi2016}. Finally, together with the Hall 
effect and ohmic resistivity, ambipolar diffusion is also relevant in 
proto-planetary discs, which are only partially ionized. In this case, the 
combination of these three non-ideal MHD effects can influence the development 
of the turbulence due to the magneto-rotational instability in such 
objects~\citep{Bai2015}, thus affecting the accretion rate on to the central 
star and the angular momentum transport within the disc~\citep{Lesur2014, 
Gressel2015, Bethune2017}.

Ohmic resistivity is also important under various circumstances. In particular, 
it allows for magnetic reconnection, a change of topology of magnetic field 
lines that is prevented in ideal MHD due to flux conservation. At the 
reconnection points, ohmic resistivity generates intense Joule dissipation, 
which may power the heating of the solar corona~\citep{Parker1983} or eruptive 
events in the Sun~\citep[see, e.g.][]{Cheng2017}. \FM{A crucial difficulty in 
the study of magnetic reconnection in simulations is due to the introduction of 
numerical resistivity, which is inherent to any discretization procedure. This 
non-physical resistivity may yield to  reconnection phenomena that are entirely 
numerical in nature, substantially affecting the reliability of the simulations. 
This is particularly severe in the low-resistivity regime, which is usually the 
case in the modelling of real systems and that thus requires very high 
resolution to properly model the (small) spatial scales over which resistive 
effects are important.}

The presence of ohmic resistivity may also render unstable otherwise stable 
configurations through the development of tearing instability 
modes~\citep{Furth1963}. Another effect of a non-zero resistivity in the gas is 
the shortening of the decay time of long-term MHD turbulence in molecular 
clouds~\citep{Basu2010}. Moreover, ohmic resistivity is a key physical process 
in the studies of the formation of discs around protostellar 
objects~\citep{Krasnopolsky2010}. In this case, it can help in alleviating the 
so-called magnetic braking catastrophe, which is the suppression of the formation 
of rotationally supported discs in simulations modelling low-mass star formation 
in ideal MHD due to the high efficiency of angular momentum transport by the 
magnetic field. Indeed, this process seems to be effective on 
small scales~\citep{Dapp2010}, but to allow for the formation of larger 
circumstellar discs, other mechanisms, such as turbulent 
reconnection~\citep{Santos-Lima2012}, have been proposed. Ohmic resistivity can 
also affect the efficiency and the mass loading of magnetically driven outflows 
in star-forming clouds~\citep{Machida2007, Matsushita2017}, by weakening or even 
suppressing them compared to ideal MHD studies~\citep{Hennebelle2011, 
Seifried2012}. Here ohmic resistivity weakens the coupling between the magnetic 
fields and the gas in regions where the field dissipation, resulting from finite 
resistivity, is effective. The reduced coupling causes the inability of magnetic 
fields to drive outflows, which, on the other hand, are present even for weakly 
magnetized configurations in the ideal MHD case~\citep[see 
again][]{Matsushita2017}. \FM{However, the extent of these effects is uncertain 
and still debated. These uncertainties are associated to the difficulty in 
computing the exact value of the resistivity coefficient, which strongly depends 
on the detailed chemical composition and ionization state of the gas, in 
molecular clouds \citep[see, e.g.][]{Nakano2002}.}

Given the importance of non-ideal MHD processes, it is not surprising that many 
numerical implementations have been developed to include them in MHD 
simulations. The techniques adopted are very different, and 
single-fluid~\citep[e.g.][]{MacLow1995, Li2011, Masson2012}, or 
multi-fluid~\citep[e.g.][]{Falle2003, Tilley2011} approaches, with a variety of 
time integration techniques, have been used. In this paper, we resort to a 
single-fluid approach and focus on the implementation of the ohmic resistive 
terms in the moving-mesh code \arepo~\citep{Arepo}. We describe such an 
implementation for the Powell divergence cleaning and constrained transport 
(CT) MHD schemes. For both schemes we present an explicit and implicit time 
integration method for the treatment of the ohmic terms. 

The paper is organized as follows. In Section \ref{sec:eq}, we describe the 
schemes that we have adopted to include the ohmic resistivity terms in \arepo, 
differentiating between the explicit (Section~\ref{sec:explicit}) and implicit 
time integration (Section~\ref{sec:implicit}) cases. In Section \ref{sec:tests} 
we test our implementation on a variety of test problems. In Sections 
\ref{sec:reconnection} and \ref{sec:collapse} we present first non-ideal MHD 
applications by studying magnetic reconnection and the gravitational collapse 
of a rotating magnetized cloud, respectively. Finally, in Section 
\ref{sec:conclusions} we summarize our results.

\section{Methods}\label{sec:eq}

We implement the ohmic diffusion term in \arepo\ for two different numerical
MHD techniques. The first one~\citep{Pakmor2011, Pakmor2013} evolves the MHD
equations using the~\citet{Powell1999} 8-wave approach to control divergence
errors. The second method~\citep{Mocz2014, Mocz2016} implements the CT 
technique in \arepo, which has the advantage of enforcing the
$\nabla\cdot\boldsymbol{B} = 0$ constraint to machine precision.  The CT scheme
in \arepo\ evolves the cell-centred magnetic vector potential rather than a
face-centred magnetic field. For the implementation of the ohmic diffusion term
we restrict ourselves to a constant gas resistivity although this can easily be 
extended to the case of a spatially varying resistivity.  Finally, for each MHD
scheme, we present an explicit and implicit time integration method of the
ohmic diffusion terms, \FM{discussing only the MHD equations directly affected 
by the introduction of such terms,} as described in the following subsections. 
\FM{The interested reader can find a complete description of the MHD treatment in \arepo\ 
in method papers cited above.}

\subsection{Explicit time integration}\label{sec:explicit}

In the limit of spatially constant gas resistivity $\eta$ the induction
equation is given by\footnote{Throughout the paper, we express magnetic field
intensities in the Lorentz--Heaviside system of units.}
\begin{equation}
\frac{\partial\boldsymbol{B}}{\partial t} - 
\nabla\times(\boldsymbol{v}\times\boldsymbol{B}) - \eta\nabla^2\boldsymbol{B} = 0,
\label{eq:induction}
\end{equation}
or in terms of the vector potential $\boldsymbol{B} = \nabla \times
\boldsymbol{A}$, under the Coulomb gauge $\nabla \cdot \boldsymbol{A} \equiv
0$:
\begin{equation}
\frac{\partial\boldsymbol{A}}{\partial t} -
(\boldsymbol{v}\times\boldsymbol{B}) - \eta\nabla^2\boldsymbol{A} = 0.
\label{eq:inductionA}
\end{equation}
A non-zero resistivity $\eta$ further modifies the energy conservation equation to
\begin{equation}
\frac{\partial (\rho e)}{\partial t} + \nabla \cdot \left\{(\rho e + p) \boldsymbol{v}  - (\boldsymbol{v} \cdot \boldsymbol{B}) \boldsymbol{B}
+ \eta(\boldsymbol{J}  \times \boldsymbol{B}) \right\} = 0.
\label{eq:energy}
\end{equation}
In the previous equations $\rho$ is the gas density, $e$ the gas total energy 
per unit mass, $P$ the gas pressure, $\boldsymbol{v}$ the gas velocity, 
$\boldsymbol{B}$ the magnetic field, $\boldsymbol{J} = \nabla \times 
\boldsymbol{B}$, and the term $\eta(\boldsymbol{J}  \times \boldsymbol{B})$ 
represents the heat added to the system due to the dissipation of the magnetic 
field through ohmic resistivity.

Equations (\ref{eq:induction})-(\ref{eq:energy}) can be integrated in time in
an explicit way by adding the contribution of the ohmic diffusion terms to the
ideal MHD fluxes. We first focus on the induction equations. The diffusive
terms have the form 
$ \nabla \cdot \boldsymbol{F_{\rm d}}$, where 
\begin{equation}
 \boldsymbol{F_{\rm d}} = 
\begin{cases}
-\eta\nabla\boldsymbol{B}\\
 -\eta\nabla\boldsymbol{A}.
\end{cases}
\end{equation}
For a finite volume discretization the flux across a face shared by the 
mesh generating points $i$ and $j$ becomes after the application of Gauss' theorem
\begin{equation}
 \boldsymbol{F_{\rm d}} = 
\begin{cases}
-\eta\displaystyle\frac{\boldsymbol{B_i} - \boldsymbol{B_j}}{r_{ij}} a_{ij}\\
-\eta\displaystyle\frac{\boldsymbol{A_i} - \boldsymbol{A_j}}{r_{ij}} a_{ij},
\end{cases}
\label{eq:inductionflux}
\end{equation}
where $\boldsymbol{B_i}$, $\boldsymbol{A_i}$ are the time-extrapolated values
of the magnetic field or the magnetic vector potential of cell $i$,
$r_{ij}$ is the distance between the mesh-generating points and $a_{ij}$ is the
area of the face. The expressions of equation~(\ref{eq:inductionflux}) are then
added to their ideal MHD counterpart before the flux limiting procedure and the
time evolution of the system is applied.

For the ohmic heating term in the energy equation~(\ref{eq:energy}) the procedure is
similar. \FM{To achieve second order accuracy in the computation of the heat flux 
across a given face shared between the mesh generating points $i$ and $j$,
we first compute the term $\eta(\boldsymbol{J} \times \boldsymbol{B})$ for the two cells and we subsequently take 
their average as the resulting second-order flux}. The expression for the heat flux is 
\FM{thus} given by
\begin{equation}
 \eta\displaystyle\frac{(\boldsymbol{J_i}\times\boldsymbol{B_i}) + (\boldsymbol{J_j}\times\boldsymbol{B_j})}{2}
 \cdot\frac{\boldsymbol{r_{ij}}}{r_{ij}}a_{ij},
 \label{eq:heatflux}
\end{equation}
with the symbols having the same meaning as in equation~(\ref{eq:inductionflux}).
\FM{Please note that other second-order discretizations are possible for these terms. For instance, the cross product 
can be taken \textit{after} the average values of $\boldsymbol{J}$ and $\boldsymbol{B}$ are 
evaluated. However, tests on the propagation of an Alfv\'en wave (see Sec.~\ref{sec:alfven} for the set-up of the test) 
show that the results obtained with this latter scheme are equivalent to the ones given by 
equation~(\ref{eq:heatflux}).}

The relative simplicity of explicit schemes has made them a popular choice 
in most of the available implementations of non-ideal MHD terms 
\citep[e.g.][]{Masson2012}. However, the major drawback of explicit schemes is 
the rather restrictive time-step criterion that must be imposed for the scheme to 
be numerically stable. We enforce this by limiting the time-step of any given gas 
cell to
\begin{equation}
 \Delta t = \min\left(\Delta t_{\rm MHD}, \frac{\xi \Delta r^2}{\eta}\right),
 \label{eq:explicittstep}
\end{equation}
where $\Delta t_{\rm MHD}$ is the time-step computed for the ideal MHD part of
the calculation and the second term is the diffusive time-step that is
composed of a pre-factor $\xi = 0.2$, the fiducial cell radius $\Delta r$, computed as
the radius of the sphere having the same volume as the Voronoi cell (or circle having the same area 
for two-dimensional (2D) configurations; in case of 1D Voronoi tessellations 
it is the cell size), and the
ohmic diffusion coefficient $\eta$. The quadratic dependence on the cell size,
contrary to the linear dependence in the case of the ideal MHD
timestep criterion, renders the explicit non-ideal MHD scheme computationally expensive
for high-resolution simulations.

\subsection{Implicit time integration}\label{sec:implicit}
The intrinsic timestep limitations of explicit time integration methods can
be avoided by employing an implicit scheme that does not request such a
stringent timestep criterion. We follow the implementation presented
in~\citet{Kannan2016,Kannan2017}, where an implicit scheme for anisotropic heat diffusion
has been presented.  The implementation of ohmic diffusion is simplified by the fact
that the ohmic diffusion equations are isotropic such that many of the aspects
described in \citet{Kannan2016}, like the slope limiting procedure of the
transverse diffusion fluxes, are not required in our case.

We start from the discretized form of equation~(\ref{eq:induction}) -- the case of 
equation~(\ref{eq:inductionA}) follows naturally by replacing the magnetic field with the vector potential -- 
in a finite volume sense by considering only the diffusive terms.
After applying Gauss' theorem for cell $i$ this can be cast into the form 
\begin{equation}
\frac{\partial\boldsymbol{B_i}}{\partial t} = \frac{\eta}{V_i}\sum_{j \neq i} \frac{\boldsymbol{B_j} - \boldsymbol{B_i}}{r_{ij}}a_{ij},
\label{eq:inductionimpl}
\end{equation}
where the index $j$ runs over all the neighbours of cell $i$ and the meaning of the 
symbols is the same as in the previous equations ($V_i$ is the volume of the $i$-th cell).

To advance equation~(\ref{eq:inductionimpl}) in time we use two methods. The first one is a first-order backwards
Euler discretization, which we can write as
\begin{equation}
\frac{\boldsymbol{B_i}^{t+\Delta t} - \boldsymbol{B_i}^t}{\Delta t} = \frac{\eta}{V_i}\sum_{j \neq i} 
\frac{\boldsymbol{B_j}^{t+\Delta t} - \boldsymbol{B_i}^{t+\Delta t}}{r_{ij}}a_{ij}.
\label{eq:backeuler}
\end{equation}
To solve equation~(\ref{eq:backeuler}) we recast it in the form
\begin{equation}
\begin{split}
\boldsymbol{B_i}^{t+\Delta t} - \Delta t \sum_{j} 
M_{ij}(\boldsymbol{B_j}^{t+\Delta t} - \boldsymbol{B_i}^{t+\Delta t}) = \boldsymbol{B_i}^t,
\end{split}
\label{eq:backeuler2}
\end{equation}
where $M_{ij}$ is a matrix with elements
\begin{equation}
M_{ij} = \begin{cases}
\displaystyle\frac{\eta a_{ij}}{V_i r_{ij}} & {\rm if} \,\,\, i \neq j\\
0 & {\rm if} \,\,\, i = j 
\end{cases}.
\end{equation}
Equation~(\ref{eq:backeuler2}) is a linear vector equation for the
three components of the magnetic field. We only focus on a generic component,
but the same procedure applies similarly to the other components as well. We 
rewrite equation~(\ref{eq:backeuler2}) for a generic component of the field in the
$i-$th cell $B_i$ as 
\begin{equation}
\begin{split}
B_i^{t+\Delta t} - \Delta t \sum_j 
M_{ij}(B_j^{t+\Delta t} - B_i^{t+\Delta t}) = B_i^t,
\end{split}
\label{eq:backeuler3}
\end{equation}
which, following the same procedure discussed in~\citet{Kannan2016},
can also be written in the form
\begin{equation}
\begin{split}
\sum_{j} \left[\delta_{ij} \left(1+\Delta t \sum_{k} M_{ik} \right) - \Delta t M_{ij} \right] B_j^{t + \Delta t} = B_i^t , \\
\end{split}
\label{eq:implicitfinal}
\end{equation}
which is in the generic matrix form 
\begin{equation}
\boldsymbol{C} \boldsymbol{B} = \boldsymbol{B}_0.
\end{equation}
\begin{figure*}
\centering
\includegraphics[width=0.33\textwidth]{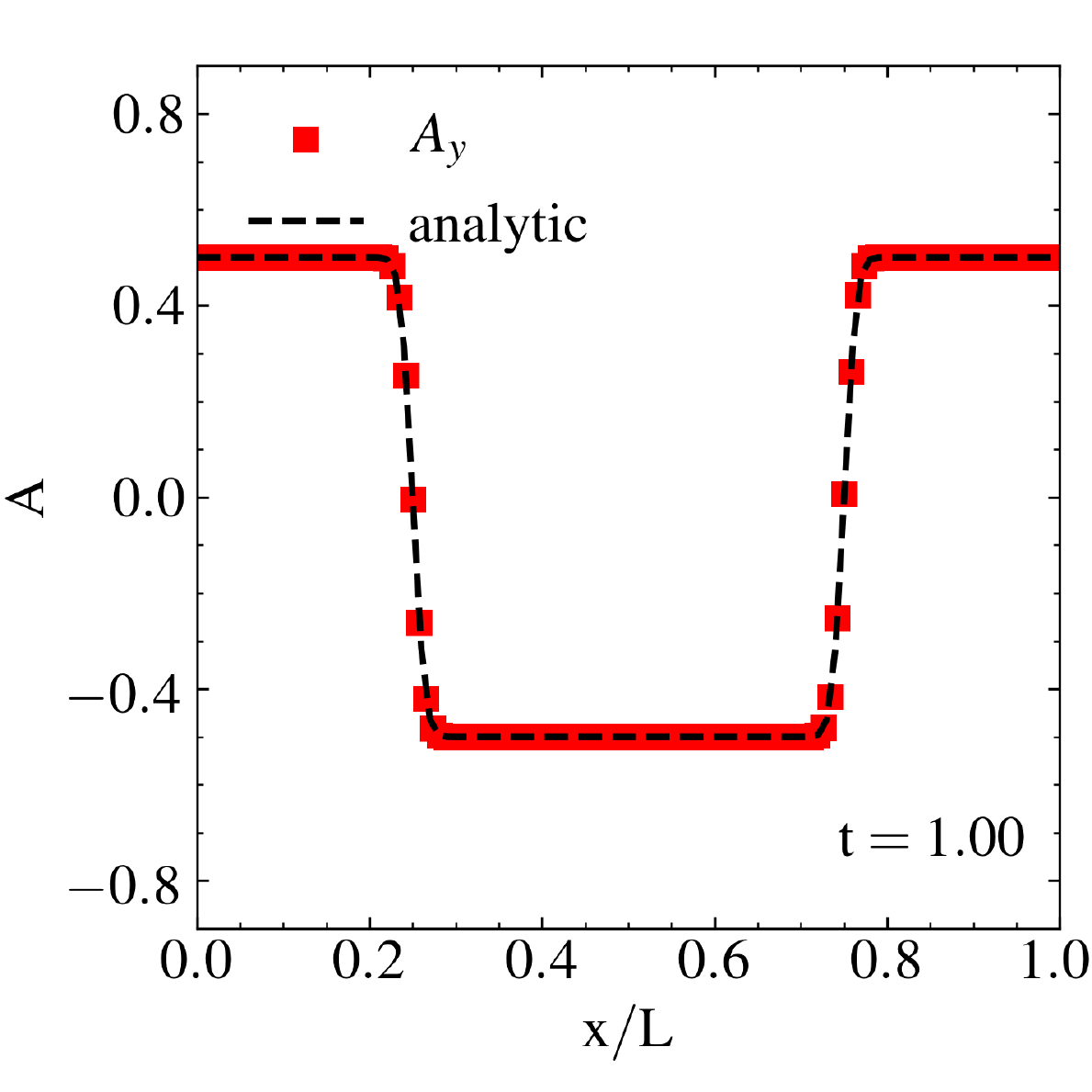}
\includegraphics[width=0.33\textwidth]{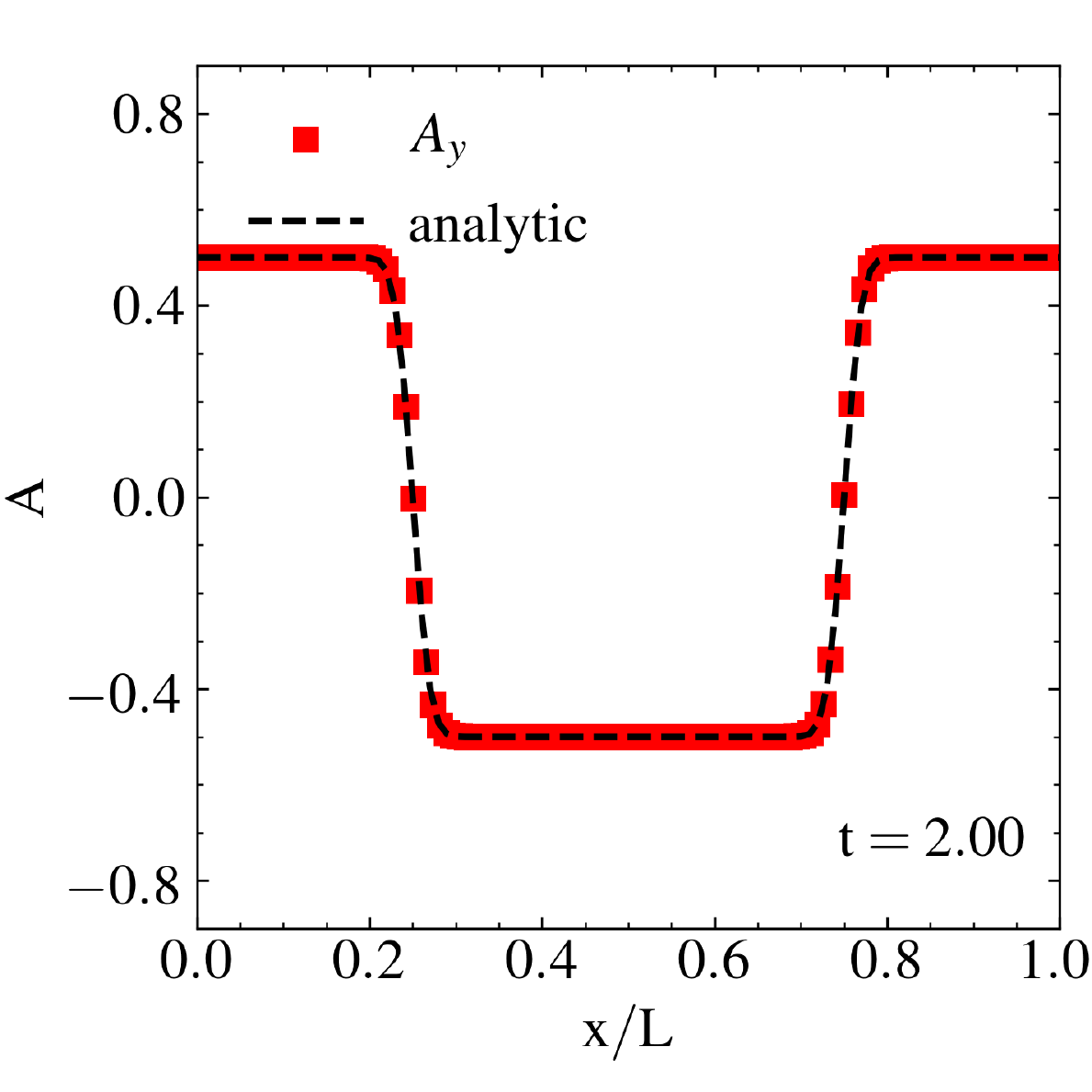}
\includegraphics[width=0.33\textwidth]{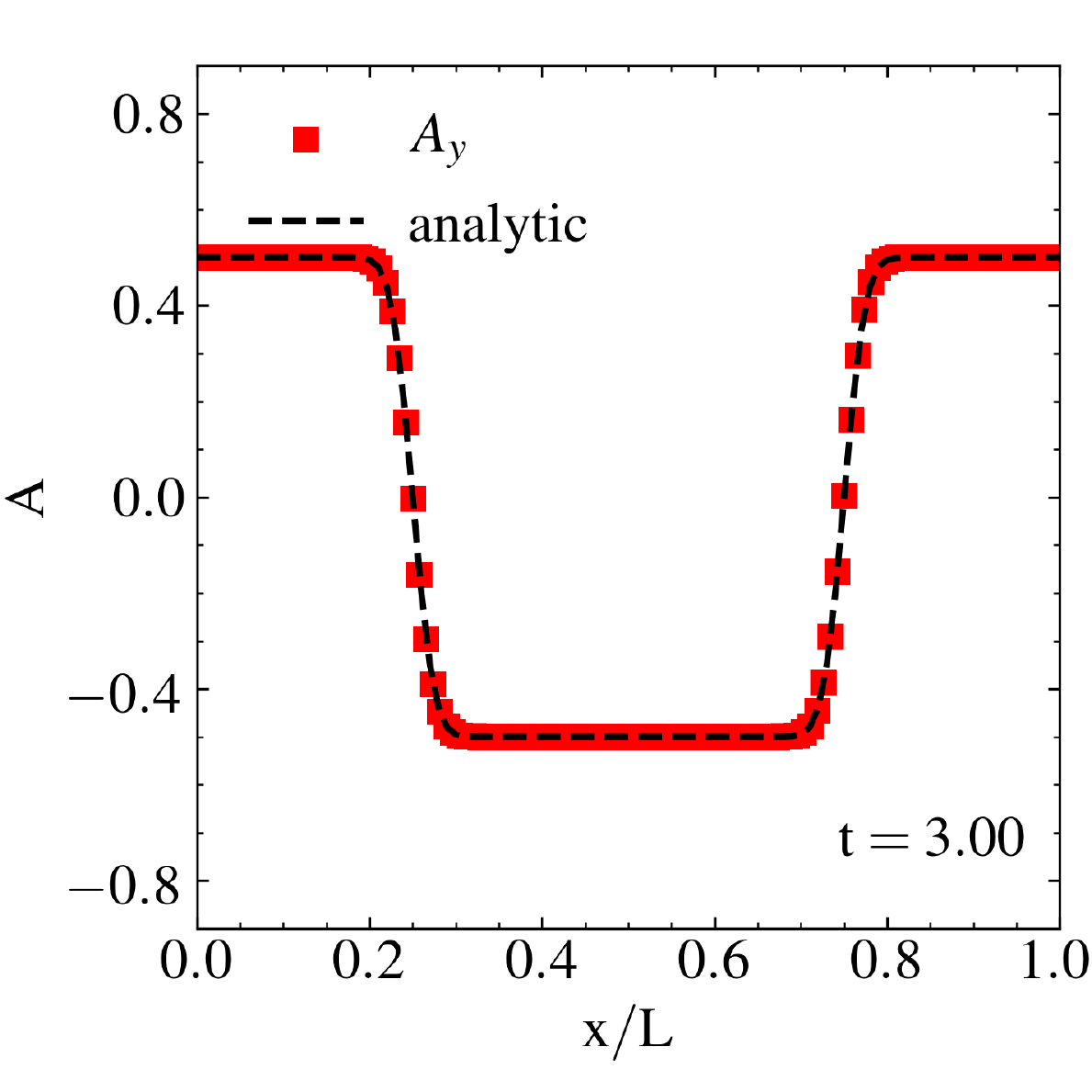}
\includegraphics[width=0.33\textwidth]{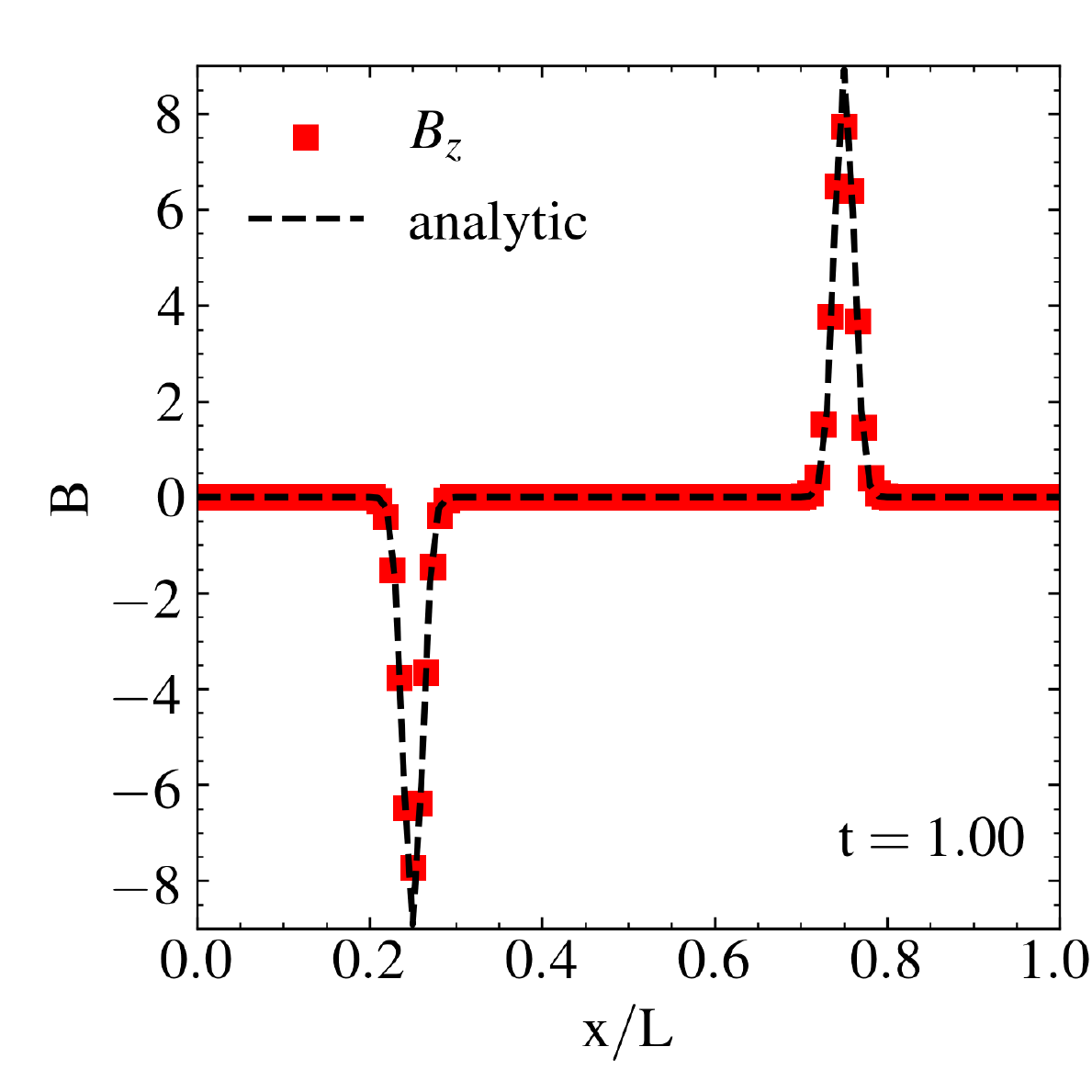}
\includegraphics[width=0.33\textwidth]{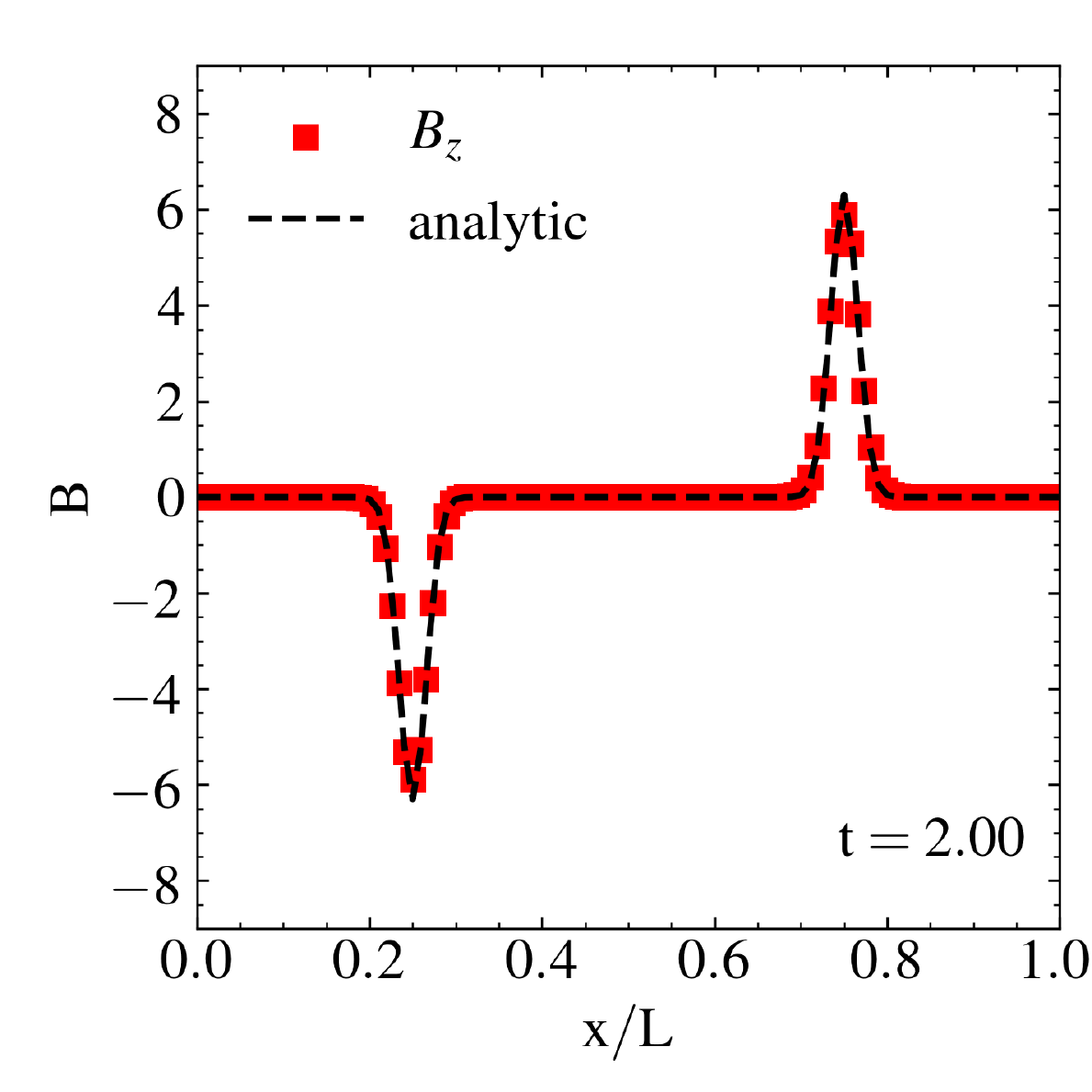}
\includegraphics[width=0.33\textwidth]{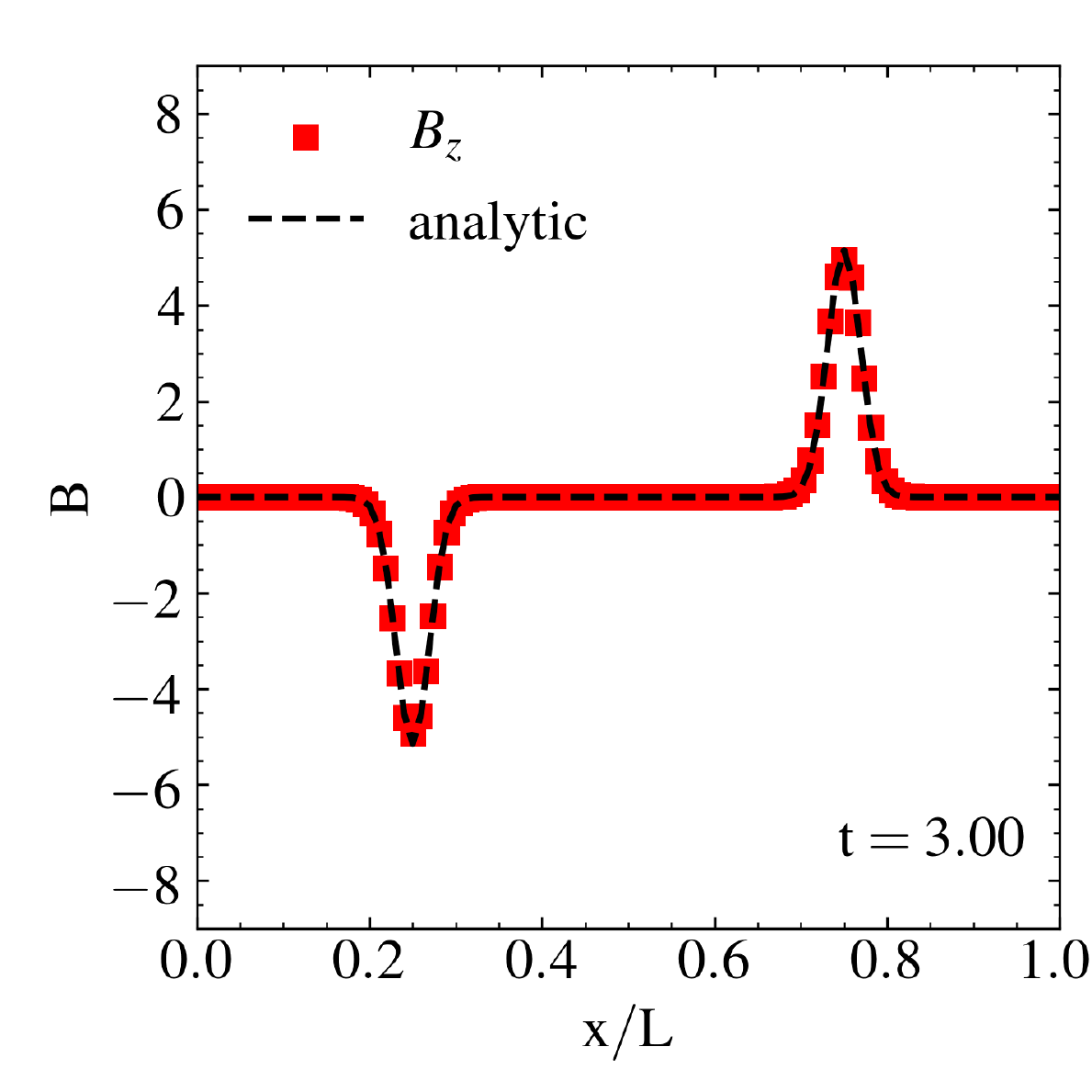}
\caption{Time evolution of the diffusion of a 1D Gaussian magnetic field pulse
with the implicit CT scheme. Both the evolution of the magnetic vector potential 
(top rows) and the associated magnetic field (bottom rows) are shown and compared to the 
analytic solution (dashed line). Time increases from the left-hand to right-hand side and is in 
units of the initial time ($t_0 = 10^{-3}$).
}
\label{fig:1Dpulse}
\end{figure*}
This linear system can efficiently be solved via standard linear parallel 
solvers. To this end, we employ the {\sc hypre}\footnote{\url{http://acts.nersc.gov/hypre}} 
library with the the generalized minimal residual (GMRES) iterative method~\citep{Saad1986} and an 
\FM{algebraic} multigrid pre-conditioner \FM{\citep{Henson2002}}. We 
use a tolerance limit of $\epsilon_{\rm tol} = 10^{-10}$ for the GMRES solver.
\FM{It can be easily shown that the matrix $\boldsymbol{C}$ is 
(strictly) diagonally dominant. In the peculiar case of a structured (and 
static) mesh, in which the volume of each resolution element remains the same, 
the matrix is also symmetric and positive definite. This is the usual configuration in 
most of the test problems (see Section~\ref{sec:tests}). In these configurations the matrix 
$\boldsymbol{C}$ will be well-conditioned and indeed convergence is 
reached after a few ($\sim 4$ maximum) iterations of the GMRES solver. The 
conditioning properties of the matrix (and in particular strict diagonal 
dominance) are independent of the size of the time step. A direct estimation of
the condition number of the matrix is difficult, but it can 
be assumed that for more distorted mesh topologies the matrix will become progressively 
more ill-conditioned. However, we would like to note that Arepo employs mesh 
regularization techniques that prevent unwanted and excessive mesh twisting and 
tangling \citep[see][]{Arepo}, which therefore should also limit 
the magnitude of the condition number of the matrix $\boldsymbol{C}$}. 

For improved 
accuracy we have also implemented a second-order Crank--Nicholson 
scheme~\citep{Crank1947}. This can efficiently be implemented by considering 
\begin{equation}
\begin{split}
B_i^{t+\Delta t} - \frac{\Delta t}{2} \sum_j 
M_{ij}(B_j^{t+\Delta t} - B_i^{t+\Delta t}) = \tilde{B}_i^t,
\end{split}
\label{eq:implicitcn}
\end{equation}
where the right-hand side of equation~(\ref{eq:implicitcn}) reads
\begin{equation}
\displaystyle\tilde{B}_i^t = B_{i}^t +  \frac{\Delta t}{2} \sum_{j} M_{ij}(B_j^t - B_i^t).
\label{eq:explicitcn}
\end{equation}
We then solve the resulting linear system with the same iterative method used 
for the first-order Euler scheme. We note that contrary to the simple backwards 
Euler, which is our default choice, the Crank-Nicholson scheme 
may induce slowly decaying oscillations to the solution if the timestep is too 
large. To avoid the appearance of this, we limit 
the timestep according to equation~(\ref{eq:explicittstep}) with safety factor $\xi < 0.5$ as derived by a von 
Neumann stability analysis. We point out that this procedure is performed only for 
the Crank-Nicholson scheme. In our simulations, should this timestep limitation become too 
computationally expensive, we resort to the more robust, but less accurate, 
Euler implicit time integration in which no timestep limitation is present. With this approach our implicit
treatment, unlike the explicit scheme, does never suffer from a too severe timestep constraint that
would otherwise prevent performing high-resolution simulations of non-ideal MHD effects.  

Finally, in the implicit integration scheme, the ohmic heating term is
directly added to the gas thermal energy before diffusing the magnetic field
or vector potential according to the equation
\begin{equation}
\frac{\partial u}{\partial t} = \eta \frac{||\boldsymbol{J}||^{2}}{\rho}, 
\end{equation}
where $\boldsymbol{J} = \nabla\times\boldsymbol{B}$, $\rho$ is the gas density and $u$
its thermal energy per unit mass.
In particular, the new value for $u_{i}$ for each cell is computed as 
\begin{equation}
u_{i}^{t + \Delta t} = u_{i}^{t} + \Delta t \eta \frac{||\boldsymbol{J}_{i}^{t}||^{2}}{\rho_{i}^{t}}.
\label{eq:impl_erg}
\end{equation}
After the new magnetic field and internal energy values have been computed, the 
gas total energy and pressure are updated \FM{self-consistently.
It is worth pointing out that the treatment for the ohmic heating term presented implies that the scheme
is not strictly conservative and this can have an effect on the evolution of the simulated systems
(see also Fig.~\ref{fig:magrecBy}).}
The implicit time 
integration schemes described above are used only on global 
time-steps~\citep[see][for details]{Kannan2016}, which in combination with a 
less restrictive limitation on the time-step renders them significantly more 
efficient than their explicit counterpart, especially for non-ideal MHD 
applications in which resistivity effects become dominant.

\section{Test Problems}\label{sec:tests}

In this section, we test the implementation of the ohmic resistivity terms in 
\arepo\ on a series of problems of increasing complexity. For each problem, we 
present the initial conditions for the magnetic field and the vector potential, 
which is required for the initialization of the CT scheme. 
\FM{In all the test problems periodic boundary conditions will be imposed. In 
the case of the CT scheme, which evolves the vector potential, while the 
magnetic field is periodic the vector potential need not be. The mean magnetic 
field in the domain leads to a discontinuity in the vector potential across 
periodic boundaries. Therefore, the treatment of the vector potential in the CT 
scheme \citep{Mocz2016} involves decomposing the vector potential into two 
parts by defining a periodic component of the vector potential and a component 
associated with the mean-field, which is discontinuous across boundaries. This 
is necessary as a discontinuity in the vector potential leads to an infinite 
value of the associated magnetic field with catastrophic effects on the runs. 
Finally, also in the case of} ohmic diffusion the mean magnetic field in the 
simulated volume is a conserved quantity originating from a \FM{(non-periodic)} 
vector potential static in time. Therefore, it is not necessary to evolve this 
part of the vector potential in time, and the \FM{associated} mean magnetic 
field is \FM{simply} added to the cell-centred field tracked by the simulation 
at the end of each CT mapping step~\citep[see][for details]{Mocz2016}.

\subsection{Gaussian pulse}

We first test the implementation of resistive MHD terms in the simplified case
where the dynamics of the gas is not followed. This is equivalent to assuming
$\boldsymbol{v}\equiv\boldsymbol{0}$ at all times. The MHD equations
then reduce to
\begin{equation}
\frac{\partial\boldsymbol{B}}{\partial t} = \eta\nabla^2\boldsymbol{B}.
\label{eq:diffusion}
\end{equation}
Mathematically, equation (\ref{eq:diffusion}) is an isotropic diffusion equation
with diffusion coefficient $\eta$ for each of the component of the magnetic  
field. A similar equation also holds for the vector potential $\boldsymbol{A}$ in the CT scheme. 

\subsubsection{1D Gaussian pulse} \label{sec:1Dpulse}
To further reduce the complexity of the problem, we first simulate the diffusion of a 1D Gaussian pulse:
\begin{equation}
\boldsymbol{B}(\boldsymbol{x}) = \delta(x)\hat{e}_{z},
\end{equation}
where $\delta(x)$ is the Dirac delta function.  The solution of this initial
value problem at time $t$ is the 1D heat kernel function
\begin{equation}
\boldsymbol{B}(\boldsymbol{x},t) = \frac{1}{\sqrt{4\pi\eta t}}\exp{\left(-\displaystyle\frac{x^2}{4\eta t}\right)}\,\hat{e}_{z}.
\label{eq:gaussian1D}
\end{equation}

To initialize this test, we sample equation (\ref{eq:gaussian1D}) with $128$ 
resolution elements at the initial time $t_0 = 10^{-3}$ and we assume
$\eta = 1$. The test is carried out on the 1D domain $[0,L]$ with
$L=4$.

For the CT scheme we adopt the following vector potential for this test
\begin{equation}
\boldsymbol{A}(\boldsymbol{x}) = \Theta(x)\hat{e}_{y},
\end{equation}
where $\Theta(x)$ is the Heaviside step function.  The solution of this initial
value problem at time $t$ is the error function
\begin{equation}
\boldsymbol{A}(\boldsymbol{x},t) = \frac{1}{2}{\rm erf}\left(\frac{x}{\sqrt{4\eta t}}\right)\hat{e}_{y}.
\label{eq:errfunct}
\end{equation}

\noindent In order to use periodic boundary conditions, and since
the ohmic diffusion operator is linear, we diffuse two of such steps by
starting from the initial conditions
\begin{equation}
\boldsymbol{A}(\boldsymbol{x}) = 
\begin{cases}\displaystyle
 \Theta(x - 0.75)\hat{e}_{y} & \mbox{if } x > 0.5\\\\
 \displaystyle
 \Theta(0.25 - x)\hat{e}_{y} & \mbox{if } x \leq 0.5.\\
\end{cases}
\label{eq:Astep}
\end{equation}
The time evolution of this vector potential gives rise to two Gaussian magnetic
fields of opposite polarity centred at $x = 0.25$ and $x = 0.75$, respectively.

Fig.~\ref{fig:1Dpulse} presents the results of this test for the initial 
conditions described in equation~(\ref{eq:Astep}) calculated with the implicit 
CT scheme\footnote{Unless otherwise stated the Crank--Nicholson scheme is used 
in this paper for the implicit time integration.} with a 1D grid of 
128 points. We chose this scheme since it formally requires the diffusion of a 
discontinuous step function for $t = 0$, and therefore better illustrates the 
robustness of our implementation. All the other implementations of ohmic 
diffusion perform equally well in this test problem.

\begin{figure}
\centering
\includegraphics[width=0.485\textwidth]{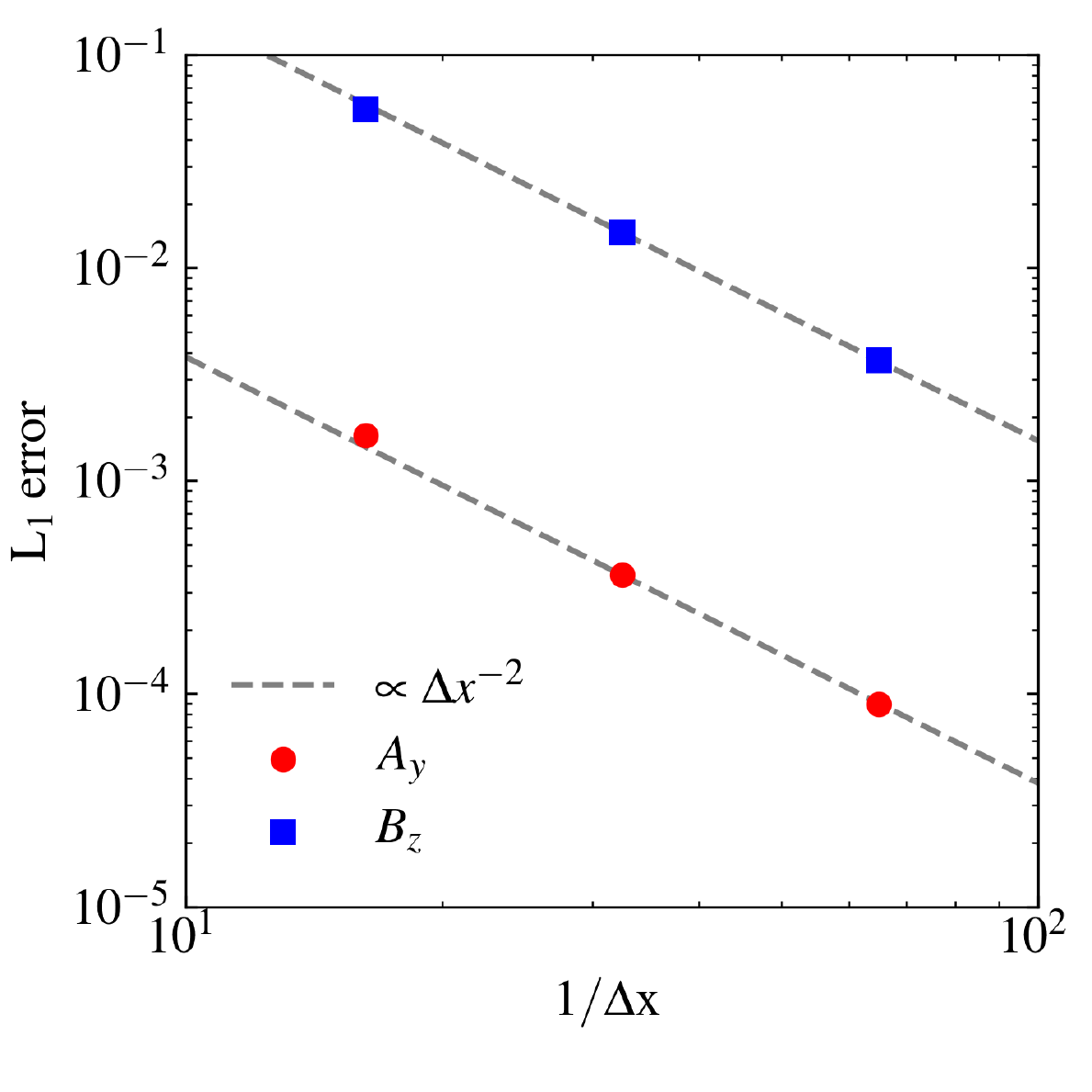}
\caption{$L_1$ norm of the error as a function of resolution for the vector 
potential (red circles) and magnetic field (blue squares) of the 1D diffusion 
test, performed with the implicit CT scheme, at time $t = 4 \times t_0$. The 
grey dashed lines represent the expected scaling for a second-order scheme. 
} 
\label{fig:1Dpulseconv}
\end{figure}

The panels show the evolution of the vector potential
(top rows) and associated magnetic field (bottom rows) at different times in units of the
initial time $t_0 = 10^{-3}$, indicated in the bottom right-hand corner of each
panel. Red squares represent the numerical solution, whereas the black dashed
lines represent the analytic solution. Our implementation correctly
captures the evolution of the vector potential and the associated magnetic
field even at the relatively low resolution used in this test problem. Only at the
locations of the maximum and minimum magnetic field (at $x=0.75$ and $x=0.25$,
respectively) the numerical values of the field are slightly underestimated
with respect to the analytic solution, which however can be cured by adopting
a higher resolution. This underestimation of the magnetic field intensities is less
pronounced or absent altogether at later times.

\begin{figure*}
\centering
\includegraphics[width=0.33\textwidth]{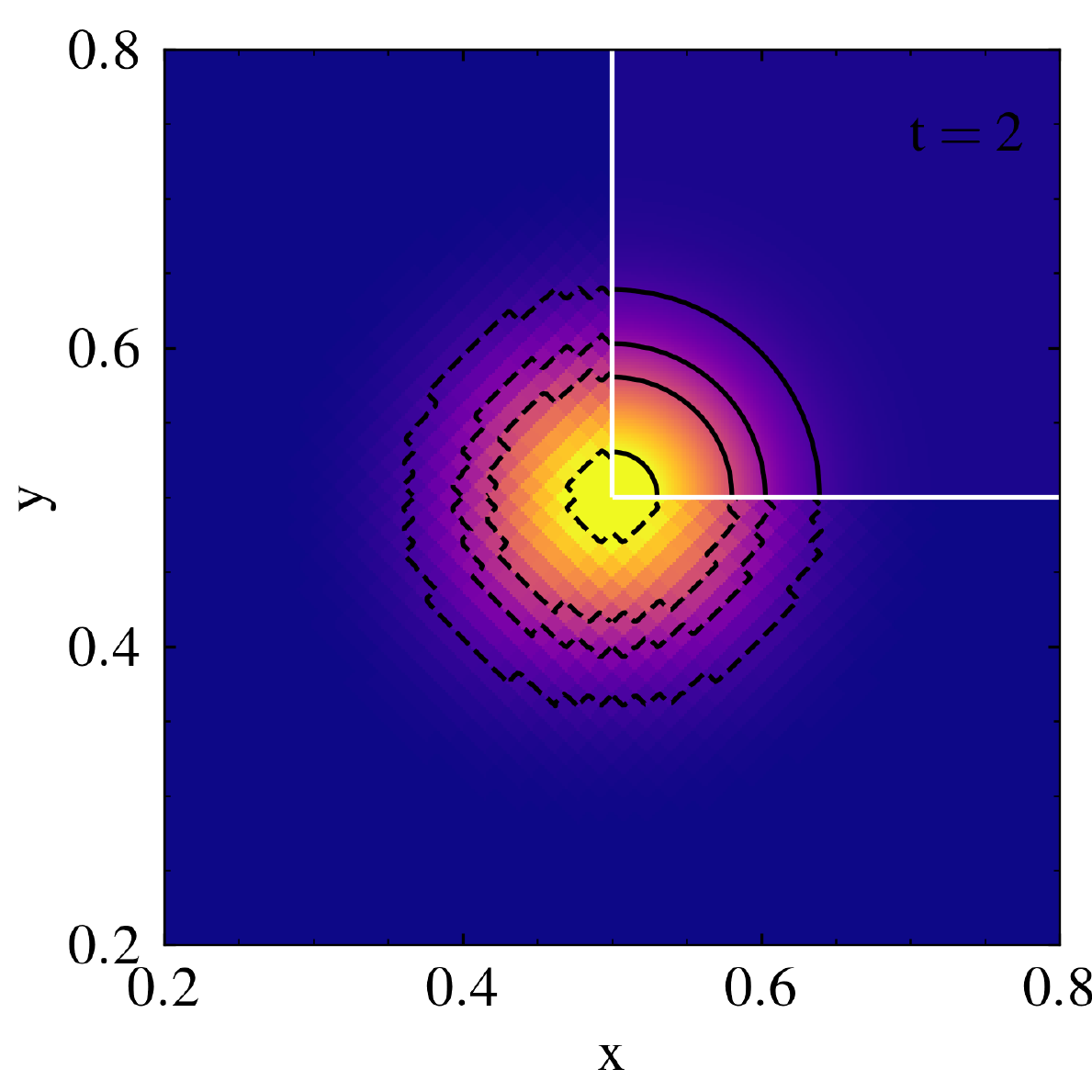}
\includegraphics[width=0.33\textwidth]{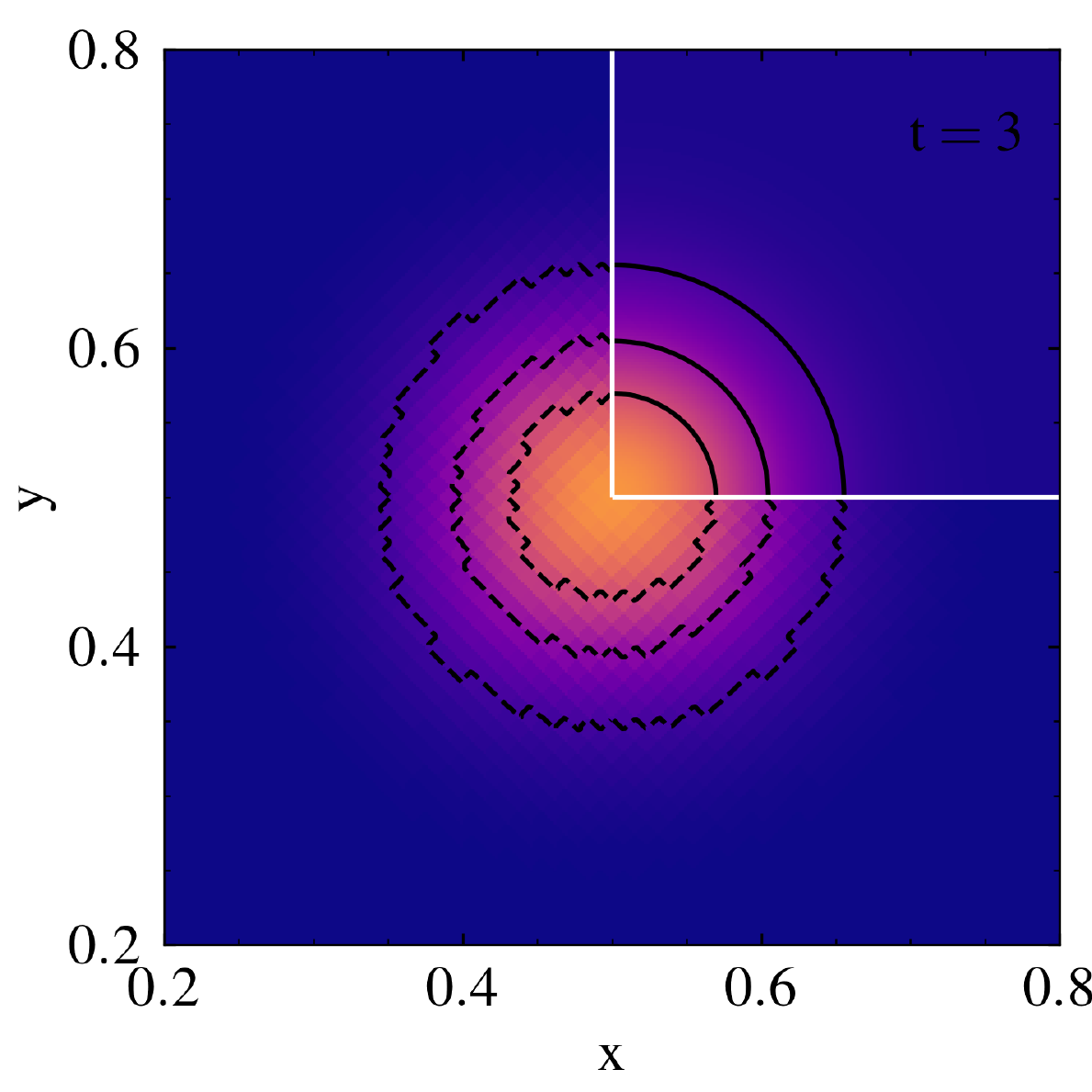}
\includegraphics[width=0.33\textwidth]{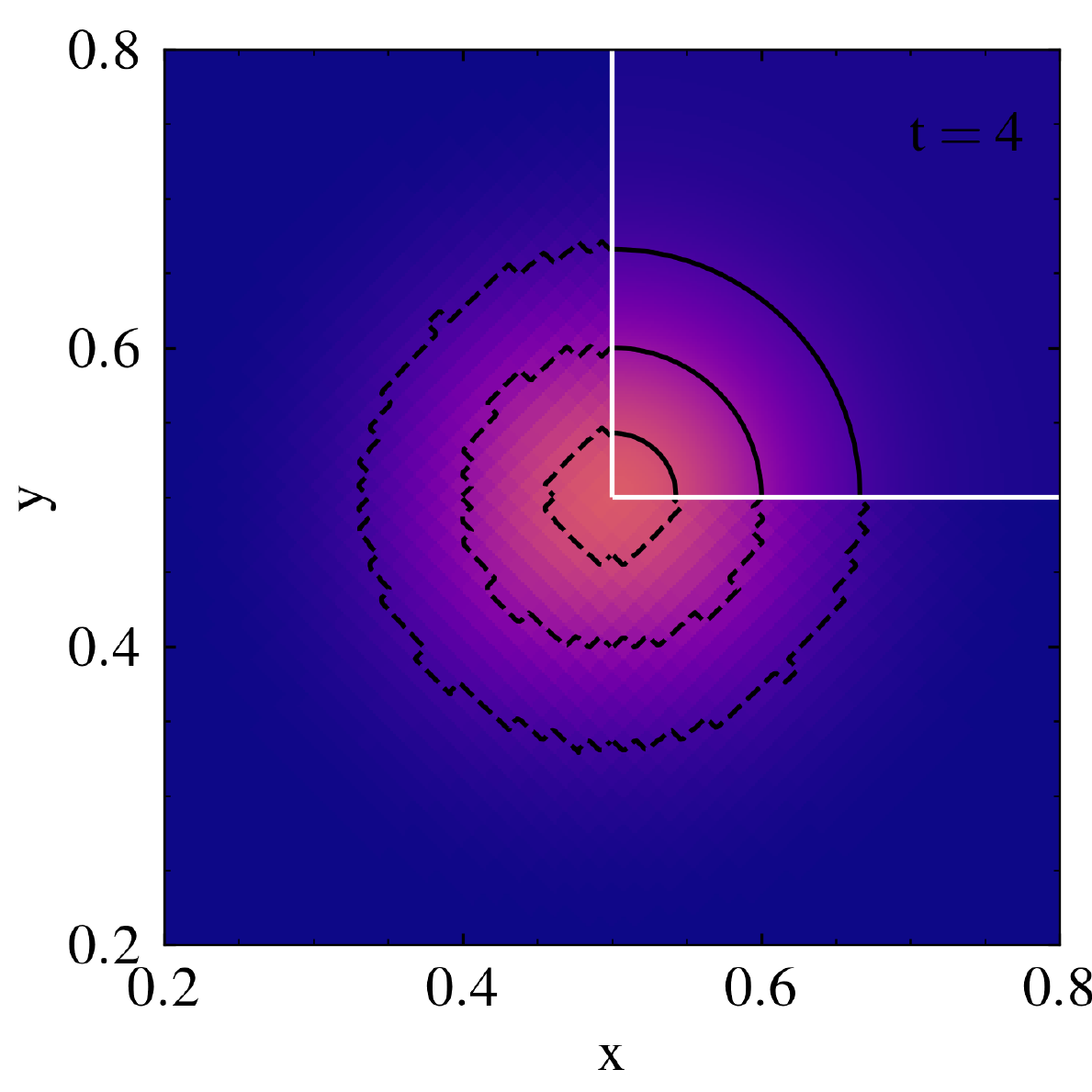}
                                       
\centering                             
\includegraphics[width=0.33\textwidth]{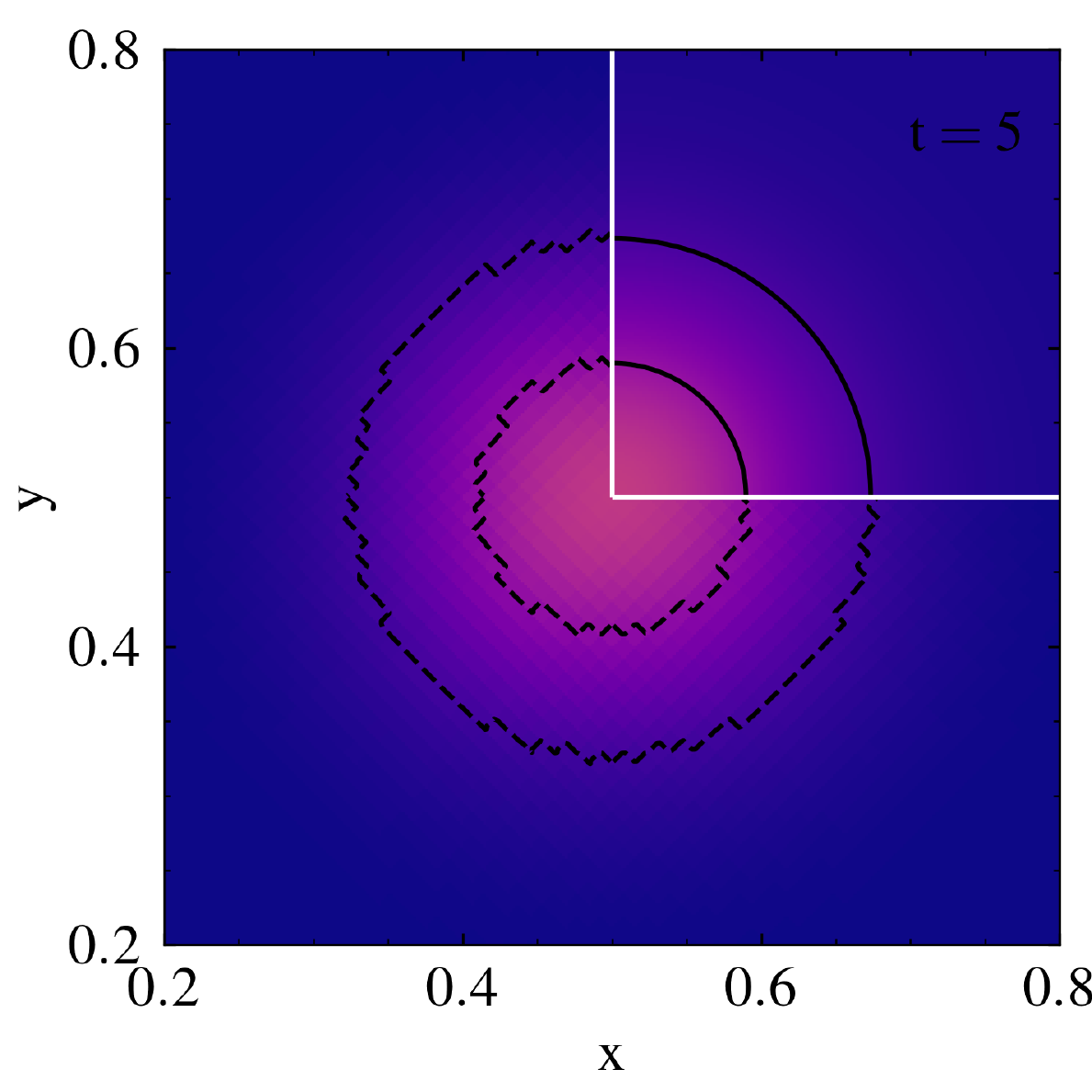}
\includegraphics[width=0.33\textwidth]{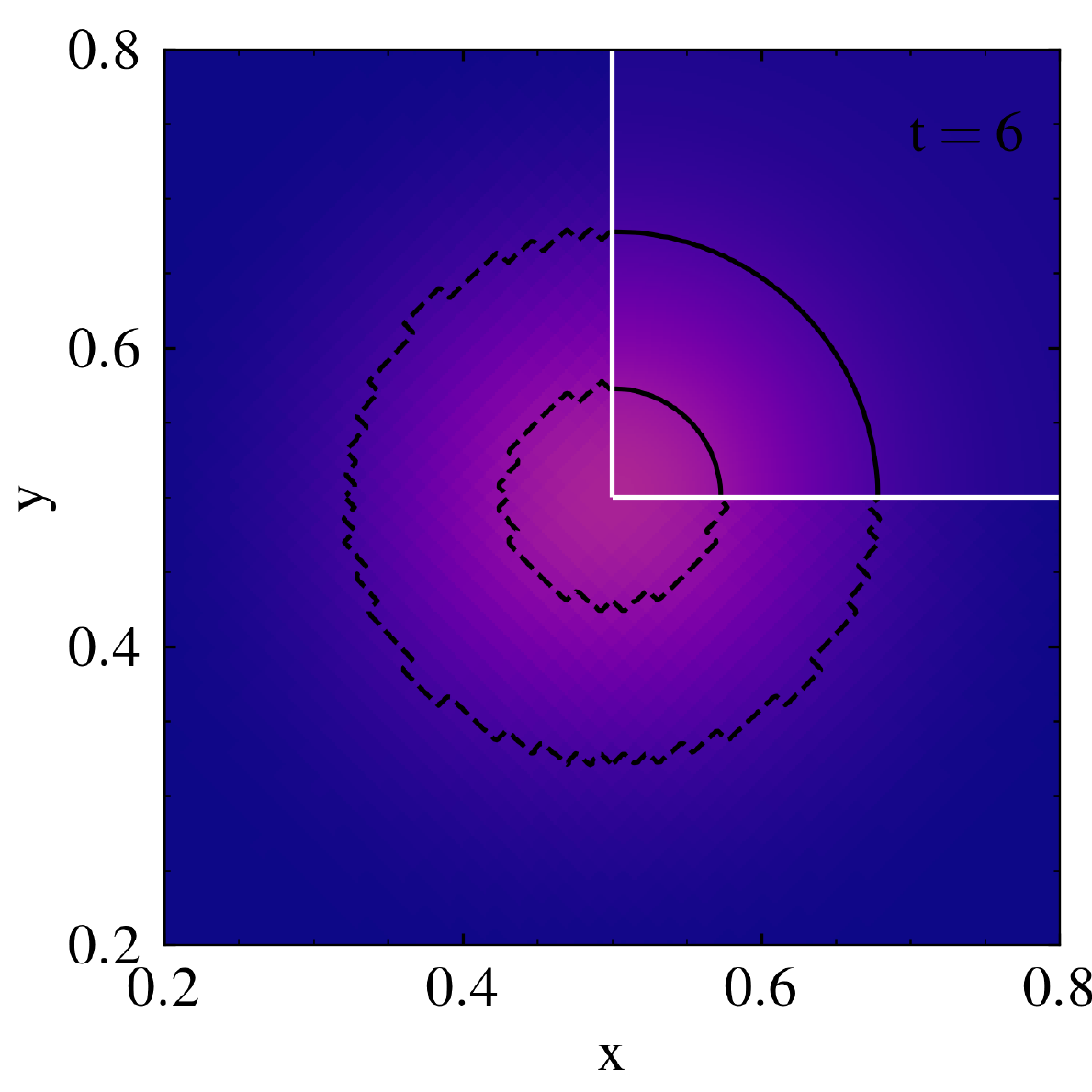}
\includegraphics[width=0.33\textwidth]{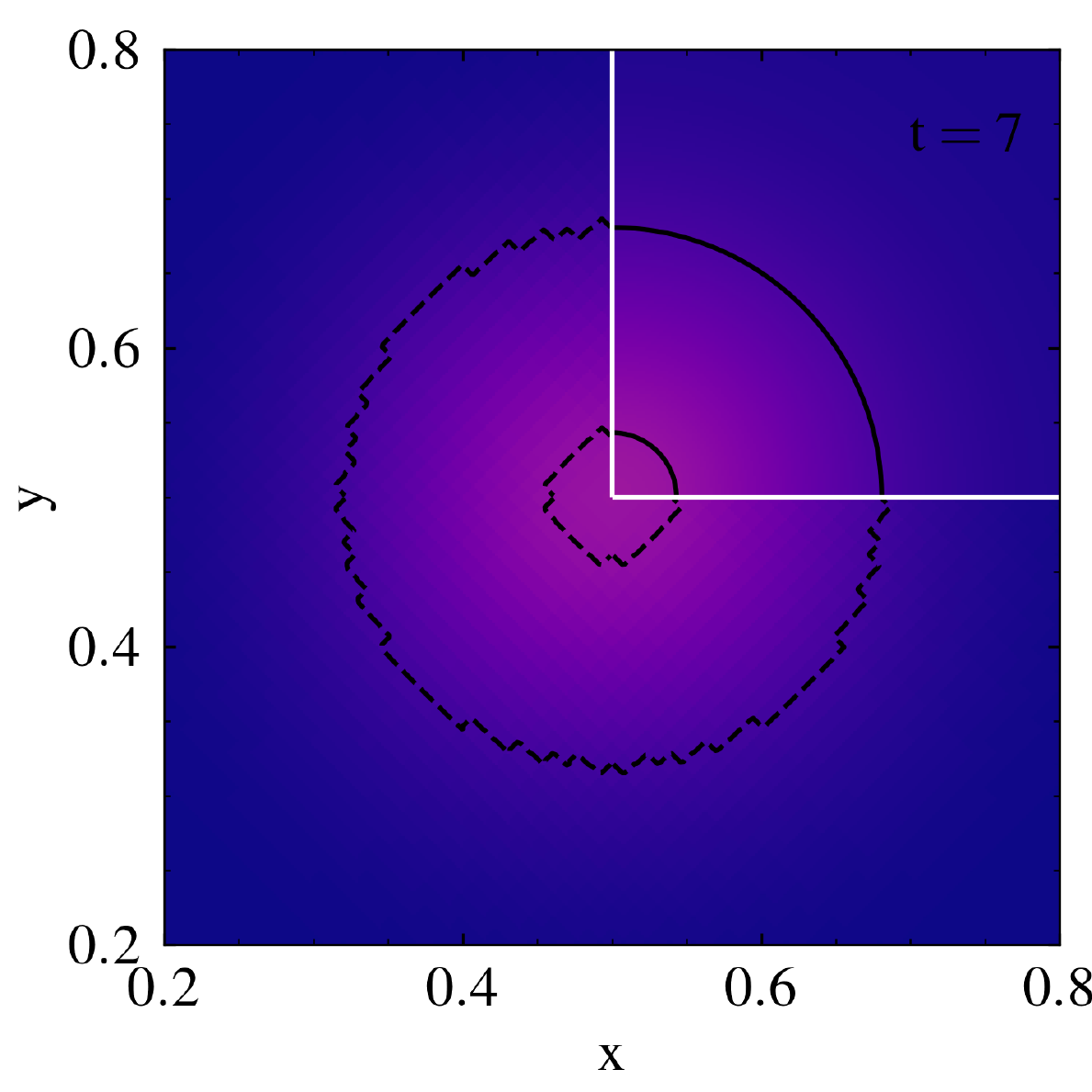}
\caption{Time evolution of the diffusion of a 2D Gaussian magnetic field pulse
evolved with the implicit Powell scheme. Contour levels are located at 10, 5, 3, and 1, from the innermost 
to the outermost. The upper right-hand quadrant of each panel shows the evolution of the analytic solution 
of this problem (see equation~\ref{eq:gaussian}). Time, in units of the initial time $t_0 = 10^{-3}$, 
increases from the left to right-hand side and from the top to bottom as indicated in the legend.
}
\label{fig:2Dpulse}
\end{figure*}

In Fig.~\ref{fig:1Dpulseconv} we assess more quantitatively the performance of our scheme by
showing the $L_p$ error computed as~\citep{Pakmor2016}
\begin{equation}
L_{p} = \displaystyle\frac{1}{V}\left(\sum_{i=0}^{N_{\rm cells}} |f_{i}|^{p} V_{i}\right)^{1/p},
\label{eq:Lerror}
\end{equation}
for the results presented in Fig.~\ref{fig:1Dpulse}. In 
equation~(\ref{eq:Lerror}), $V$ is the total simulated volume, $V_i$ is the 
volume of the $i-$th cell, and $f_{i}$ is the difference between the analytic 
and numerical solution in the cell $i$. In Fig.~\ref{fig:1Dpulseconv} we show 
the $L_1$ error ($p = 1$), for both the vector potential (red squares) and the 
magnetic field (blue circles), as a function of the mesh resolution expressed as 
the inverse of the mean cell size $1/\Delta x$. We note that we 
adopt these choices for stating the resolution in all similar figures quantifying the convergence of our 
schemes that we will present below. The mean cell size $\Delta x$ can be
computed as the radius of a sphere (circle) having the same volume (area) of a 
given cell for 3D (2D) configurations depending on 
the problem analysed. We note that finer resolution 
corresponds to larger values of $1/\Delta x$.
The grey dashed line represents the second-order scaling 
of the $L_1$ error expected for our schemes. We find that in this test 
problem the $L_1$ error follows exactly the scaling predicted for second-order 
convergence. Since the magnetic field is a derived quantity in the CT 
scheme~\citep[see][]{Mocz2016}, the amplitude of the $L_1$ error is larger than 
for the vector potential whose evolution is directly followed.

\subsubsection{2D Gaussian pulse}

\begin{figure}
\centering
\includegraphics[width=0.485\textwidth]{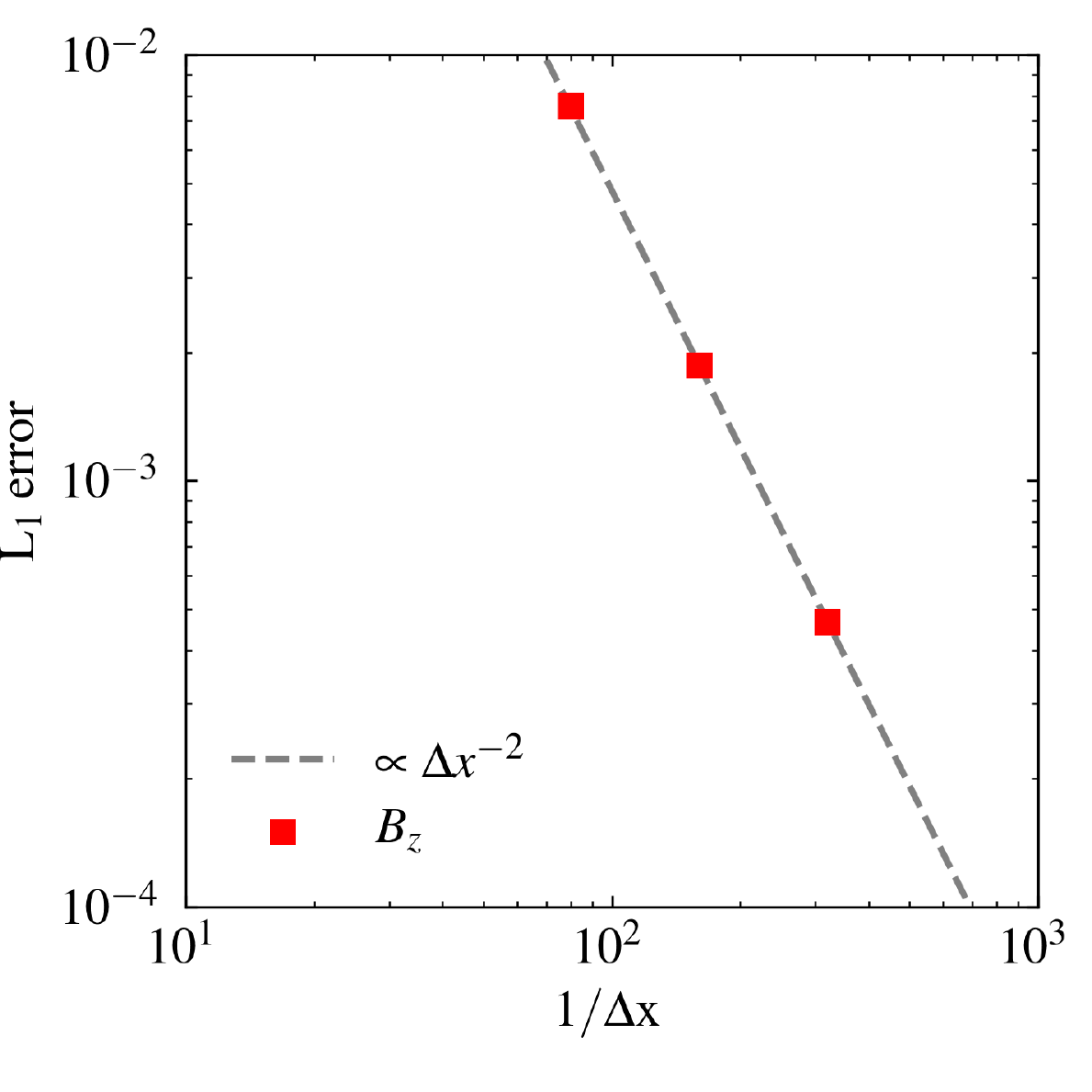}
\caption{$L_1$ norm of the error as a function of resolution for the magnetic field in the 2D 
diffusion test at time $t = 4 \times t_0$ performed with the implicit Powell scheme. 
The grey dashed line represents the expected scaling for a second-order scheme.
}
\label{fig:2Dpulseconv}
\end{figure}

The previous test problem assessed the accuracy of our ohmic diffusion scheme
in a 1D set-up. We now increase the dimensionality by using as initial
conditions a magnetic field of the form
\begin{equation}
\boldsymbol{B}(\boldsymbol{x}) = \delta(x)\delta(y)\hat{e}_{z}.
\end{equation}

\noindent The solution of this initial value problem at time $t$ is the
2D heat kernel
\begin{equation} 
\boldsymbol{B}(\boldsymbol{x},t) = \frac{1}{4\pi\eta 
t}\exp{\left(-\displaystyle\frac{x^2 + y^2}{4\eta t}\right)}\,\hat{e}_{z}.
\label{eq:gaussian} 
\end{equation} 
To initialize the
simulation, we sample equation (\ref{eq:gaussian}) at the initial time $t_0 =
10^{-3}$, and we assume $\eta = 1$. The test is carried out on the 2D domain
$[0,1]\times[0,1]$. \FM{The domain is partitioned in cells via a structured Voronoi mesh 
in which the mesh generating points are arranged in a 2D rhombic lattice. The lattice is 
built by using two interleaved Cartesian meshes of $32^2$ points, separated by half the cell 
spacing, for a total of $2\times 32^2$ resolution elements.}

An explicit expression for the vector potential corresponding to the magnetic
field presented in equation (\ref{eq:gaussian}) can be found in polar
coordinates
\begin{equation}
\boldsymbol{A}(\boldsymbol{x},t) = -\frac{1}{2\pi R}\exp{\left(-\displaystyle\frac{R^2}{4\eta t}\right)}\,\hat{e}_{\varphi},
\label{eq:A2Dgausspolar}
\end{equation}
or equivalently in Cartesian coordinates
\begin{equation}
\boldsymbol{A}(\boldsymbol{x},t) = -\frac{y\hat{e}_{x} - x\hat{e}_{y}}{2\pi(x^2+y^2)}\exp{\left(-\displaystyle\frac{x^2+y^2}{4\eta t}\right)}.
\label{eq:A2Dgausscart}
\end{equation}
To initialize the test for the CT scheme, we 
use again \FM{the same mesh of} $2\times 32^3$ resolution elements on the 2D domain
$[0,1]\times[0,1]$ \FM{adopted for the Powell scheme}. Equation~(\ref{eq:A2Dgausscart}) is sampled on this mesh at the 
the initial time $t_0 = 10^{-3}$, and we assume $\eta = 1$.
 
Fig.~\ref{fig:2Dpulse} illustrates the result of this test for the initial 
conditions described in equation~(\ref{eq:gaussian}) calculated with the 
implicit Powell scheme. We chose here the implicit Powell scheme, instead of the more complex implicit CT scheme, to demonstrate that our non-ideal MHD implementation
is also working for the cleaning scheme.
Again, we note that all the other schemes applied to this test problem essentially give the 
same results. The panels show the evolution of the magnetic field at different times 
(in units of the initial time $t_0 = 10^{-3}$) indicated in the 
top right-hand corner of each panel. The colour map shows the values of the field 
mapped linearly in the range [0; 10], whereas the contour lines are placed at the 
values 1, 5, 5 and 10 (from the outside in). The upper right-hand quadrant of each 
panel shows the analytic solution obtained from equation~(\ref{eq:gaussian}), 
whereas in the rest of the plot the numerical solution is presented.

The implicit treatment of ohmic diffusion with the Powell
scheme is able to correctly capture the evolution of the magnetic field
intensity with time. The diffusion of the field is visible in the panels as a
decrease of the central magnetic field strength as a function of time. The contour
levels clearly illustrate this trend. In particular, the highest contour
shrinks in size with time, as the magnetic field diffuses out, and 
disappears in the second panel. Similarly, the second highest contour shrinks in size 
and disappears at time $t = 5\times t_0$.  We note that, contrary
to the computation of the analytic solution, no smoothing of the simulation values has
been applied to produce this figure; i.e. the magnetic field values of the
cell closest to any given pixel has been assigned to that pixel. This has been done
on purpose to show the structure of the underlying rhombic mesh. The structure is made
more evident by the shape of the contour levels in the quadrants displaying the
numerical solution, which unlike the smooth circular analytic contours, present
a jagged shape along the cell boundaries. However, their spatial position is in
excellent agreement with the analytic expectations.

Fig.~\ref{fig:2Dpulseconv} presents the $L_1$ error as a function of the mesh resolution for the implicit
Powell scheme investigated in this test. The second-order convergence of the scheme, as indicated by 
the grey dashed line, is clearly visible. All the other schemes implemented in this
work show the same behaviour \FM{when they are coupled with a second-order accurate
time integrator}.

\subsection{Alfv\'en waves}\label{sec:alfven}
We now test our implementation of the ohmic diffusion term in the presence of
gas dynamics by studying the evolution of a circularly polarized Alfv\'en wave.
The resistivity term in equations (\ref{eq:induction}) and
(\ref{eq:inductionA}) causes the amplitude of the wave to decay exponentially,
whereas the ohmic dissipation term added to the energy equation~(\ref{eq:energy})
leads to an increase of the thermal energy content of the gas via Joule heating.
We test two cases: a progressive wave propagating along the negative $z$-direction
(Section~\ref{sec:progressive}) and a stationary wave obtained as the superposition
of two progressive waves propagating again along the $z-$axis but in opposite
directions (Section~\ref{sec:standing}). Both tests are presented for the implicit
CT scheme since the additional step needed to reconstruct the magnetic field from the 
diffused vector potential makes it a more complex problem to test compared to the Powell method. 

\begin{figure*}
\centering
\includegraphics[width=0.33\textwidth]{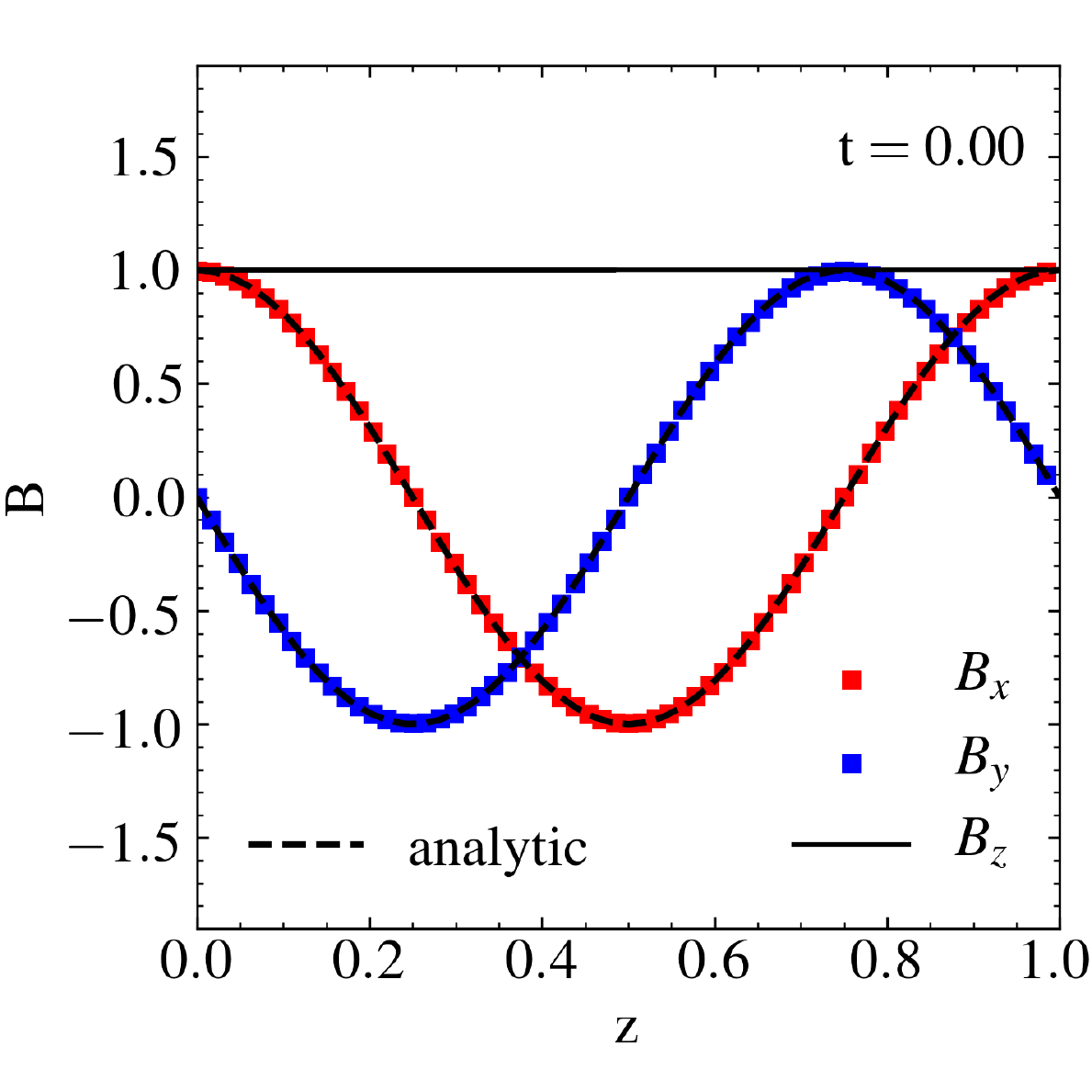}
\includegraphics[width=0.33\textwidth]{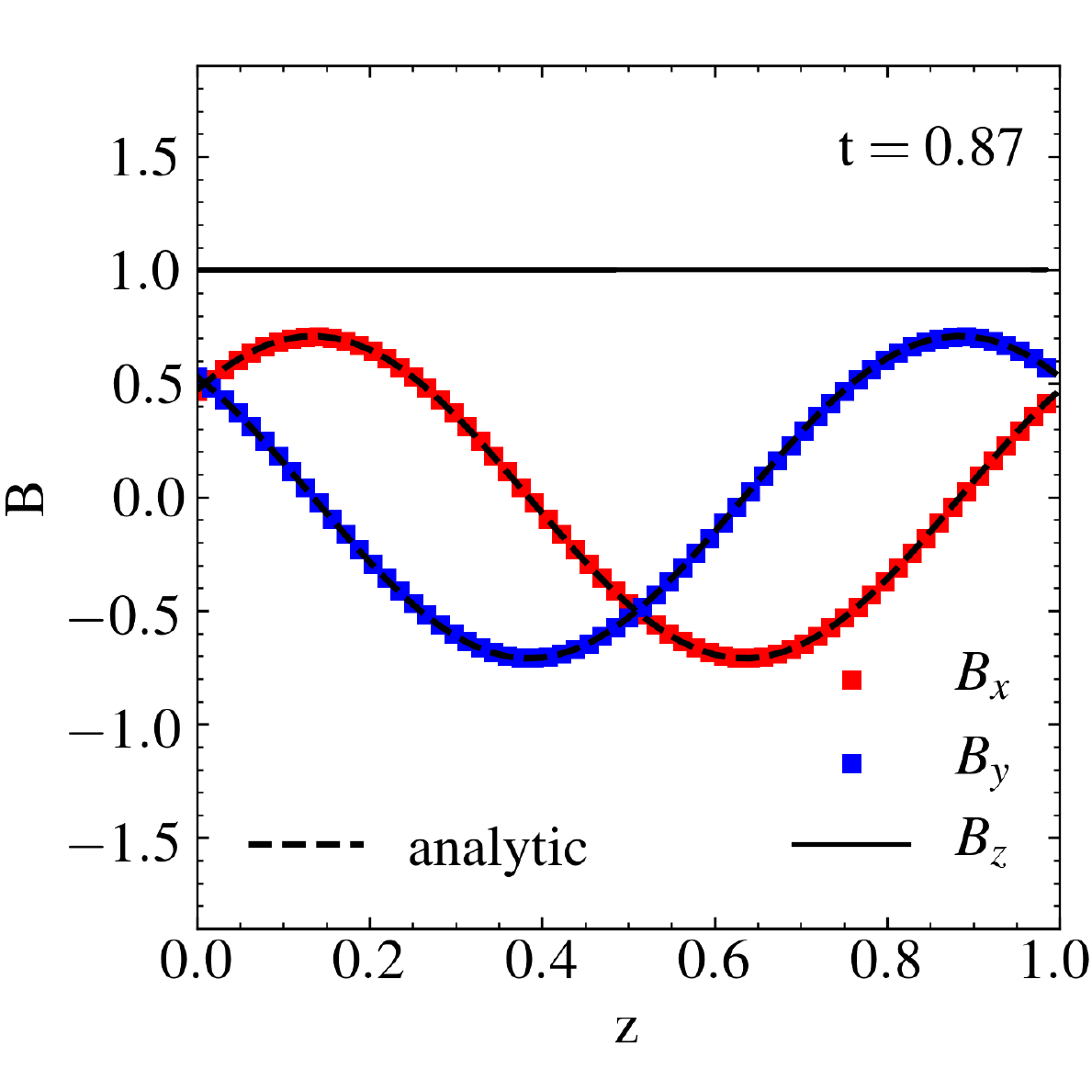}
\includegraphics[width=0.33\textwidth]{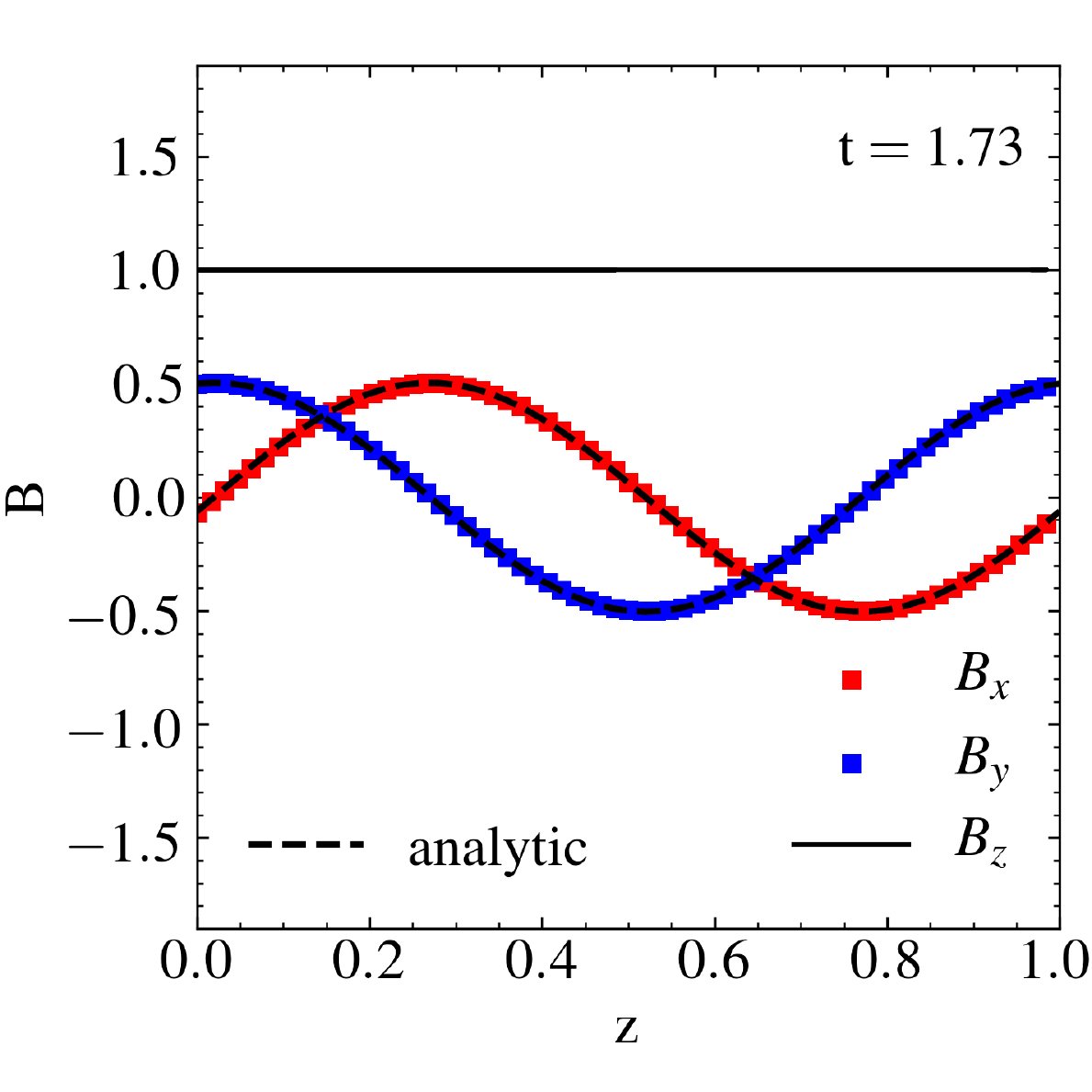}
\includegraphics[width=0.33\textwidth]{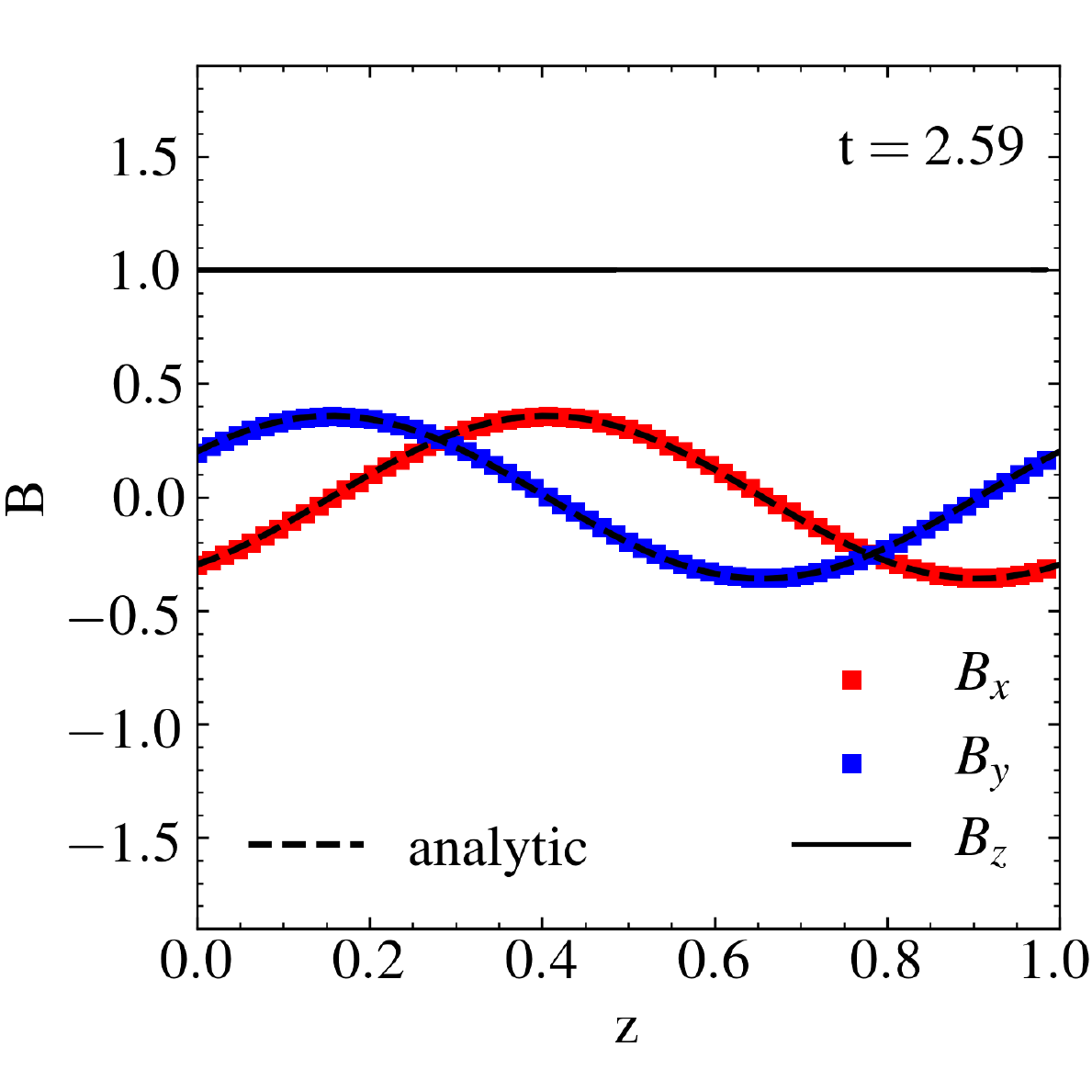}
\includegraphics[width=0.33\textwidth]{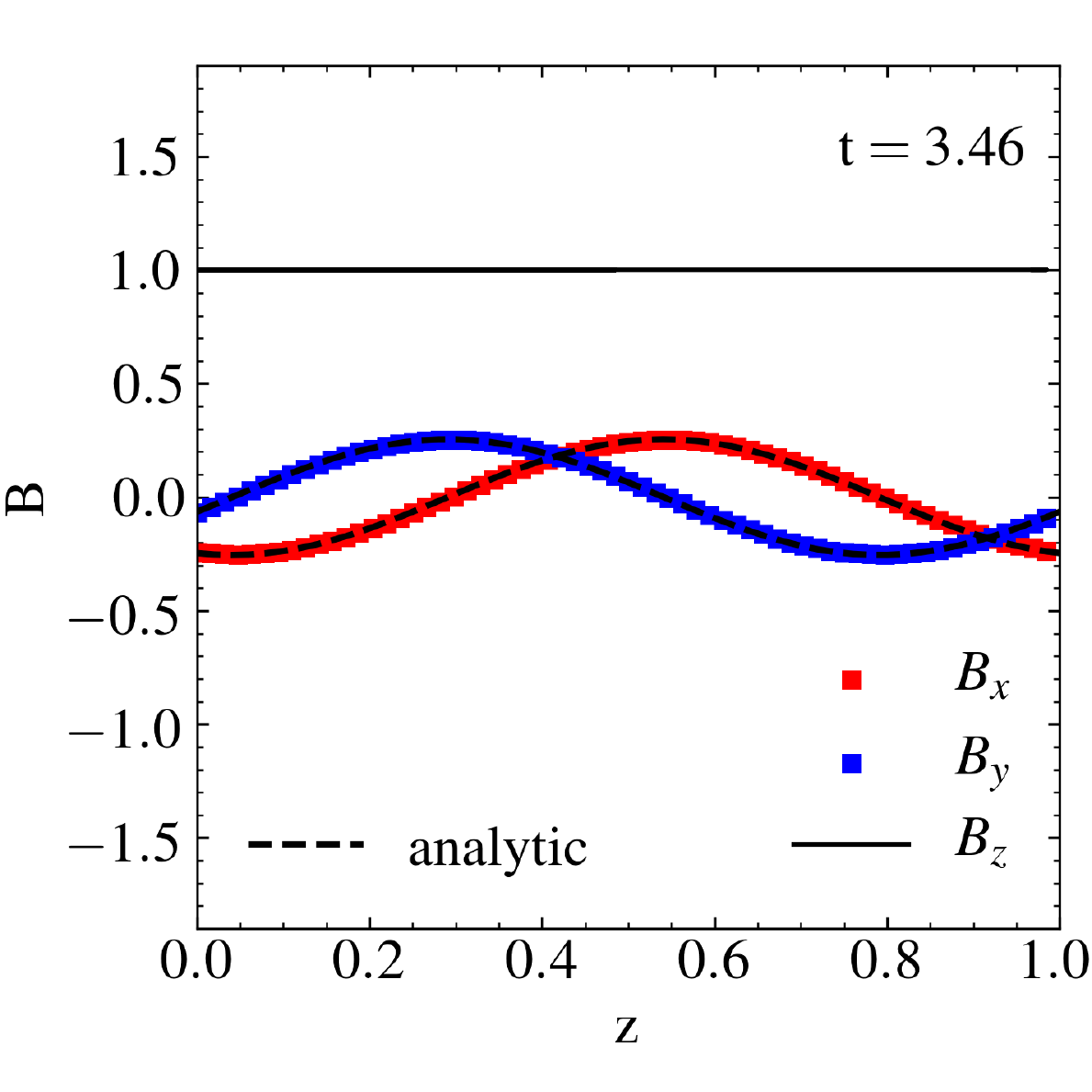}
\includegraphics[width=0.33\textwidth]{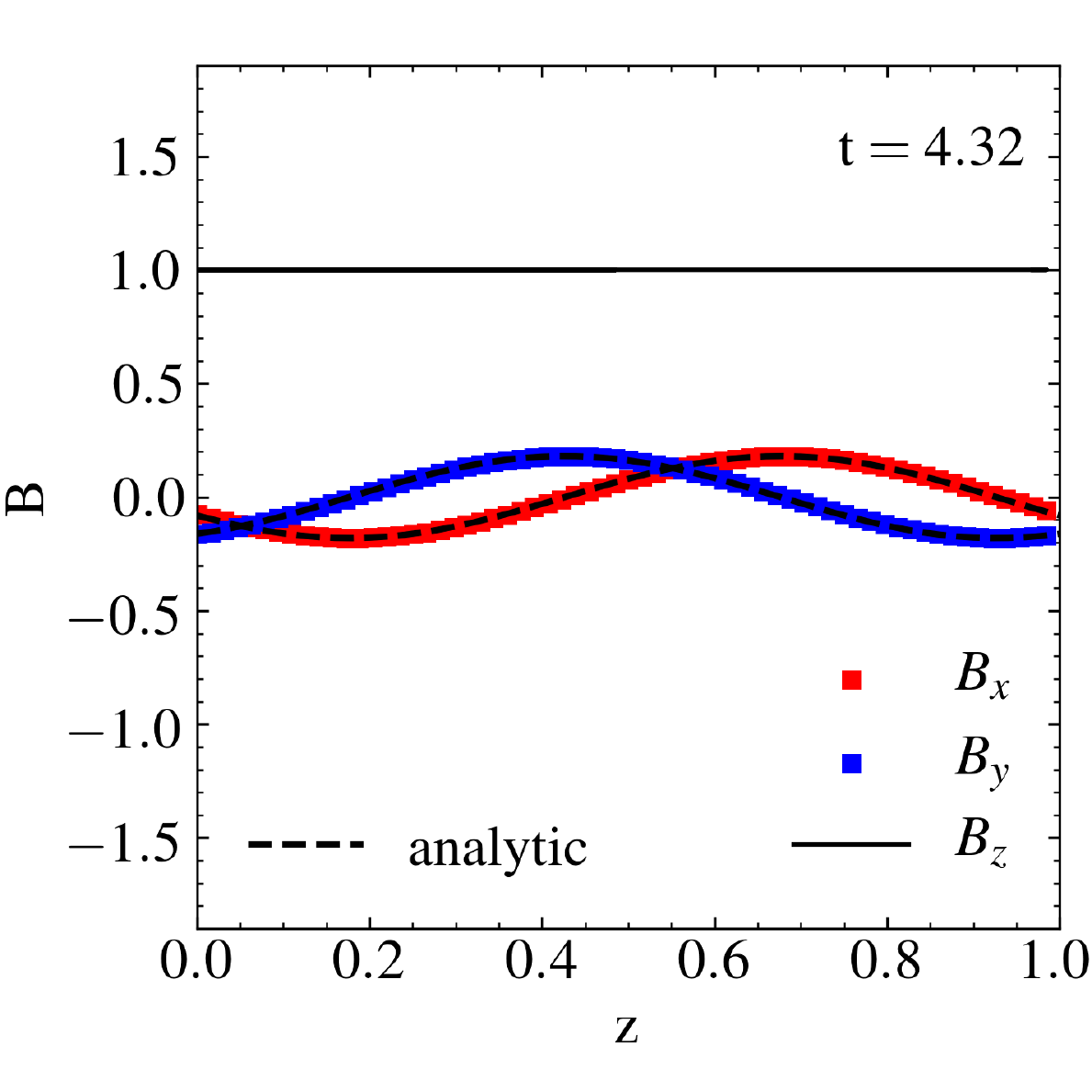}
\caption{Time evolution of a progressive Alfv\'en wave in the presence of
ohmic diffusion simulated with the implicit CT scheme. The panels show the evolution of the three components of the magnetic field 
(coloured symbols and black solid line) contrasted to the analytic solution (dashed line). The exponential decay in
amplitude of the wave is clearly visible. The direction of wave propagation is the negative
$z$-axis.
}
\label{fig:progalfven}
\end{figure*}

\subsubsection{Progressive wave}\label{sec:progressive}
We follow~\citet{Masson2012} to initialize this test problem.  In the case of a
progressive wave, we evolve the following initial conditions in a 3D periodic
domain of side length $L=1$,
\begin{equation}
 \boldsymbol{B}(\boldsymbol{x}) = \delta B[\cos(kz)\hat{e}_x - \sin(kz)\hat{e}_y] + B_0\hat{e}_z,
 \label{eq:Balf}
\end{equation}
\begin{align}\displaystyle
 \nonumber\boldsymbol{v}(\boldsymbol{x}) = \delta v\{&[\omega_i\cos(kz) - \omega_r\sin(kz)]\,\hat{e}_x
 \\- &[\omega_i\sin(kz) + \omega_r\cos(kz)]\,\hat{e}_y\},
 \label{eq:valf}
\end{align}
with
\begin{align}
 \nonumber &\delta v = \frac{k B_0}{\rho_0 \omega^2}\delta B, \qquad \omega^2 = \omega_r^2+\omega_i^2,
 \qquad \omega_r = -\frac{k^2\eta}{2}, \\ &\omega_i = \sqrt{(k v_A)^2 - \omega_r^2}, \qquad
 k = 2\pi, \qquad v_A = \frac{B_0}{\sqrt{\rho_0}}.
 \label{eq:definitions}
\end{align}
The wave will be evolving as
\begin{align}
 \nonumber\boldsymbol{B}(\boldsymbol{x},t) = e^{\omega_r t}\delta B[\cos(kz + \omega_i t)\,\hat{e}_x - \sin(kz + \omega_i t)\hat{e}_y]
 \\+B_0\hat{e}_z,
 \label{eq:evBalf}
\end{align}
\begin{align}\displaystyle
 \nonumber\boldsymbol{v}(\boldsymbol{x},t) = e^{\omega_r t}\delta v &\,\times \\ 
 \nonumber\{&[\omega_i\cos(kz + \omega_i t) - \omega_r\sin(kz + \omega_i t)]\,\hat{e}_x
 \\ - &[\omega_i\sin(kz + \omega_i t) + \omega_r\cos(kz + \omega_i t)]\,\hat{e}_y\},
 \label{eq:evvalf}
\end{align}
which is a planar, circularly polarized Alfv\'en wave in a clockwise direction
from the source perspective.

The previous equations demonstrate how the amplitude of the wave is decaying
exponentially at a rate equal to $\omega_r$. The rate is faster for larger
values of the resistivity $\eta$. Moreover, the frequency of the wave is
decreased due to the resistivity in the system,
which implies a lower propagation speed compared to the Alfv\'en speed. As a
result of ohmic dissipation, the gas internal energy is expected to grow 
alongside an increase of the gas thermal pressure that can be described by~\citep[see again][]{Masson2012}
\begin{equation}
P(t) = 1 + (\gamma - 1) k^2\delta B^2 \eta \frac{e^{2\omega_r t} - 1}{2\omega_r},
 \label{eq:evPalf}
\end{equation}
with $\gamma = 5/3$ being the ratio of the specific heats of the gas. We note
that $\omega_r$ is a negative quantity. Therefore the increase in pressure
reaches a maximum formally for $t \to +\infty$.  This situation corresponds to
the total dissipation of the initial magnetic and kinetic energy contained in
the wave to thermal energy due to Joule heating. Moreover, the absence of any
spatial dependence in the pressure expression implies that the heating is
uniform throughout the simulated domain.

For the CT scheme, the periodic part of the vector potential originating the magnetic 
field of the progressive Alfv\'en wave (\ref{eq:evBalf})--(\ref{eq:evvalf}) is
given by 
\begin{equation}
 \displaystyle\boldsymbol{A}(\boldsymbol{x}) = \delta B\left[\frac{\cos(kz)}{k}\hat{e}_x - \frac{\sin(kz)}{k}\hat{e}_y\right]
 \label{eq:Aalf}
\end{equation}
and evolves as
\begin{align}
 \nonumber\displaystyle\boldsymbol{A}(\boldsymbol{x},t) = e^{\omega_r t}&\delta B\,\times \\
 &\left[\frac{\cos(kz + \omega_i t)}{k}\,\hat{e}_x - \frac{\sin(kz + \omega_i t)}{k}\hat{e}_y\right].
 \label{eq:evAalf}
\end{align}
The mean magnetic field is represented in this set-up by the $z$-component of
equation (\ref{eq:Balf}).  We initialize the simulation by assuming a uniform
initial density $\rho_0 = 1$ and pressure $P_0 = 1$, a guide field in the $z-$direction $B_0 = 1$,
$\delta B = 1$ and a resistivity $\eta=2\times10^{-2}$. All other quantities can be
derived from relations~(\ref{eq:definitions}). For the mesh generating points
of the Voronoi tessellation, we use a cubic body-centred lattice, \FM{composed by two interleaved Cartesian 
meshes of $32^3$ points and separated by half the cell spacing, for a total of} $2\times 32^3$
resolution elements.

In Fig.~\ref{fig:progalfven}, we present the results of this test problem for the 
implicit CT scheme on a static mesh. We show the amplitude of the two transverse 
components of the magnetic field (coloured squares) and of the guide field (black solid 
line) at different times, indicated in the top right-hand corner of each panel. The 
simulation results are compared to the analytic expectations, indicated by the 
dashed black lines in each panel. The simulation is run approximately for five 
periods of the wave. This is a time-scale over which the effect of ohmic 
dissipation is particularly noticeable. 

\begin{figure*}
\centering
\includegraphics[width=0.45\textwidth]{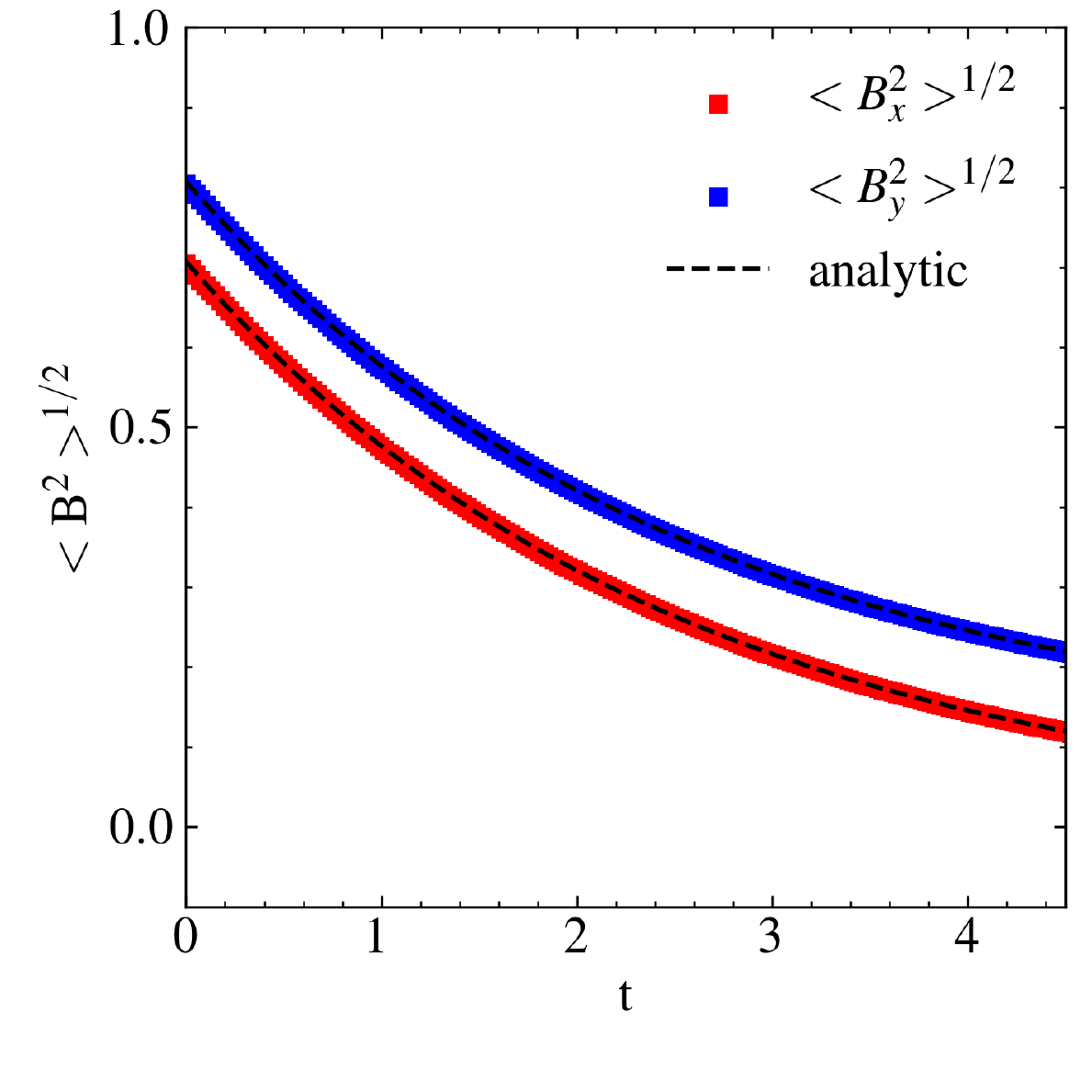}
\includegraphics[width=0.45\textwidth]{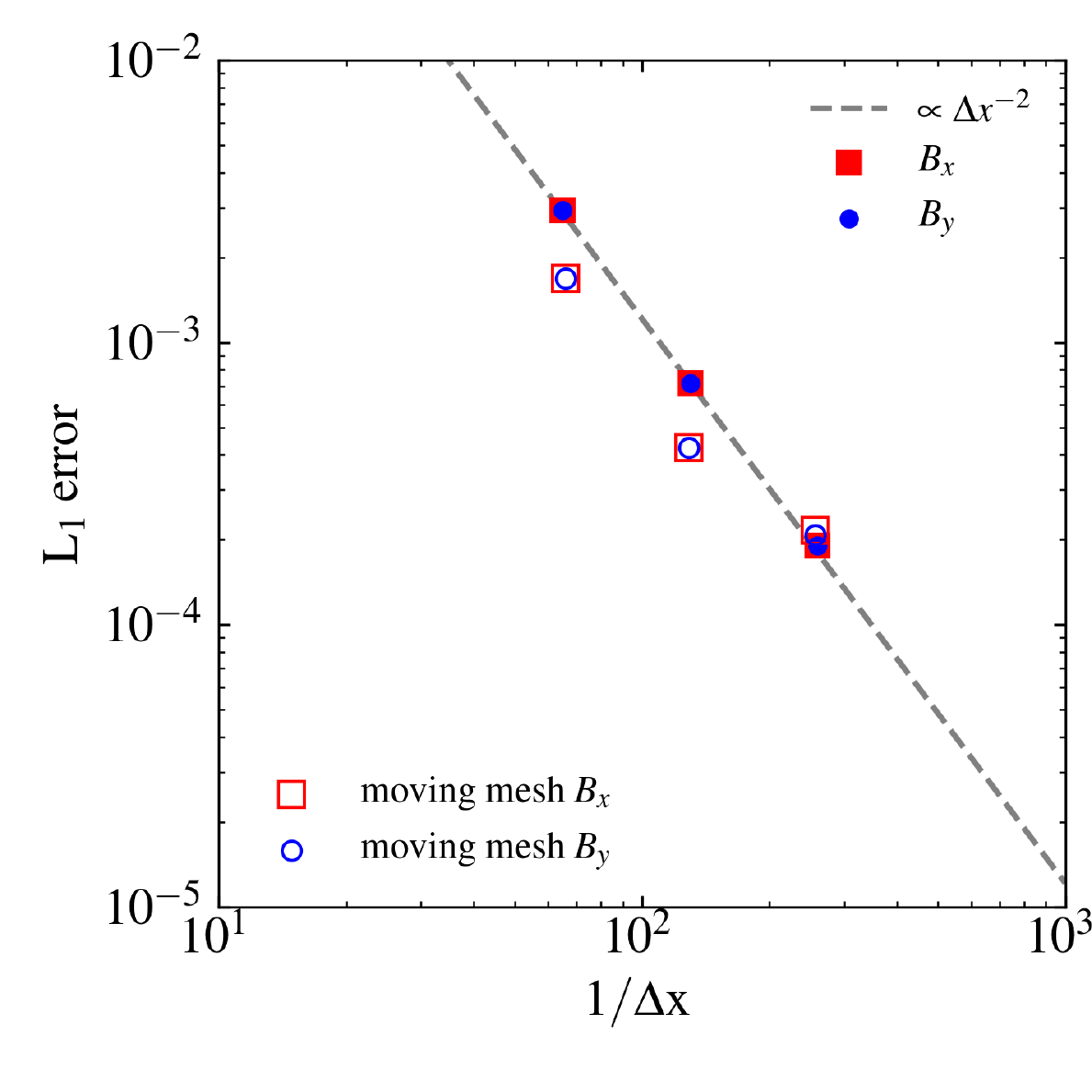}
\caption{\textit{Left-hand panel}: Time evolution of the average rms intensity of the 
transverse components of the magnetic field for the progressive Alfv\'en wave test simulation 
with the implicit CT scheme. The panel shows the evolution of these components (coloured squares) 
contrasted to the analytic solution (dashed line). The $y-$component of the magnetic 
field is offset from its true value to improve clarity. The exponential decay in 
the amplitude of the magnetic field is clearly visible. \textit{Right-hand panel}: $L_1$ norm of 
the error as a function of resolution for the progressive Alfv\'en wave tests 
run with the implicit CT scheme at time $t = 0.74$. Different coloured symbols 
show the error of the individual components of the magnetic field as indicated in the 
legend, whereas the grey dashed line represents the expected scaling for a second-order 
scheme. Open symbols show the results obtained for the implicit Powell 
scheme run on a moving-mesh configuration. At high resolution the convergence becomes slower 
than second-order due to a significantly distorted mesh.
}
\label{fig:alfvenavgergprog}
\end{figure*}

The numerical results agree with the analytic expectations for this test. In 
particular, the guide field in the $z-$direction is not affected by the ohmic 
dissipation thus staying at its initial strength. The two transverse components, 
instead, clearly show an exponential decay in their amplitude, such that at the 
final time their maximum values are about a quarter of their 
initial amplitude. Also noticeable is the propagation of the wave towards 
decreasing values of the coordinate $z$. No phase offset is apparent in this 
test between the numerical values of the solution and the analytic estimates.

In the left-hand panel of Fig.~\ref{fig:alfvenavgergprog} we quantify the 
exponential decay of the magnetic field by showing the time evolution of the 
volume-weighted rms values of the two transverse components of the magnetic field for the implicit CT scheme. 
This quantity gives an indication of the magnetic energy density contained \FM{within the simulated box in each
of the magnetic field components,} which is dissipated \FM{by} ohmic resistivity. 
The $y-$component
is offset by $0.1$ from its true value to improve the clarity of the plot.

For the initial conditions used in this experiment we expect analytically that
the mean rms values decrease exponentially in a characteristic time-scale
$\omega_r$, from an initial amplitude of $\sqrt{2}$. This trend is recovered in Fig.~\ref{fig:alfvenavgergprog}, where the simulation results (coloured
squares) overlap well with the analytic expectations (black dashed line). 

In the right-hand panel of Fig.~\ref{fig:alfvenavgergprog} we show the $L_1$ error in 
the two transverse magnetic field components (coloured symbols) as a function of 
the simulation resolution for the implicit CT scheme at $t = 0.74$. The grey 
dashed line shows the expected scaling for second-order convergence. The open 
coloured symbols indicate the results obtained for this test problem for the 
implicit Powell scheme run on a moving-mesh configuration in which the mesh 
generating points are free to move with the fluid motion. The figure clearly 
demonstrates the quadratic decrease of the $L_1$ error of the numerical solution 
with increasing resolution, thus signalling that even in this more complex case 
where gas dynamics must be taken fully into account, our implementations of the 
ohmic terms in \arepo\ perform as expected. 

\subsubsection{Stationary wave}\label{sec:standing}

\begin{figure*}
\centering
\includegraphics[width=0.33\textwidth]{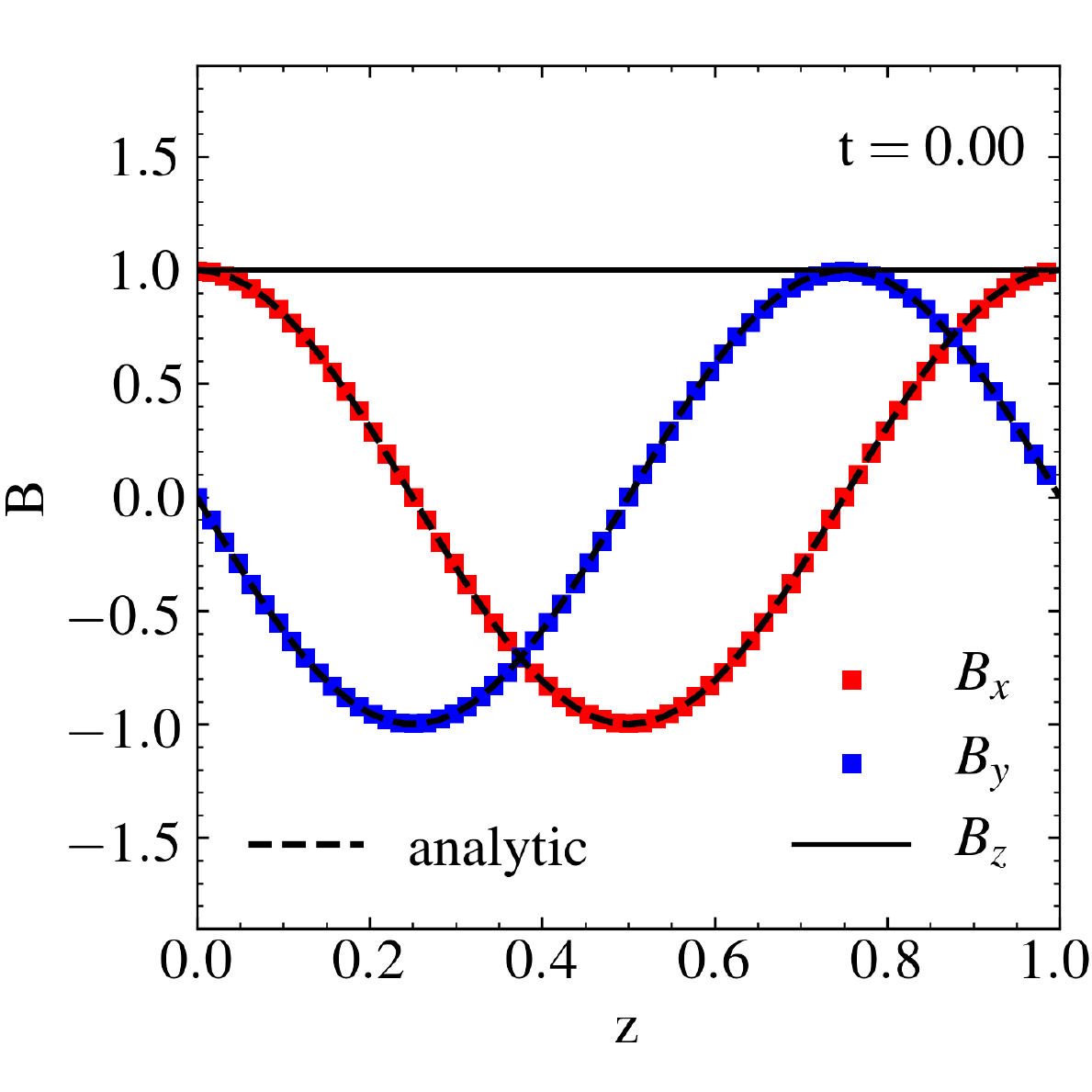}
\includegraphics[width=0.33\textwidth]{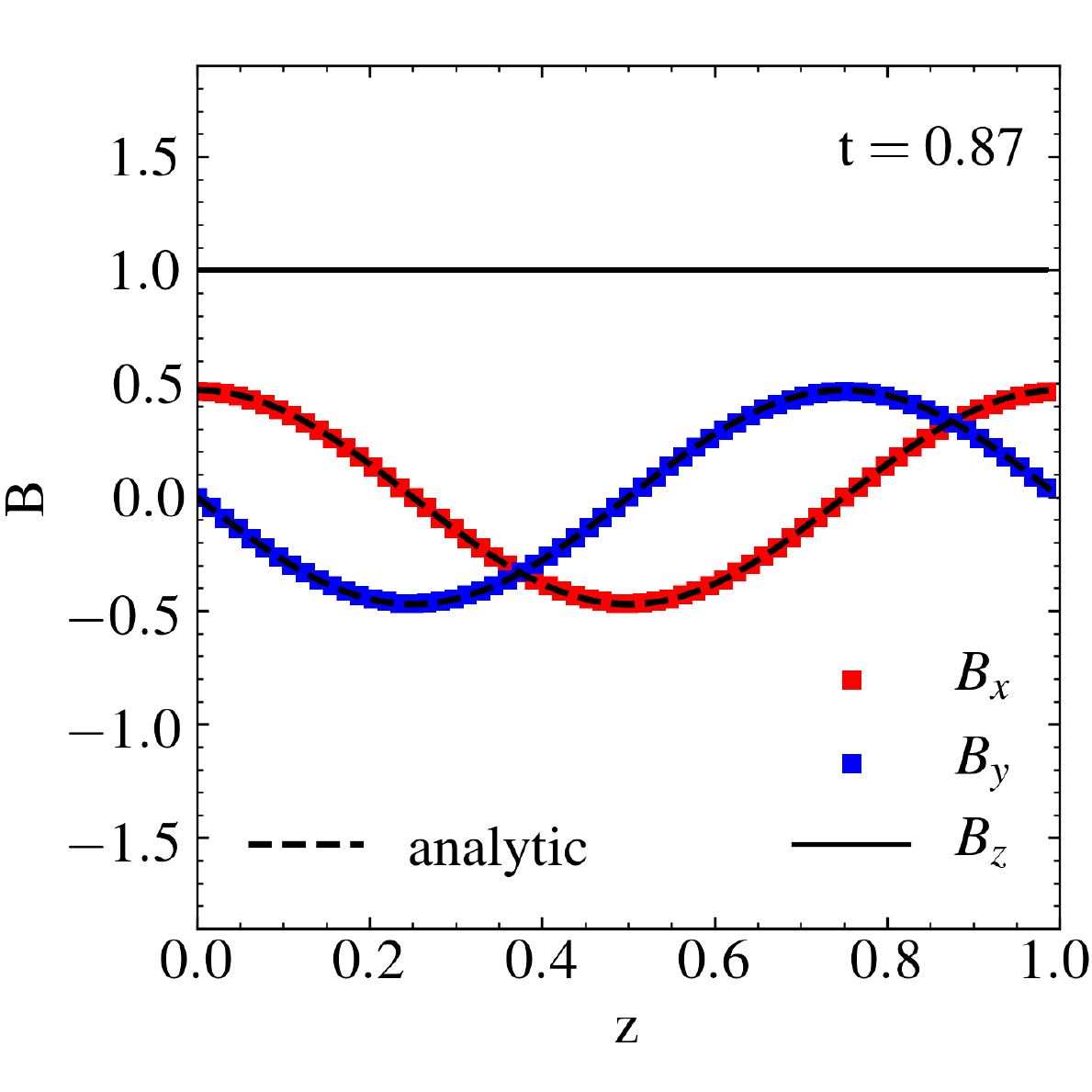}
\includegraphics[width=0.33\textwidth]{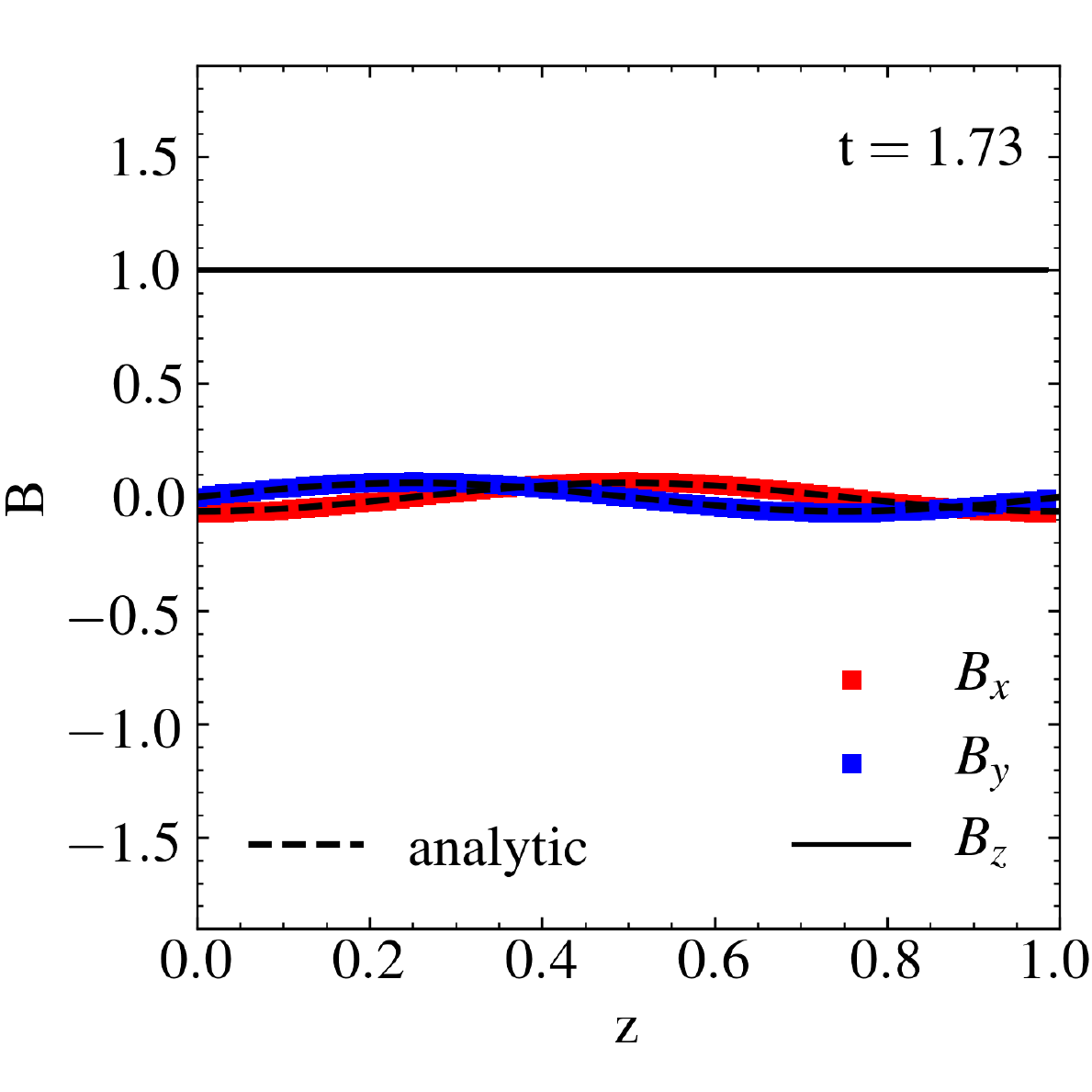}
\includegraphics[width=0.33\textwidth]{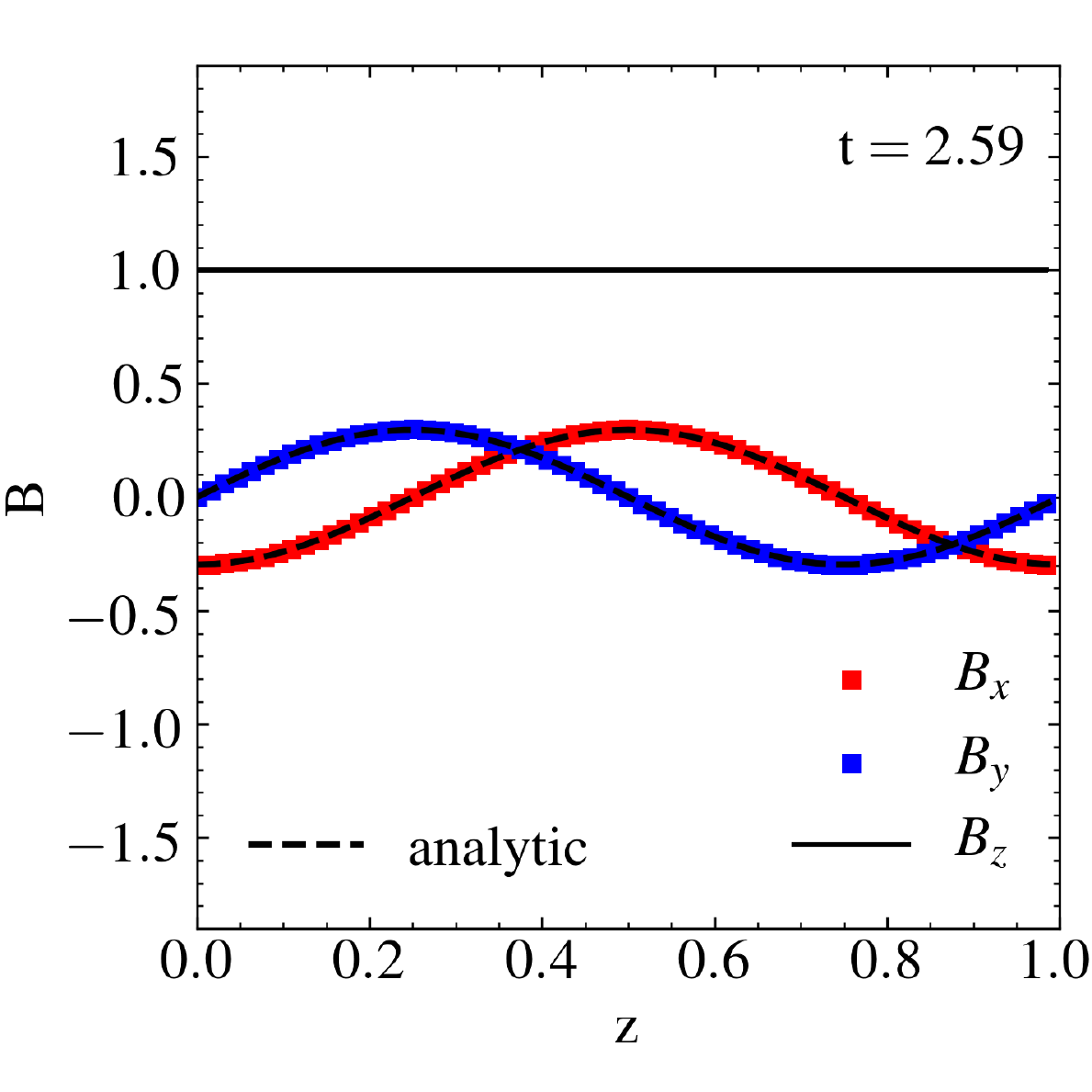}
\includegraphics[width=0.33\textwidth]{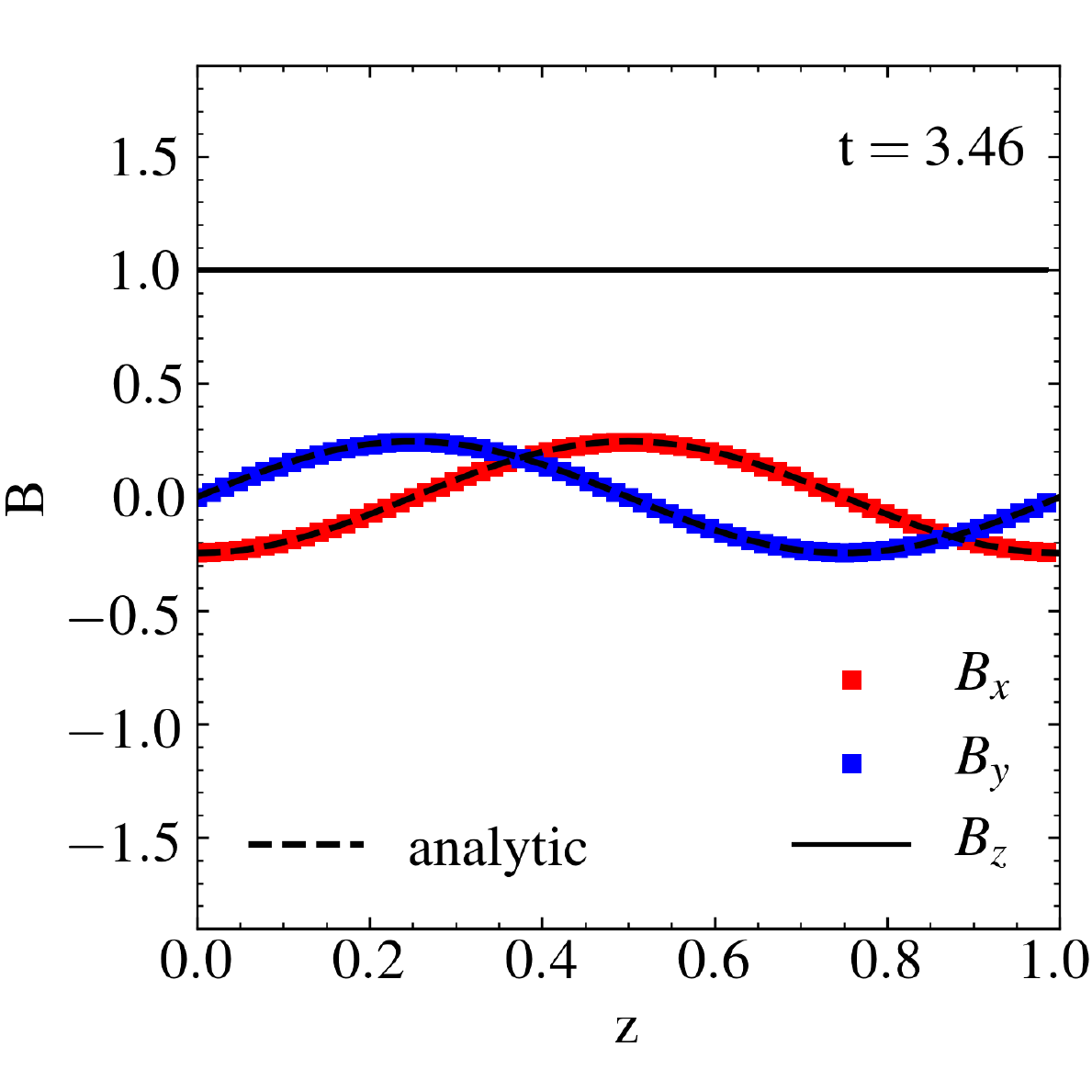}
\includegraphics[width=0.33\textwidth]{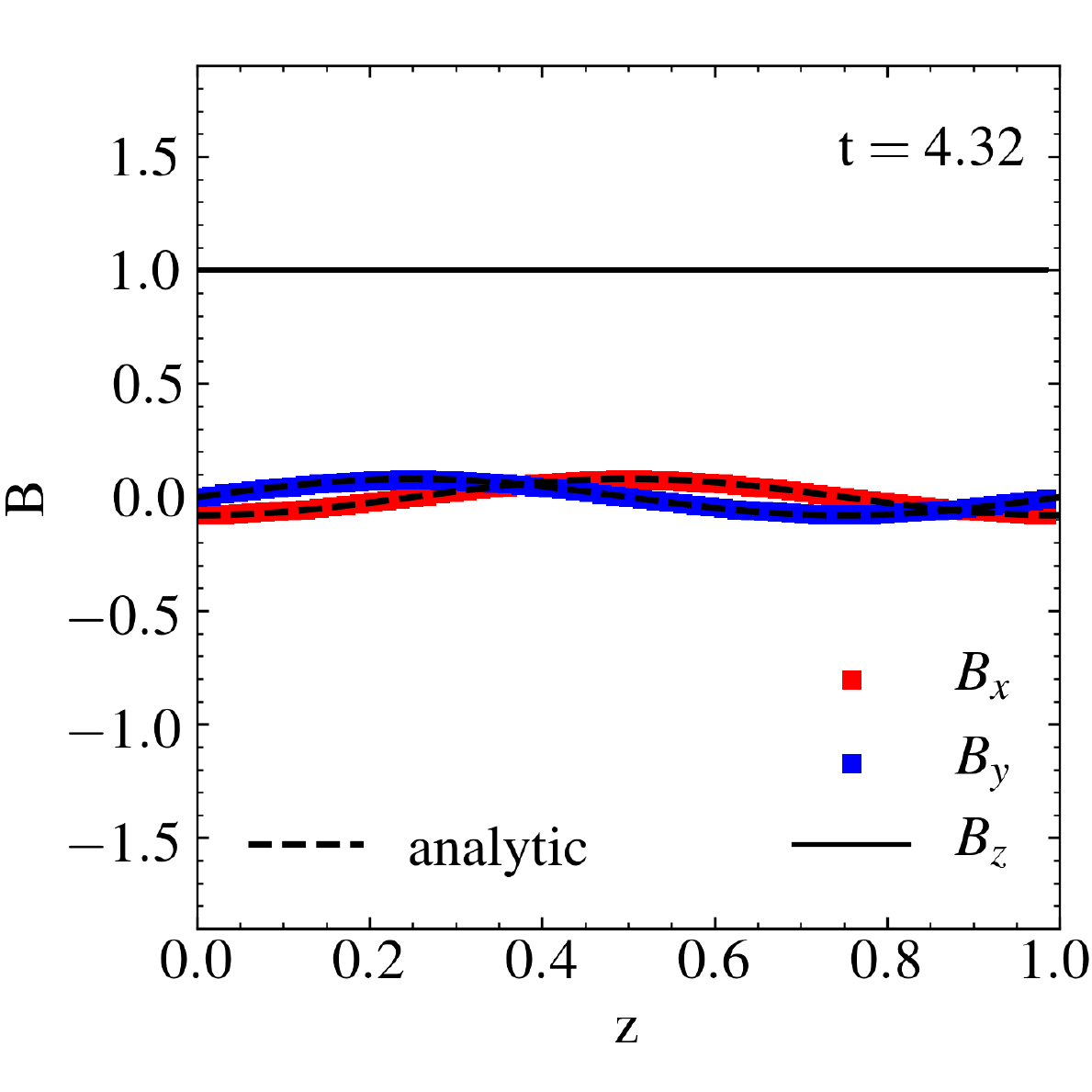}
\caption{Time evolution of a stationary Alfv\'en wave in the presence of
ohmic diffusion with the implicit CT scheme. The panels show the evolution of the three components of the magnetic field 
(coloured symbols and black solid line) contrasted to the analytic solution (dashed line). The exponential decay in
amplitude of the wave is clearly visible.
}
\label{fig:statalfven}
\end{figure*}

The case of a stationary wave is obtained by linearly combining two progressive
waves with equal weights ($1/2$) as described by equations~(\ref{eq:Aalf}),
(\ref{eq:Balf}) and (\ref{eq:valf}) propagating in opposite direction and thus
with opposite $\omega_i$. This results in
\begin{equation}
 \boldsymbol{B}(\boldsymbol{x}) = \delta B[\cos(kz)\hat{e}_x - \sin(kz)\hat{e}_y] + B_0\hat{e}_z,
 \label{eq:Balfst}
\end{equation}
\begin{equation}\displaystyle
 \boldsymbol{v}(\boldsymbol{x}) = -\delta v[\omega_r\sin(kz)\,\hat{e}_x
 +\omega_r\cos(kz)\,\hat{e}_y],
 \label{eq:valfst}
\end{equation}
with all the symbols defined by equation~(\ref{eq:definitions}). The wave will
be evolving as
\begin{align}
 \nonumber\boldsymbol{B}(\boldsymbol{x},t) = e^{\omega_r t}\delta B[&\cos(kz)\cos(\omega_i t)\,\hat{e}_x 
 \\- &\sin(kz)\cos(\omega_i t)\,\hat{e}_y] + B_0\,\hat{e}_z,
 \label{eq:evBalfst}
\end{align}
\begin{align}\displaystyle
 \nonumber\boldsymbol{v}(\boldsymbol{x},t) & = -e^{\omega_r t} \delta v\,\times \\
 \nonumber\{&[\omega_i\sin(kz)\sin(\omega_i t) + \omega_r\sin(kz)\cos(\omega_i t)]\,\hat{e}_x
  \\+ &[\omega_i\cos(kz)\sin(\omega_i t) + \omega_r\cos(kz)\cos(\omega_i t)]\,\hat{e}_y\}.
 \label{eq:evvalfst}
\end{align}
We note that contrary to the previous case, the spatial and temporal
dependences are separated such that the wave does not propagate. In particular,
the location of the knots of the wave -- where the magnetic field and velocity
amplitude are zero -- does not change with time. Only the amplitude of the
wave is decaying exponentially at a rate equal to $\omega_r$, as in the
progressive case. Again, due to the ohmic dissipation, the gas internal energy
increases and the gas thermal pressure evolves as~\citep[see][]{Masson2012}
\begin{align}
\nonumber P(t) & = 1 + \frac{(\gamma - 1)}{4} k^2\delta B^2 \eta 
\left\{\frac{e^{2\omega_r t} - 1}{\omega_r}\right. \\
& \left.+ e^{2\omega_r t} \left[\frac{\omega_r \cos(2\omega_i t) + \omega_i \sin(2\omega_i t)}{\omega^2}\right] - 
\frac{\omega_r}{\omega^2}\right\}.
 \label{eq:evPalfst}
\end{align}
As in the progressive wave, $\omega_r$ is a negative quantity, which implies
that for $t\to +\infty$ the pressure reaches a maximum value once the initial
magnetic field is totally dissipated by resistive effects. The heating rate of
the gas is independent of the position in this case as well.

\begin{figure*}
\centering
\includegraphics[width=0.45\textwidth]{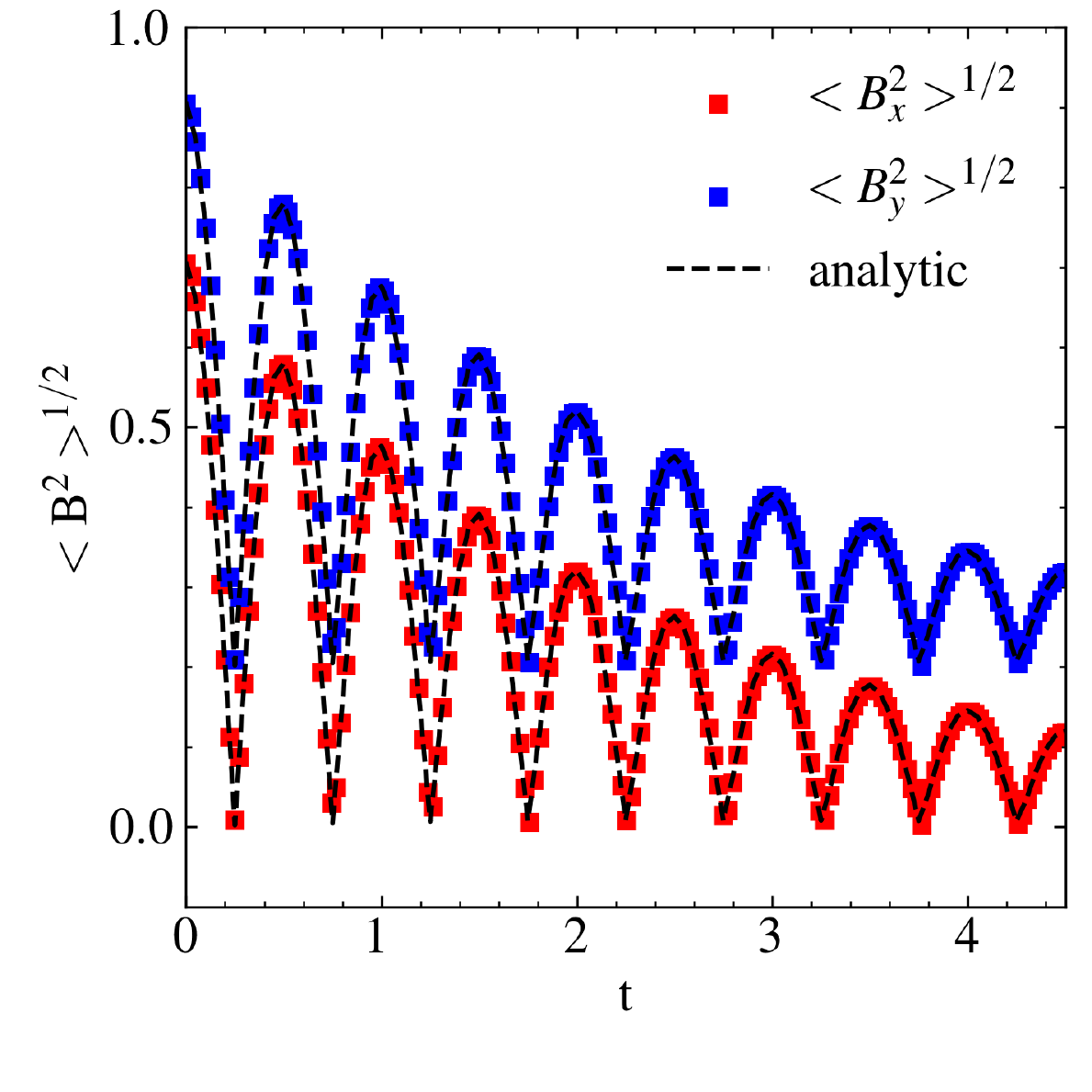}
\includegraphics[width=0.45\textwidth]{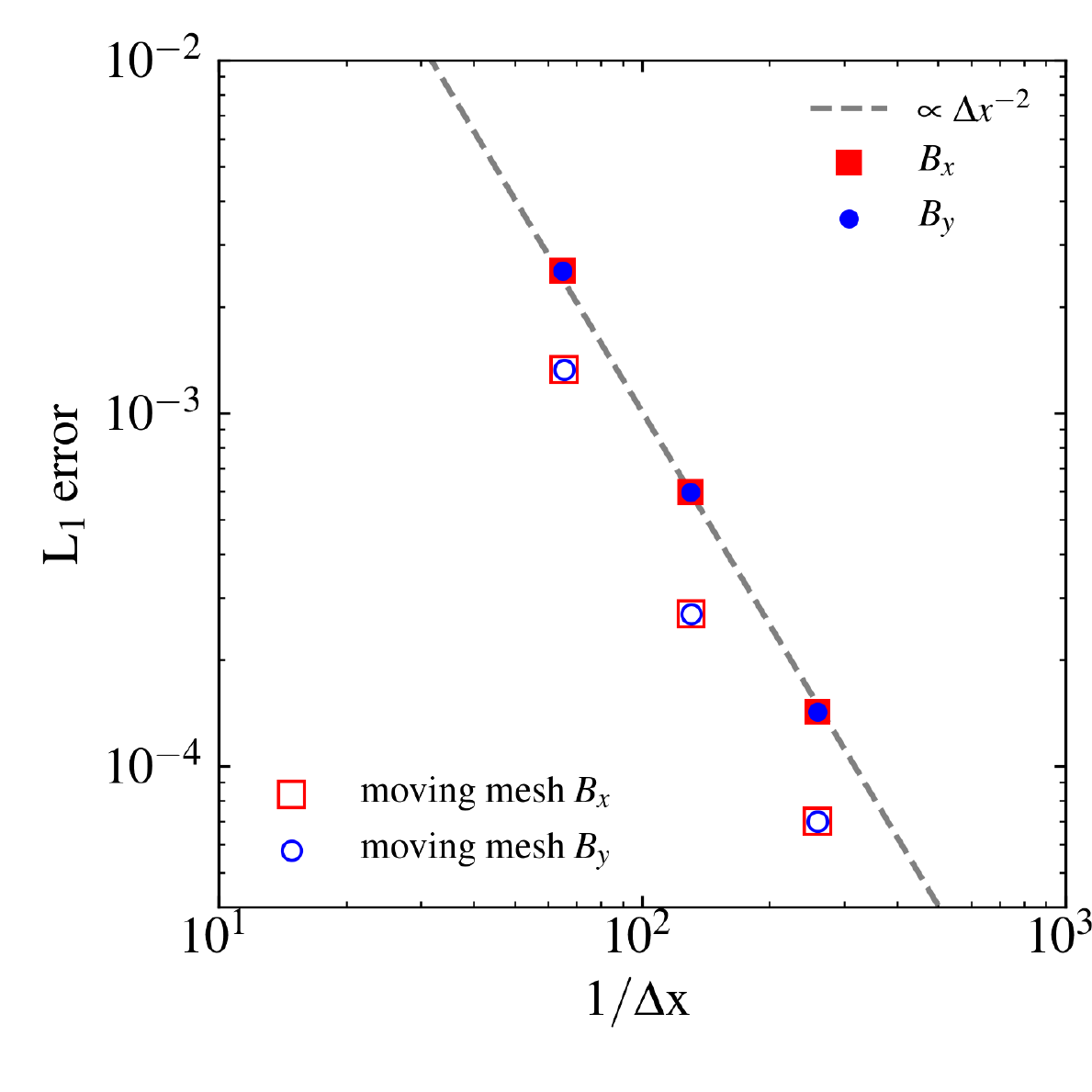}
\caption{\textit{Left-hand panel}: Time evolution of the average rms intensity of the 
transverse components of the magnetic field for the stationary Alfv\'en wave test simulation 
with the implicit CT scheme. The panel shows the evolution of this quantity 
contrasted to the analytic solution (dashed line). The $y-$component of the magnetic 
field is offset from its true value to improve clarity. The exponential decay in 
the amplitude of the magnetic field, modulated by a cosine function, is clearly 
visible. \textit{Right-hand panel}: $L_1$ norm of the error as a function of resolution for 
the stationary Alfv\'en wave tests at time $t = 0.74$. Different coloured symbols 
show the error of the individual components of the magnetic field as indicated in the 
legend, whereas the grey dashed line represents the expected scaling for a second-order
scheme. Open symbols show the results obtained for the implicit Powell 
scheme run on a moving-mesh configuration. 
}
\label{fig:alfvenavgergstat}
\end{figure*}

For the CT scheme, the (periodic) vector potential originating a stationary
Alfv\'en wave can be expressed as
\begin{equation}
 \displaystyle\boldsymbol{A}(\boldsymbol{x}) = \delta B\left[\frac{\cos(kz)}{k}\hat{e}_x - \frac{\sin(kz)}{k}\hat{e}_y\right],
 \label{eq:Aalfst}
\end{equation}
and its evolution is given by
\begin{align}
 \nonumber\boldsymbol{A}(\boldsymbol{x},t) &= e^{\omega_r t}\delta B\,\times \\ 
 &\left[\frac{\cos(kz)\cos(\omega_i t)}{k}\,\hat{e}_x - \frac{\sin(kz)\cos(\omega_i t)}{k}\hat{e}_y\right].
 \label{eq:evAalfst}
\end{align}
The mean magnetic field is also represented in this set-up by the $z$-component of
equation (\ref{eq:Balfst}).  The same set-up as in the progressive case is used
in this test problem as well as for what concerns the values of both the initial
gas properties and the grid geometry. 

In Fig.~\ref{fig:statalfven}, we present the results of this test for the 
implicit CT scheme on a static mesh. As in the previous case, we show the amplitude of the two transverse 
components of the magnetic field (coloured squares) and of the guide field (black solid 
line) at different times, shown in the top right-hand corner of each panel. The 
analytic solution is indicated by the dashed black lines in each panel. The simulation 
is run again approximately for five periods of oscillation of the wave to give ample time 
for ohmic diffusion to act.

This figure demonstrates that the numerical results agree very well 
with the analytic solution. As expected no change is visible in the guide field 
in the $z-$direction, which remains at the initial strength. On the other hand, 
the amplitude of the two transverse components decays exponentially as a 
function of time. At the final time displayed for this test problem they only 
reach one tenth of their initial amplitude. This fact might appear surprising at 
first, given that the time-scale for dissipation $\omega_r$ is the same as in 
the progressive case \FM{($e^{\omega_r t} \approx 0.18$ for $t = 4.32$)}. However, a closer inspection of equation~(\ref{eq:evBalfst})
reveals that the magnetic field amplitude is further modulated by a $\cos(\omega_i t)$ term
that accounts for this discrepancy.

The modulation due to this cosine term can be seen more easily if the 
mean energy content of the magnetic field is plotted as a function of time. We present 
this in the left-hand panel of Fig.~\ref{fig:alfvenavgergstat}, where the time evolution of 
the volume-weighted mean rms values of the two transverse components of the magnetic
field is shown for the implicit CT scheme. 
The $y$-component
of the field is offset by $0.2$ from its true value to improve the clarity of the plot. We 
expect an exponential decay of the field amplitude on a characteristic 
time-scale $\omega_i$ starting from an initial amplitude of $\sqrt{2}$. It is 
evident from the figure that both the numerical (coloured squares) and analytical 
(black dashed line) solutions follow this expected trend and that they are in 
agreement with one another. In addition to the exponential decay, the modulation 
of the $\cos(\omega_i t)$ term is clearly visible as oscillations in the time 
evolution of the magnetic field rms values.

Finally, in the right-hand panel of Fig.~\ref{fig:alfvenavgergstat} we present the 
$L_1$ error in the two transverse magnetic field components (coloured symbols) as a 
function of the simulation resolution for this set-up at $t = 0.74$. The grey 
dashed line indicates the scaling for second-order convergence, whereas the open 
coloured symbols show the results obtained for this test problem for the 
implicit Powell scheme run on a moving-mesh configuration. As in the progressive 
case, the convergence is second-order accurate. The plot also demonstrates that our implementation 
performs well when gas dynamics has to be followed to model self-consistently 
the evolution of the simulated system.

\section{Magnetic reconnection}\label{sec:reconnection}

\begin{figure*}
\centering
\includegraphics[width=0.315\textwidth]{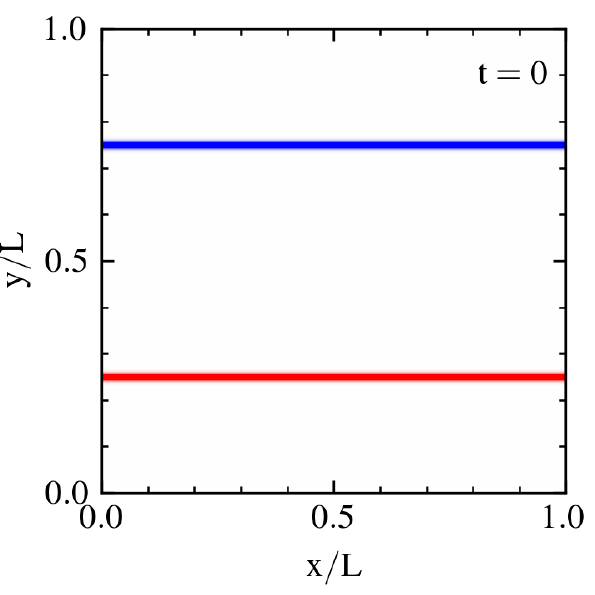}
\includegraphics[width=0.315\textwidth]{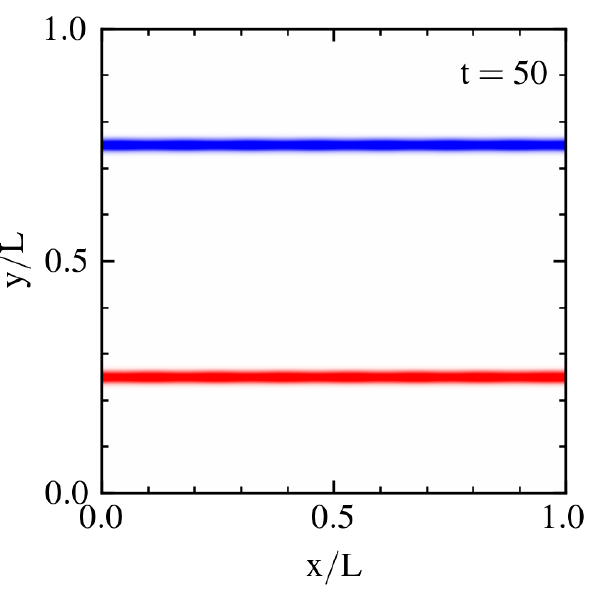}
\includegraphics[width=0.315\textwidth]{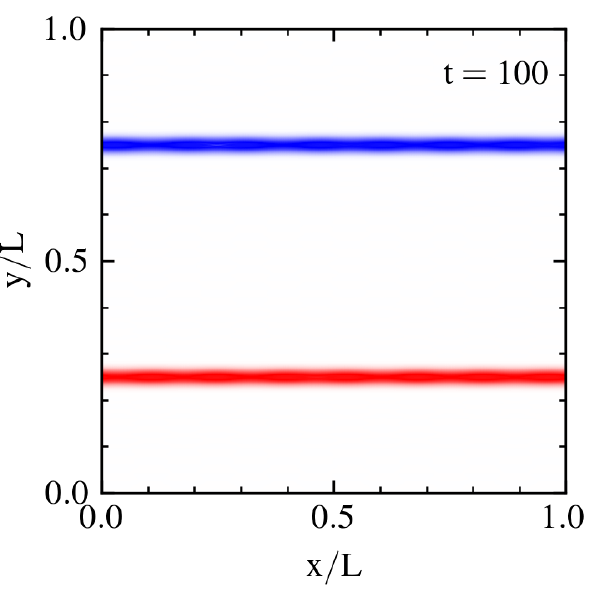}
\includegraphics[width=0.315\textwidth]{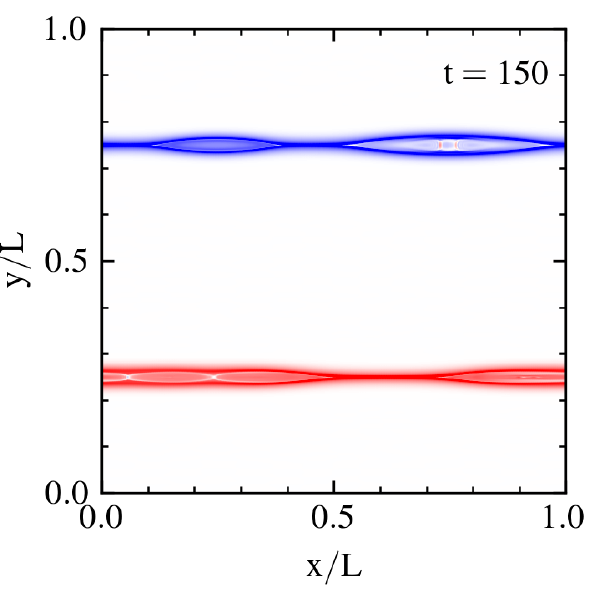}
\includegraphics[width=0.315\textwidth]{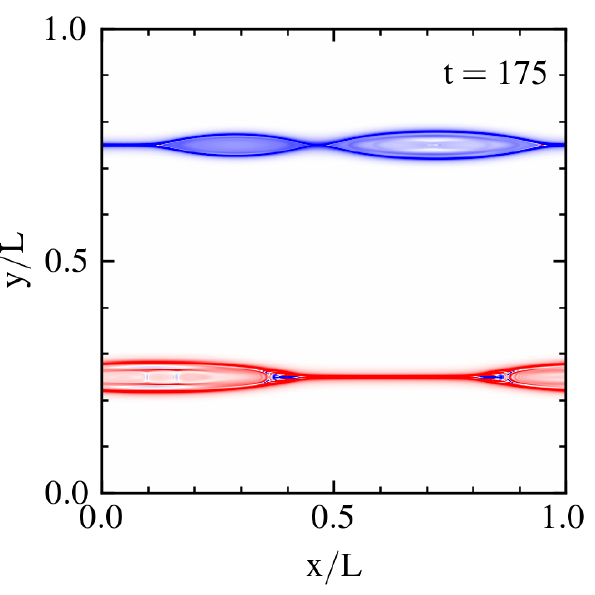}
\includegraphics[width=0.315\textwidth]{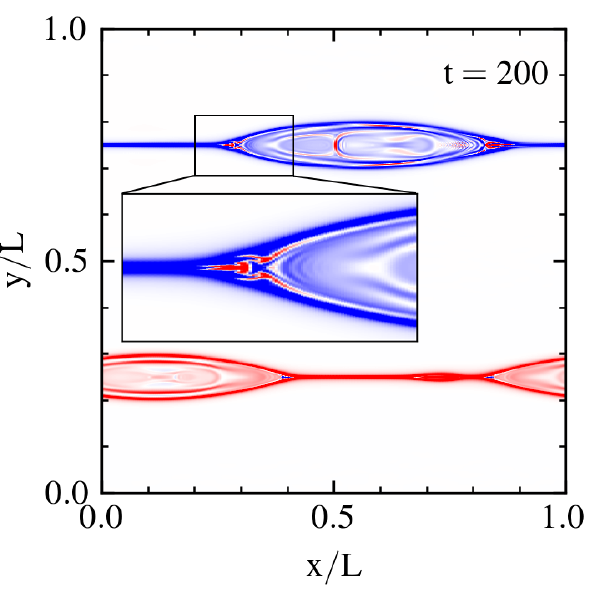}
\includegraphics[width=0.315\textwidth]{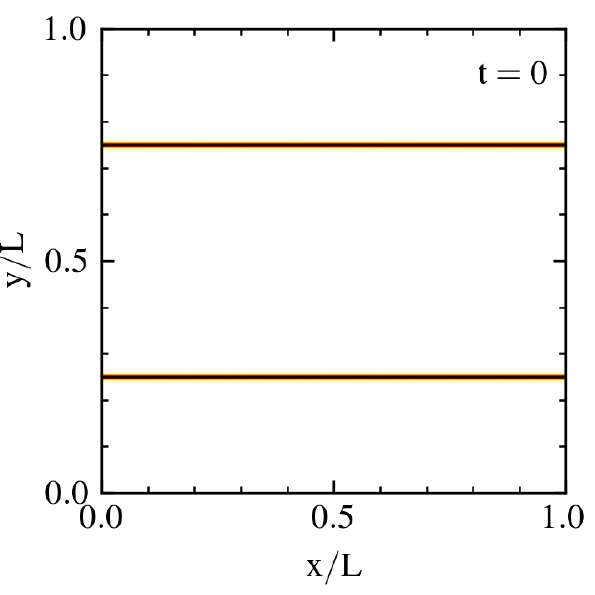}
\includegraphics[width=0.315\textwidth]{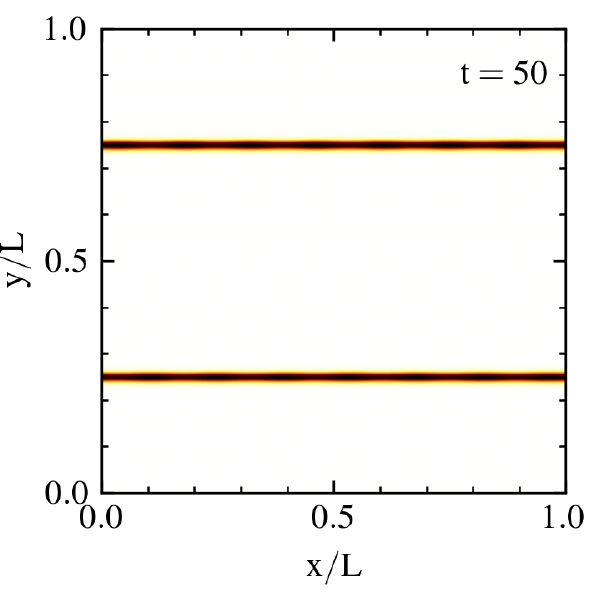}
\includegraphics[width=0.315\textwidth]{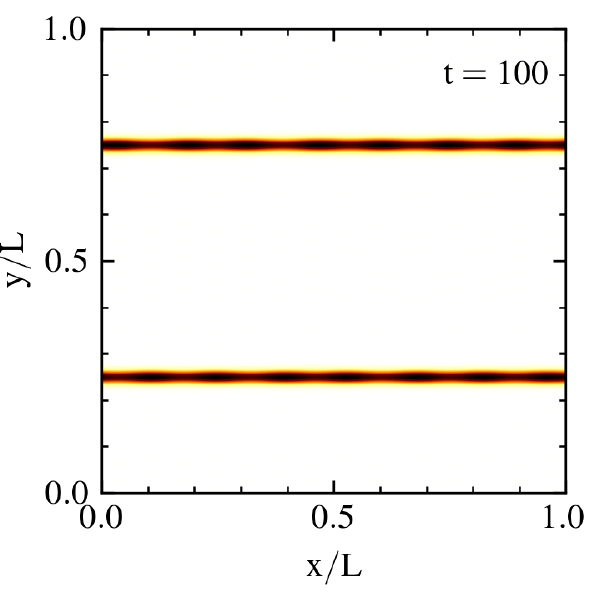}
\includegraphics[width=0.315\textwidth]{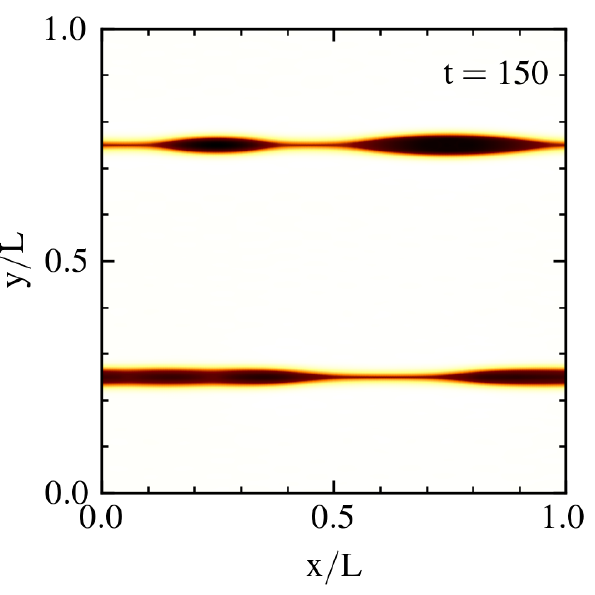}
\includegraphics[width=0.315\textwidth]{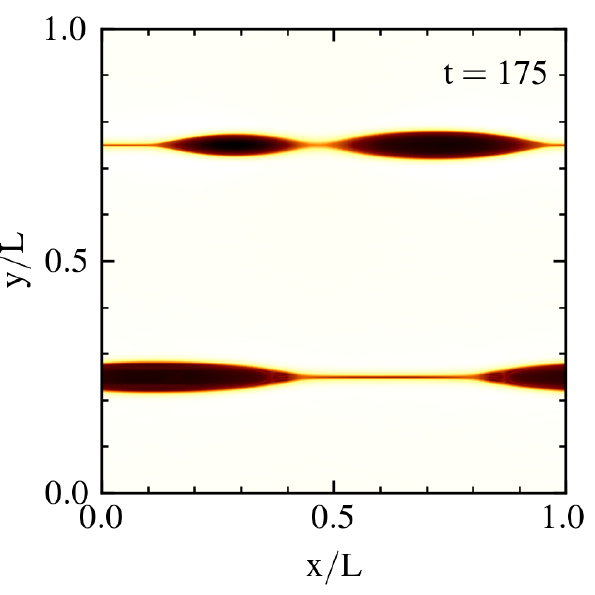}
\includegraphics[width=0.315\textwidth]{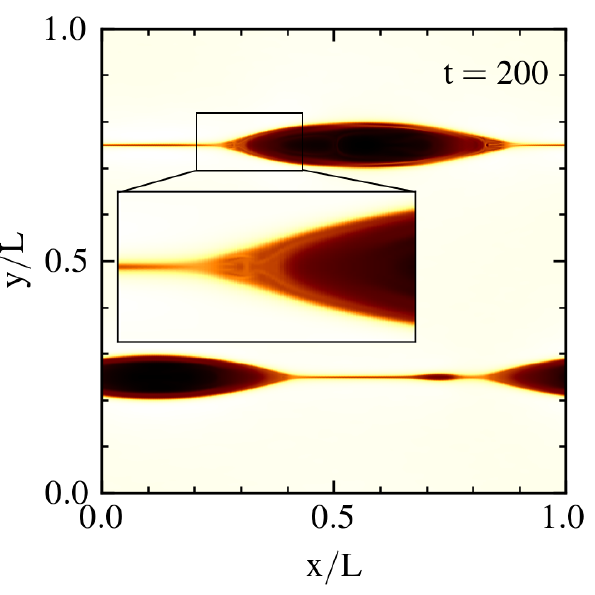}
\caption{Time evolution of the out-of-plane current density $J_z$ (top panels) 
and gas thermal pressure (bottom panels) of the magnetic reconnection simulation
performed with the implicit CT scheme. 
Each snapshot has been taken at the time (normalized to $t_A$) indicated in each 
panel. Note the development of the X-point reconnection regions for times $t \gsim 
100\times t_A$ in the $J_z$ snapshots, where the topology of the magnetic field 
is modified. The insets in the last panels show a magnified portion of the upper 
magnetic island.
}
\label{fig:magnreconn}
\end{figure*}

In this section, we present a first application of our ohmic resistivity 
implementation exploring the effects of magnetic reconnection. Magnetic 
reconnection is the rearrangement of the magnetic field topology that occurs in 
highly conducting plasmas with finite resistivity. During the reconnection phase, the energy that is present in the 
magnetic field can be rapidly converted into thermal and kinetic energy of the 
plasma. Therefore, this mechanism has been widely proposed as the key process 
that lies at the heart of eruptive events in the 
Sun~\citep{Zhu2016,Cheng2017,Seaton2017} or the heating of its 
corona~\citep[][see also~\citealt{Klimchuk2006} and references 
therein]{Parker1983}.

To study this process, we simulate the 
so-called tearing instability~\citep[][]{Furth1963}. In this configuration, 
magnetic fields of opposite polarity are connected by a thin current sheet. 
Upon perturbing this configuration, reconnection of the field is triggered, which 
eventually leads to the formation of magnetic islands with increasing size that 
eventually coalesce~\citep[][and references therein for numerical work done on the
instability]{Landi2012}.

To simulate the tearing instability we use an adapted version of the initial conditions presented 
in~\citet{Landi2008}. In particular, we 
use a 2D domain with side length $L_x = L_y = L = 6\pi$, 
which we simulate with $1024\times3072$ resolution elements. The larger number 
of resolution elements in the $y$-direction is necessary to resolve the steep 
gradients across the current sheets. The gas density is uniform and set 
to $\rho_0 = 1$. The initial conditions for this test start with a 
so-called~\citet{Harris1962} current sheet configuration, which is an 
equilibrium solution for ideal MHD equations (i.e. when the resistivity $\eta$ 
is put to zero). To employ periodic boundary conditions throughout (\FM{see discussion
at the beginning of Section~\ref{sec:tests}}), we 
use two of such current sheets of opposite polarity that are placed in the 
computational domain as 
\begin{equation}
 \boldsymbol{B}(y) = 
 \begin{cases}
 \displaystyle B_0\tanh\left[\delta \left(y - \frac{3 L_y}{4}\right)\right]\hat{e}_x & \mbox{if } y > \displaystyle\frac{L_y}{2}\\\\
 \displaystyle B_0\tanh\left[\delta \left(\frac{L_y}{4} - y\right)\right]\hat{e}_x & \mbox{if } y \leq \displaystyle\frac{L_y}{2},\\ 
 \end{cases}
 \label{eq:harrisB}
\end{equation}
where $B_0$ is the amplitude of the magnetic field at large distances from
the current sheet and $\delta = 10$ is its characteristic thickness.
Equilibrium is ensured by the condition
\begin{equation}\displaystyle
 P + \frac{||\boldsymbol{B}||^2}{2} = {\rm const},
\end{equation}
in which the gas thermal pressure $P$ counterbalances its magnetic counterpart. 
This condition can be rewritten as 
\begin{equation}\displaystyle
 P(y) = \frac{\beta + 1 - ||\boldsymbol{B}||^2}{2},
 \label{eq:harrisP}
\end{equation}
and $\beta$ can be interpreted as the ratio between
thermal and magnetic pressure in the plasma at large distances from the current sheet(s). 
We fix $\beta = 5$ in our runs, so magnetic fields are dynamically important in this
set-up. 
We then perturb this equilibrium solution by adding a component in velocity as
\begin{equation}
 \boldsymbol{v}(x,y) =
 \begin{cases}
 \displaystyle\epsilon\frac{\tanh\left[\delta \left(y - \displaystyle\frac{3\,L_y}{4}\right)\right]}{\cosh\left[\delta \left(y - \displaystyle\frac{3\,L_y}{4}\right)\right]}\sin(k_x x)\hat{e}_y & \mbox{if } y > \displaystyle\frac{L_y}{2}\\\\
 \displaystyle\epsilon\frac{\tanh\left[\delta \left(\displaystyle\frac{L_y}{4} - y\right)\right]}{\cosh\left[\delta \left(\displaystyle\frac{L_y}{4} - y\right)\right]}\sin(k_x x)\hat{e}_y & \mbox{if } y \leq \displaystyle\frac{L_y}{2},\\
 \end{cases}
 \label{eq:harrisv}
\end{equation}
where $\epsilon = 10^{-2}$, and $k_x = 2\pi m/ L_x$. For the wavelength of the perturbation we chose $m = 7$,
which~\citet{Landi2008} showed to be the fastest growing mode. 
We employ Alfv\'enic units so that lengths are normalized to a characteristic scale $L$, which we assume to be unity,
densities are normalized to a characteristic value $\rho_0 = 1$, magnetic fields are normalized to $B_0 = 1$, velocities are normalized 
to the Alfv\'en velocity $c_A =B_0/\sqrt{\rho_0}$, and times are normalized to $t_A = c_A / L$. 
The system is evolved up to the final time $t=250\,t_A$ with a resistivity 
$\eta = 2\times 10^{-4}$. 

For the CT scheme, a periodic vector potential that gives rise to the magnetic field in equation~(\ref{eq:harrisB}) is given by 
\begin{equation}
 \boldsymbol{A}(y) = 
 \begin{cases}
 \displaystyle\frac{B_0}{\delta}\ln\cosh\left[\delta \left(y - \frac{3L_y}{4}\right)\right]\hat{e}_z & \mbox{if } y > \displaystyle\frac{L_y}{2}\\\\
 \displaystyle\frac{B_0}{\delta}\left\{C - \ln\cosh\left[\delta \left(\frac{L_y}{4} - y\right)\right]\right\}\hat{e}_z & \mbox{if } y \leq \displaystyle\frac{L_y}{2},\\
 \end{cases}
 \label{eq:harrisA}
\end{equation}
where $C=2\ln\cosh\left(\delta L_y/4\right)$ is chosen to ensure the continuity of the vector potential at $y = L_y/2$. In the configuration that we have used in
this test problem, the average magnetic field is zero.

\begin{figure}
\centering
\includegraphics[width=0.46\textwidth]{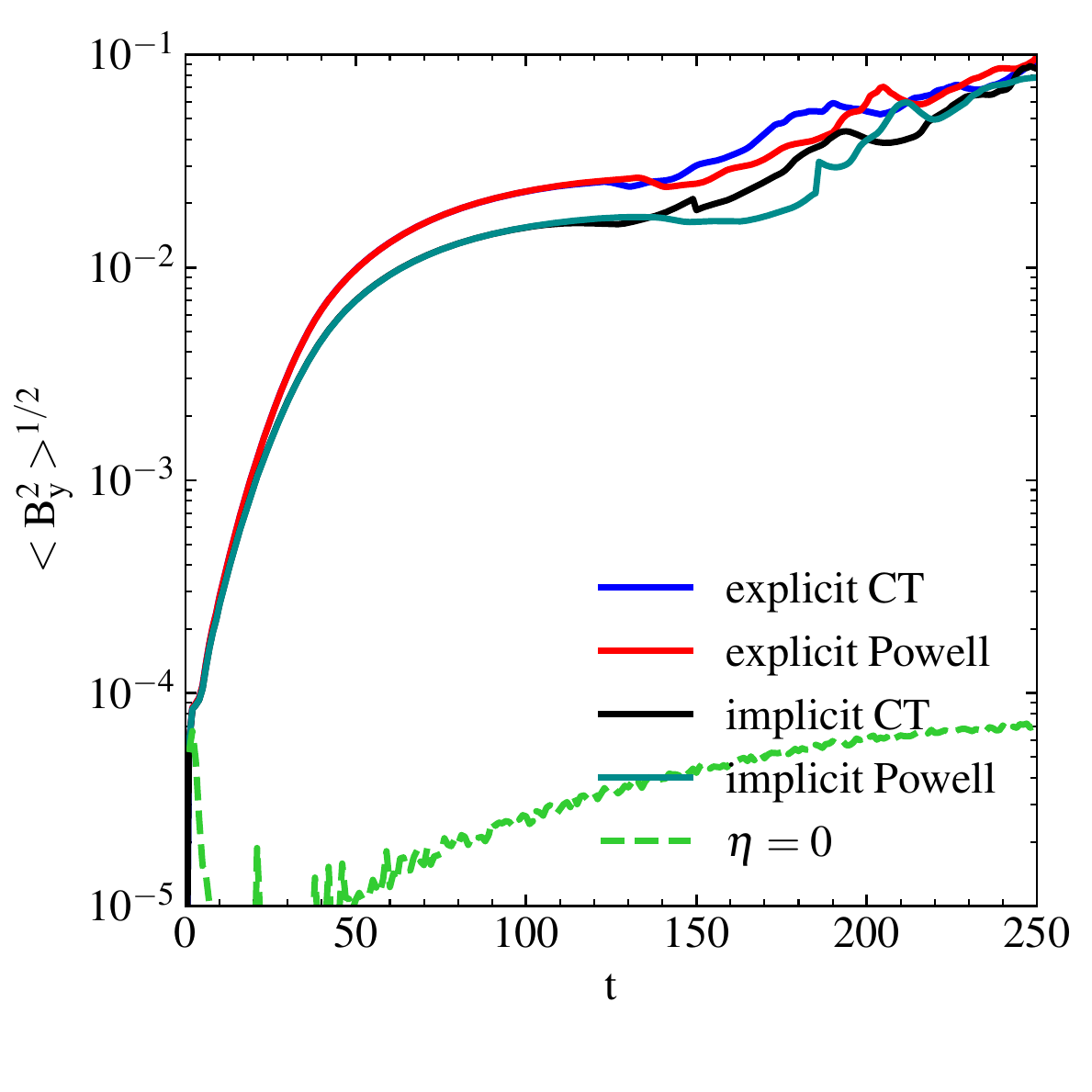}
\caption{Volume-weighted average magnetic field along the $y$-direction 
as a function of time for the magnetic reconnection simulation (tearing instability)
performed with different implementations of ohmic diffusivity as indicated in the legend.
After an exponential increase at early times, the growth rate decreases sensibly after 
$t \sim 50\times t_{A}$, although the field value keeps increasing steadily.
Overall, the Powell and CT schemes agree quite well in their predictions. However,
the implicit implementation predicts lower magnetic field values at times 
$t \gsim 25\times t_{A}$. The discrepancy is about a factor of 2 \FM{at $t \approx 150\, t_A$ to then reduce} at the end of the
examined time span. \FM{The green dashed line shows the evolution of the $B_y$ rms amplitude 
for a non-resistive plasma. Note that in this case the growth rate of the field is
much more reduced compared to the resistive simulation and the tearing instability 
does not develop although numerical reconnection is present to some extent.}
}
\label{fig:magrecBy}
\end{figure}

We present the results of this calculation in Fig.~\ref{fig:magnreconn} for the 
implicit CT scheme on a static mesh. We point out that all our other schemes yield essentially the same results (see also 
Fig~\ref{fig:magrecBy}). In the top six panels we show the time evolution of the 
out-of-plane current density vector $J_z = \nabla\times\boldsymbol{B}$ at the 
time indicated in the top right-hand corner. The bottom six panels are the analogous 
figure for the evolution of the gas thermal pressure. At early times, it is 
evident how the gradient in the gas thermal pressure, which reaches its maximum 
values at the locations of the current sheets, balances the opposite gradient in 
the magnetic pressure -- magnetic fields are zero at the sheet location, 
reaching their maximum amplitude far away from it (i.e. for $|y| \gg 1/\delta$). The 
thickness of the current sheets, indicated by the size of the coloured regions 
where $J_z$ is not zero, slowly increases with time due to the presence of ohmic 
diffusion. At around $t = 100\, t_A$ the linear perturbation added to the 
velocity also starts to be noticeable in $J_z$ with its characteristic $m = 7$ 
pattern. At $t = 150\,t_A$ the instability has fully developed in the non-linear 
regime and X shaped regions in $J_z$ are present. In these regions magnetic 
reconnection operates, changing the topology of the magnetic field, an effect 
that it is not possible in the ideal regime, and reorienting its direction from 
the $x$- to the $y$-axis. These reconnection points divide the current sheets in 
topological islands that coalesce at later times. The evolution of the pressure 
follows a trend akin to the current density, with similar morphological 
features. In the region where the current dissipation is maximal, i.e. mostly 
inside magnetic islands, the maximum of the pressure is also reached due to the 
intense associated ohmic heating.

In Fig.~\ref{fig:magrecBy} we present for all numerical schemes the time 
evolution of the volume-weighted rms values of the $B_y$ component as a proxy 
for the evolution of the instability. %
\begin{figure*}
\centering
\includegraphics[width=0.99\textwidth]{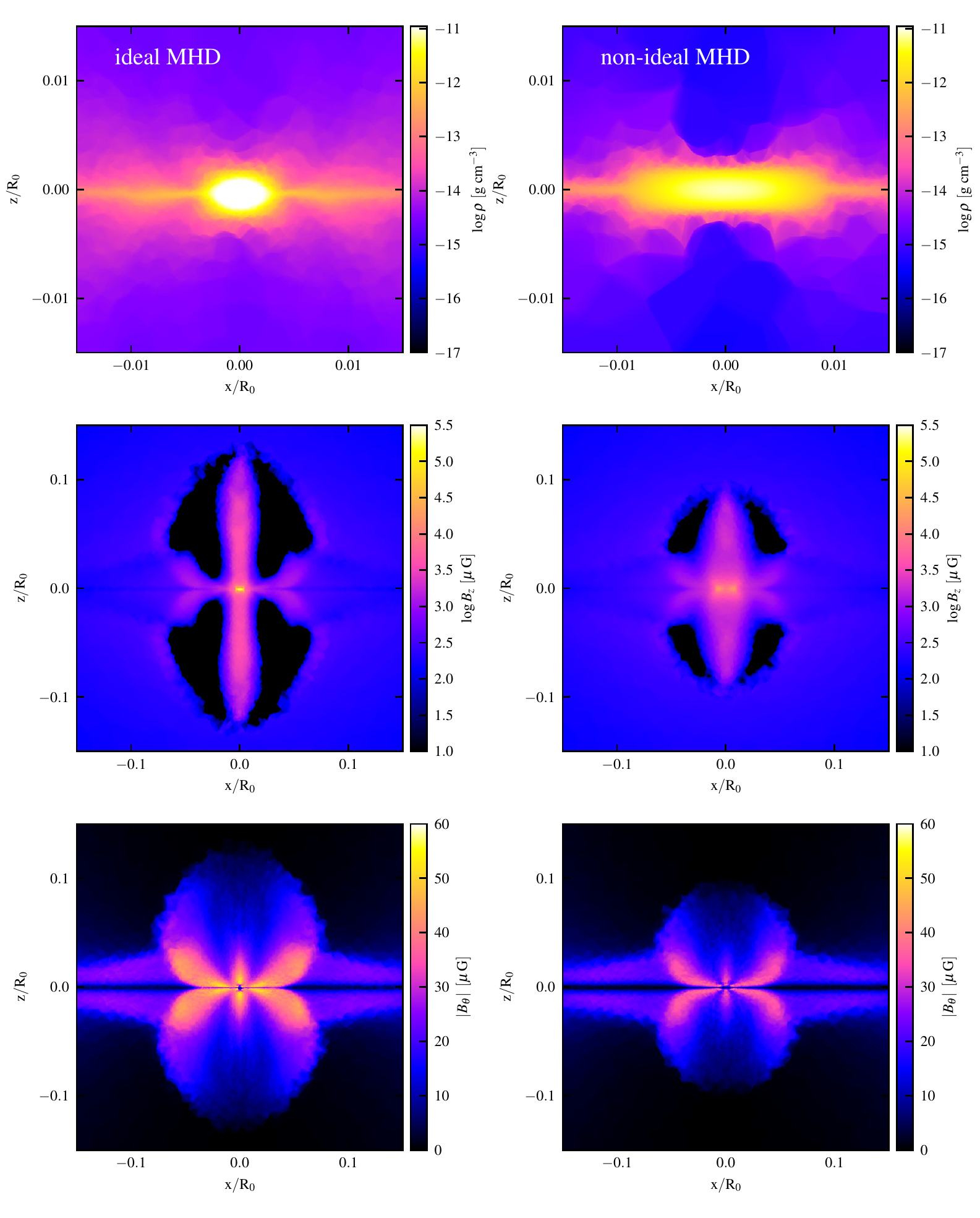}
\caption{Collapse of a magnetized cloud in the ideal (left-hand column) and non-ideal 
(right-hand column) MHD case. The panels show a slice (of depth equal to $0.2$ times
the side length of the projection) through the centre of the 
simulated domain in the $xz-$plane. The top row shows a zoom-in of the volume-weighted gas density 
on the central region ($0.03\,R_0$), where most of the mass of the cloud has collapsed, whereas 
the central and the bottom rows display the density-weighted magnetic field in the $z-$ and 
azimuthal directions on a larger scale ($0.3\, R_0$), respectively. The main 
effect of ohmic diffusivity in the calculation is to reduce the strength of 
magnetically-driven outflows (and of the global magnetic field strength) and to 
favour the formation of a larger disc-like structure in the central regions. 
All the panels are displayed at $t = 1.13\,t_{\rm ff}$.}
\label{fig:collapse}
\end{figure*}
The fraction of magnetic energy in the $y$-component of the field at the initial 
time is zero, so its evolution reflects the growth of the instability and the 
amount of reconnection occurring in the system. It is evident that in the linear 
regime of the instability (at very early times) the $B_y$ rms value increases 
exponentially. \FM{At very early times ($t \sim t_A$) there is a wiggle in the $B_y$ 
rms amplitude, which is likely due to our choice of perturbing the $y$-component
of the gas velocity only rather than \textit{both} gas velocity and magnetic field 
$y$-components based on an analytic solution of the tearing instability mode 
\citep[e.g.][]{Rembiasz2017}. For} $t \simeq 
50\, t_A$, the growth rate decreases sensibly, although the average $B_y$ field 
keeps steadily increasing. In general, the Powell and the CT schemes give 
consistent results across all the examined time span. There is a difference in 
the final values of the $B_y$ component between the explicit and implicit time 
integration, with the latter giving consistently lower values after a time of $t 
\gsim 25\,t_A$ has elapsed. The difference reaches a maximum of about a factor 
of $2$ at late times \FM{($t \approx 150\,{t_A}$) to then reduce at the 
end of the simulated time span}. This trend is an indication that the implicit schemes are 
slightly more diffusive than their explicit counterparts. We ascribe this 
behaviour to the first-order \FM{and non-strictly conservative} treatment of the 
Joule heating term in the implicit schemes (see equation~\ref{eq:impl_erg}). 
To \FM{investigate whether this could be the cause of the observed difference}, 
we reran the magnetic reconnection test with the explicit schemes, but using the 
same first-order treatment for the Joule heating term as in the implicit 
implementation. We find \FM{a closer} agreement in the evolution of the rms 
$B_y$ values in this case between explicit and implicit schemes, thus confirming 
that the additional diffusivity is caused by the treatment of the Joule term. 
\FM{The green dashed line shows the evolution of the $B_y$ rms amplitude 
for a non-resistive plasma, in which $\eta$ has been fixed to zero. Theoretically,
the tearing instability, and the associated magnetic reconnection, can not develop 
in this configuration. However, any code introduces a finite amount of numerical 
resistivity due to the discretization of the equations governing the system. This numerical 
resistivity, which is dependent on resolution, can lead to magnetic reconnection that is entirely 
numerical in nature, and if it is large enough trigger the onset of the instability. 
It can be appreciated from the figure that there is an increase of the $B_y$ rms 
amplitude due to this effect. However, the growth is much more reduced compared to the resistive 
simulation and the tearing instability does not develop at the resolution presented here.
This is not the case for lower resolution realizations of this set-up, in which the amplitudes reached
by the $B_y$ field are comparable to the ones obtained with the onset of the tearing instability mode.
In particular, degrading the resolution by a factor of 4 in both directions leads to the 
development of the instability for purely numerical reasons.}
Summarizing, these results illustrate the ability of our implementations to 
handle complex non-ideal MHD applications, which include ohmic resistivity. 

\section{Magnetized cloud collapse}\label{sec:collapse}

As another application of our scheme, we study next the gravitational collapse of a magnetized 
sphere and compare the outcome of simulations performed in the ideal and 
non-ideal MHD case. This system represents an important astrophysical problem as this 
set-up can be considered as an idealized model of the formation of a protostar. 

The initial conditions for this problem are taken from~\citet{Pakmor2011}, which 
are an adaptation of those presented in~\citet{Hennebelle2008b}. They consist of 
a spherical cloud of uniform density with a radius of $R_0 = 0.015\,{\rm pc}$. The 
cloud is embedded in a more tenuous atmosphere with a small transition region 
at the boundary. The initial mass of the cloud is $1\,M_\odot$, which implies an 
initial density of $4.8\times 10^{-18}\,{\rm g\,cm^{-3}}$. With this initial 
density the free-fall time is $3\times10^{4}\,{\rm yr}$. The atmosphere surrounding the 
cloud is $100$ times less dense than the cloud. At the beginning of the calculation the 
gas in the cloud rotates as a rigid body with a period of $4.7\times10^{5}\,{\rm yr}$. 
The  simulation domain is a box of side length $0.06\,{\rm pc}$ and is filled with a 
uniform magnetic field with a strength of $30\,\muG$ directed in the same 
direction of the angular momentum of the gas. The gas follows a barotropic 
equation of state given by \citep[see][]{Hennebelle2008b} 
\begin{equation}
 P = \rho c_0^2 \sqrt{1 + (\rho/\rho_c)^{4/3}},
 \end{equation}
where $c_0 = 0.2\,\kms$ and $\rho_c = 10^{-13}\,{\rm g\,cm^{-3}}$. 
Inflow/outflow boundary conditions are applied at all sides of the domain. We 
start the simulation with a Cartesian mesh with $128^3$ cells, but we 
allow for the refinement of gas cells whose free-fall time-scale becomes smaller 
than $10$ times its sound-crossing time-scale. With this criterion, we basically 
resolve the local Jeans length with at least $10$ resolution elements. To avoid an 
excessive number of gas cells as the simulation progresses, we limit their 
volume to a minimum value of $5\times10^{-17}\,{\rm pc^{3}}$, which is 
equivalent to an effective resolution of $16384^3$ resolution elements 
\citep[see][]{Hennebelle2008b, Pakmor2011}. In the simulation with ohmic 
resistivity, performed with the explicit Powell scheme, we use a 
spatially constant resistivity $\eta = 10^{18}\,{\rm cm^{2}\,s^{-1}}$.
We note that this calculation is meant to be an idealized collapse model, and we
therefore do not account for the variation of resistivity with gas properties
\FM{(such as chemical composition and ionization state), a task that is non-trivial 
and outside the scope of this paper}. However, the chosen resistivity value is appropriate 
for densities $n \gsim 10^{12}\,{\rm cm^{-3}} \simeq 1.67\times 10^{-12}\, {\rm g\,cm^{-3}}$,
assuming a fully hydrogen composition \citep[see also][Fig. 1]{Machida2007}. 
These densities are reached in the regions surrounding the protostar in our set-up.

\FM{The choice of the explicit Powell scheme was also adopted on the basis that the ratio 
between the resistive and the CFL time-step (see equation~\ref{eq:explicittstep}) 
reaches a minimum value of about four at the end of the simulated time span ($1.13 
t_{\rm ff}$). So in terms of the size of the time-step, the advantages of using an 
implicit scheme are limited for this set-up. However, we would like to mention two important 
aspects: (i) the value of the resistivity might be larger than the one that we 
have adopted at higher densities \citep{Nakano2002,Machida2007}, and (ii) we have imposed a minimum size to the gas 
cells in the simulation, effectively limiting the maximum resolution that can be 
achieved. Both factors contribute to keep the ratio of time-steps large enough 
that explicit schemes are a more convenient choice with respect to implicit schemes for this 
particular set-up. Increasing the resistivity value or the resolution of the simulation 
(or a combination of both) may render implicit schemes competitive 
with explicit schemes for this calculation.}

Fig.~\ref{fig:collapse} presents the output of the simulations in the ideal 
(left-hand column) and resistive (right-hand column) cases at the final time $t = 
1.13\,t_{\rm ff}$. The rows show slices (of depth equal to $0.2$ times the side 
length of the projection) through the centre of the simulated domain in the 
$xz$-plane (the $z$-axis coincides with the cloud's rotation axis) of the 
volume-weighted gas density (top) and the density-weighted magnetic field in the 
$z$- (middle) and azimuthal (bottom) directions. The density panels display the 
results on a smaller scale ($0.03\,R_0$) compared to the magnetic field panels 
($0.3\,R_0$). In the ideal MHD case, results are similar to those found 
by~\citet{Pakmor2011}. At the centre of the domain a protostar is formed, which 
is surrounded by a disc of material. Compression of the gas due to the collapse 
has amplified the initial magnetic field to values of about $10^5\,\muG$ close 
to the protostar in the $z$-direction and to $\sim 70\,\muG$ in the azimuthal 
direction immediately above and below the mid-plane of the disc. The 
amplification of the magnetic field also causes the launching of 
magnetically driven outflows reaching distances in excess of $\sim 0.1\,R_0$ 
from the protostar in the $z$-direction. The inclusion of ohmic resistivity 
changes this picture. In particular, the amplification of the field is less 
pronounced because of the diffusive effects. As a consequence, gas outflows are 
less strong (i.e. they reach a smaller distance from the protostar) and also 
the gas distribution in the protostar region is different, featuring a more 
thick and extended disc-like structure. These results are in line with numerical 
studies of star-forming clouds highlighting the importance of ohmic diffusion on 
the transport of angular momentum \citep[e.g.][]{Dapp2010} and the generation of 
magnetically driven gas outflows \citep[e.g.][]{Matsushita2017}, and further 
validate the applicability of our non-ideal MHD schemes to complex astrophysical 
systems. \FM{We caution again that the detailed effects of ohmic resistivity on the 
cloud collapse depend on the exact value of $\eta$ as a function of the gas properties. This is 
a non-trivial task to accomplish and our simulations have not attempted 
such detailed modelling resorting, instead, to a constant value for the resistivity.}

\section{Summary and conclusions}\label{sec:conclusions}

Magnetic fields are an essential component of many physical processes that 
influence the evolution of the objects populating the Universe. Although in many 
astrophysical circumstances magnetic fields can be well modelled in the ideal MHD 
approximation, there are phenomena in which non-ideal effects such as ohmic 
resistivity, ambipolar diffusion and the Hall effect play an essential role. It 
is therefore desirable to extend the capabilities of numerical MHD codes to 
treat such non-ideal terms in order to faithfully model these phenomena. 

In this paper, we have made a step in this direction by focusing on the 
inclusion of ohmic terms, which appear in the MHD equations when the gas 
resistivity is non-zero, in the moving-mesh code \arepo. The code has two main 
approaches for treating MHD, namely a~\citet{Powell1999} 
divergence cleaning scheme and a CT method~\citep{Mocz2014,Mocz2016} that 
evolves the vector potential to ensure the $\nabla\cdot\boldsymbol{B} = 0$ 
constraint. We have implemented the resistive terms for both techniques with 
explicit and implicit time integration. This allows for a high degree of 
flexibility in treating MHD problems in which diffusivity plays a role. In 
particular, the implicit time integration treatment makes it possible to 
circumvent the restrictive time-step CFL condition ($\propto \Delta x^{-2}$) 
necessary to guarantee the stability of explicit time integration schemes for 
diffusive phenomena. These explicit schemes are adopted in many non-ideal MHD 
simulation codes~\citep[see e.g.][]{Masson2012, Mignone2012, Hopkins2017}, owing 
to their relatively simple implementation. However, the quadratic spatial 
resolution scaling of their CFL condition renders them impractical for 
high-resolution applications.

We have tested our implementation in problems of increasing physical complexity. 
We have first confirmed that the magnetic field properly diffuses, in the 
absence of any gas dynamics, in all our implementations. To this end we have 
performed a classical 1D diffusion test of a Gaussian magnetic 
field configuration recovering the expected evolution. We have also extended 
this test to a 2D configuration and found that all our 
implementations yielded the expected results. In particular, we demonstrated 
that, regardless of the scheme employed, second-order convergence is achieved.

We have then proceeded to include gas dynamics in our test problems by studying 
the decay of Alfv\'en waves due to a finite resistivity of the plasma. We have 
tested all our schemes in two different initial configurations: a progressive 
wave and a superposition of two waves travelling in opposite directions that 
give rise to a stationary wave configuration. In both cases, all the schemes 
that we have implemented recovered the expected exponential decay of the 
magnetic field strength, and showed second-order convergence also in the 
presence of gas dynamics. We note that ohmic resistivity not only causes the 
magnetic field to diffuse -- and, in particular, to decay exponentially in this 
problem -- but also increases the plasma temperature through Joule dissipation. 
In the diffusion of an Alfv\'en wave (both in the progressive and stationary 
configurations), Joule dissipation increases uniformly the gas pressure as the 
intensity of the magnetic field declines. This behaviour is captured correctly 
by our schemes, \FM{although the treatment of Joule heating is different (second 
versus first-order accurate) between explicit and implicit schemes, and this 
difference may sometimes have a more pronounced impact on the results (see 
Fig.~\ref{fig:magrecBy}).}

As a first application, we have investigated magnetic reconnection in a plasma 
configuration that develops the tearing instability~\citep{Furth1963}. The study 
of the emergence of this instability is complicated by the fact that any 
numerical scheme introduces non-physical numerical resistivity due to the 
discretization procedure. This numerical resistivity can affect the results, 
especially in the low-resistivity regime, which is interesting for the modelling 
of real systems such as the solar corona. It is therefore important that the 
level of numerical resistivity is lower than the physical resistivity that is 
considered in the calculations, which can be achieved by adopting a high enough 
resolution in the simulation. We took care of this aspect by first running a 
version of this problem with zero resistivity for increasingly high resolution 
until no instability due to numerical effects was present in the calculation. We 
then introduced physical resistivity in the system and studied its evolution. 
All our schemes were able to capture the onset and the evolution of the 
instability into to the non-linear regime. Furthermore, our simulations clearly 
showed the emergence of X shaped regions in the out-of-plane current density 
$J_z$, demonstrating that intense magnetic reconnection is occurring. These 
regions of strong magnetic reconnection divide the plasma in magnetic islands 
that eventually coalesce.

Finally, to further test our implementation on a problem directly relevant for 
astrophysical applications and in particular for star formation studies, we have 
examined the gravitational collapse of a magnetized rotating 
cloud~\citep{Hennebelle2008b}. We have demonstrated that for high-enough, but 
admissible, values of the ohmic resistivity there are visible effects on the 
density gas distribution around the emerging protostar, the amplification of 
the magnetic field due to the collapse, and the strength of the 
magnetically driven outflows. In particular, compared to the ideal MHD 
case~\citep[see also][]{Pakmor2011}, the gas in the vicinity of the protostar 
is distributed in a more thick and extended disc-like structure, the final 
magnetic field strength is lower and the resulting gas outflows are weaker and 
less extended, in broad agreement with previous non-ideal MHD 
work~\citep[e.g.][]{Dapp2010, Matsushita2017}.

To conclude, we have presented a first implementation of non-ideal MHD terms in 
the moving-mesh code \arepo. Interesting applications of the new code 
capabilities include the study of massive star formation in atomic cooling 
haloes~\citep{Becerra2015}, or the role of magnetic fields on small-scale star 
formation~\citep{Hull2017} and its correlations to supersonic turbulence in 
star-forming cores~\citep{Mocz2017}. We intend to pursue these lines of research 
in future work.

\section*{Acknowledgements}
\FM{We thank the referee for their detailed comments on the 
manuscript.} MV acknowledges support through an MIT RSC award, the support of 
the Alfred P. Sloan Foundation, and support by NASA ATP grant NNX17AG29G. RK and 
PM acknowledge support from NASA through Einstein Postdoctoral Fellowship grant 
numbers PF7-180163 (RK) and PF7-180164 (PM) awarded by the \textit{Chandra} 
X-ray Center, which is operated by the Smithsonian Astrophysical Observatory for 
NASA under contract NAS8-03060. VS acknowledges support through sub-project 
EXAMAG of the Priority Programme 1648 SPPEXA of the German Science Foundation. 
VS and RP are also supported by the European Research Council through ERC-StG 
grant EXAGAL-308037. The simulations were performed on the joint MIT-Harvard 
computing cluster supported by MKI and FAS. All the figures in this work were 
created with the {\sc matplotlib} graphics environment~\citep{Matplotlib}.

\bibliographystyle{mnras}                              
\bibliography{paper}

\begin{thebibliography}{}
\makeatletter
\relax
\def\mn@urlcharsother{\let\do\@makeother \do\$\do\&\do\#\do\^\do\_\do\%\do\~}
\def\mn@doi{\begingroup\mn@urlcharsother \@ifnextchar [ {\mn@doi@}
  {\mn@doi@[]}}
\def\mn@doi@[#1]#2{\def\@tempa{#1}\ifx\@tempa\@empty \href
  {http://dx.doi.org/#2} {doi:#2}\else \href {http://dx.doi.org/#2} {#1}\fi
  \endgroup}
\def\mn@eprint#1#2{\mn@eprint@#1:#2::\@nil}
\def\mn@eprint@arXiv#1{\href {http://arxiv.org/abs/#1} {{\tt arXiv:#1}}}
\def\mn@eprint@dblp#1{\href {http://dblp.uni-trier.de/rec/bibtex/#1.xml}
  {dblp:#1}}
\def\mn@eprint@#1:#2:#3:#4\@nil{\def\@tempa {#1}\def\@tempb {#2}\def\@tempc
  {#3}\ifx \@tempc \@empty \let \@tempc \@tempb \let \@tempb \@tempa \fi \ifx
  \@tempb \@empty \def\@tempb {arXiv}\fi \@ifundefined
  {mn@eprint@\@tempb}{\@tempb:\@tempc}{\expandafter \expandafter \csname
  mn@eprint@\@tempb\endcsname \expandafter{\@tempc}}}

\bibitem[\protect\citeauthoryear{{Bai}}{{Bai}}{2015}]{Bai2015}
{Bai} X.-N.,  2015, \mn@doi [\apj] {10.1088/0004-637X/798/2/84}, \href
  {http://adsabs.harvard.edu/abs/2015ApJ...798...84B} {798, 84}

\bibitem[\protect\citeauthoryear{{Basu} \& {Ciolek}}{{Basu} \&
  {Ciolek}}{2004}]{Basu2004}
{Basu} S.,  {Ciolek} G.~E.,  2004, \mn@doi [\apjl] {10.1086/421464}, \href
  {http://adsabs.harvard.edu/abs/2004ApJ...607L..39B} {607, L39}

\bibitem[\protect\citeauthoryear{{Basu} \& {Dapp}}{{Basu} \&
  {Dapp}}{2010}]{Basu2010}
{Basu} S.,  {Dapp} W.~B.,  2010, \mn@doi [\apj] {10.1088/0004-637X/716/1/427},
  \href {http://adsabs.harvard.edu/abs/2010ApJ...716..427B} {716, 427}

\bibitem[\protect\citeauthoryear{{Becerra}, {Greif}, {Springel}  \&
  {Hernquist}}{{Becerra} et~al.}{2015}]{Becerra2015}
{Becerra} F.,  {Greif} T.~H.,  {Springel} V.,   {Hernquist} L.~E.,  2015,
  \mn@doi [\mnras] {10.1093/mnras/stu2284}, \href
  {http://adsabs.harvard.edu/abs/2015MNRAS.446.2380B} {446, 2380}

\bibitem[\protect\citeauthoryear{{Beck} \& {Wielebinski}}{{Beck} \&
  {Wielebinski}}{2013}]{Beck2013c}
{Beck} R.,  {Wielebinski} R.,  2013, {Magnetic Fields in Galaxies}.
p.~641, \mn@doi{10.1007/978-94-007-5612-0_13}

\bibitem[\protect\citeauthoryear{{B{\'e}thune}, {Lesur}  \&
  {Ferreira}}{{B{\'e}thune} et~al.}{2017}]{Bethune2017}
{B{\'e}thune} W.,  {Lesur} G.,   {Ferreira} J.,  2017, \mn@doi [\aap]
  {10.1051/0004-6361/201630056}, \href
  {http://adsabs.harvard.edu/abs/2017A%26A...600A..75B} {600, A75}

\bibitem[\protect\citeauthoryear{{Cheng}, {Guo}  \& {Ding}}{{Cheng}
  et~al.}{2017}]{Cheng2017}
{Cheng} X.,  {Guo} Y.,   {Ding} M.,  2017, \mn@doi [Science in China Earth
  Sciences] {10.1007/s11430-017-9074-6}, \href
  {http://adsabs.harvard.edu/abs/2017ScChE..60.1383C} {60, 1383}

\bibitem[\protect\citeauthoryear{{Cox}}{{Cox}}{2005}]{Cox2005}
{Cox} D.~P.,  2005, \mn@doi [\araa] {10.1146/annurev.astro.43.072103.150615},
  \href {http://adsabs.harvard.edu/abs/2005ARA%26A..43..337C} {43, 337}

\bibitem[\protect\citeauthoryear{{Crank}, {Nicolson}  \& {Hartree}}{{Crank}
  et~al.}{1947}]{Crank1947}
{Crank} J.,  {Nicolson} P.,   {Hartree} D.~R.,  1947, \mn@doi [Proceedings of
  the Cambridge Philosophical Society] {10.1017/S0305004100023197}, \href
  {http://adsabs.harvard.edu/abs/1947PCPS...43...50C} {43, 50}

\bibitem[\protect\citeauthoryear{{Dapp} \& {Basu}}{{Dapp} \&
  {Basu}}{2010}]{Dapp2010}
{Dapp} W.~B.,  {Basu} S.,  2010, \mn@doi [\aap] {10.1051/0004-6361/201015700},
  \href {http://adsabs.harvard.edu/abs/2010A%26A...521L..56D} {521, L56}

\bibitem[\protect\citeauthoryear{{Dolag} \& {Stasyszyn}}{{Dolag} \&
  {Stasyszyn}}{2009}]{Dolag2009}
{Dolag} K.,  {Stasyszyn} F.,  2009, \mn@doi [\mnras]
  {10.1111/j.1365-2966.2009.15181.x}, \href
  {http://adsabs.harvard.edu/abs/2009MNRAS.398.1678D} {398, 1678}

\bibitem[\protect\citeauthoryear{{Dolag}, {Bartelmann}  \& {Lesch}}{{Dolag}
  et~al.}{1999}]{Dolag1999}
{Dolag} K.,  {Bartelmann} M.,   {Lesch} H.,  1999, \aap, \href
  {http://adsabs.harvard.edu/abs/1999A%26A...348..351D} {348, 351}

\bibitem[\protect\citeauthoryear{{Dolag}, {Bartelmann}  \& {Lesch}}{{Dolag}
  et~al.}{2002}]{Dolag2002}
{Dolag} K.,  {Bartelmann} M.,   {Lesch} H.,  2002, \mn@doi [\aap]
  {10.1051/0004-6361:20020241}, \href
  {http://adsabs.harvard.edu/abs/2002A%26A...387..383D} {387, 383}

\bibitem[\protect\citeauthoryear{{Dolag}, {Komatsu}  \& {Sunyaev}}{{Dolag}
  et~al.}{2016}]{dolag2016}
{Dolag} K.,  {Komatsu} E.,   {Sunyaev} R.,  2016, \mn@doi [\mnras]
  {10.1093/mnras/stw2035}, \href
  {http://adsabs.harvard.edu/abs/2016MNRAS.463.1797D} {463, 1797}

\bibitem[\protect\citeauthoryear{{Falle}}{{Falle}}{2003}]{Falle2003}
{Falle} S.~A.~E.~G.,  2003, \mn@doi [\mnras]
  {10.1046/j.1365-8711.2003.06908.x}, \href
  {http://adsabs.harvard.edu/abs/2003MNRAS.344.1210F} {344, 1210}

\bibitem[\protect\citeauthoryear{{Feretti}, {Giovannini}, {Govoni}  \&
  {Murgia}}{{Feretti} et~al.}{2012}]{Feretti2012}
{Feretti} L.,  {Giovannini} G.,  {Govoni} F.,   {Murgia} M.,  2012, \mn@doi
  [\aapr] {10.1007/s00159-012-0054-z}, \href
  {http://adsabs.harvard.edu/abs/2012A%26ARv..20...54F} {20, 54}

\bibitem[\protect\citeauthoryear{{Fermi}}{{Fermi}}{1949}]{Fermi1949}
{Fermi} E.,  1949, \mn@doi [Physical Review] {10.1103/PhysRev.75.1169}, \href
  {http://adsabs.harvard.edu/abs/1949PhRv...75.1169F} {75, 1169}

\bibitem[\protect\citeauthoryear{{Ferri{\`e}re}}{{Ferri{\`e}re}}{2001}]{Ferriere2001}
{Ferri{\`e}re} K.~M.,  2001, \mn@doi [Reviews of Modern Physics]
  {10.1103/RevModPhys.73.1031}, \href
  {http://adsabs.harvard.edu/abs/2001RvMP...73.1031F} {73, 1031}

\bibitem[\protect\citeauthoryear{{Fromang}, {Hennebelle}  \&
  {Teyssier}}{{Fromang} et~al.}{2006}]{Fromang2006}
{Fromang} S.,  {Hennebelle} P.,   {Teyssier} R.,  2006, \mn@doi [\aap]
  {10.1051/0004-6361:20065371}, \href
  {http://adsabs.harvard.edu/abs/2006A%26A...457..371F} {457, 371}

\bibitem[\protect\citeauthoryear{{Furth}, {Killeen}  \& {Rosenbluth}}{{Furth}
  et~al.}{1963}]{Furth1963}
{Furth} H.~P.,  {Killeen} J.,   {Rosenbluth} M.~N.,  1963, \mn@doi [Physics of
  Fluids] {10.1063/1.1706761}, \href
  {http://adsabs.harvard.edu/abs/1963PhFl....6..459F} {6, 459}

\bibitem[\protect\citeauthoryear{{Gressel}, {Turner}, {Nelson}  \&
  {McNally}}{{Gressel} et~al.}{2015}]{Gressel2015}
{Gressel} O.,  {Turner} N.~J.,  {Nelson} R.~P.,   {McNally} C.~P.,  2015,
  \mn@doi [\apj] {10.1088/0004-637X/801/2/84}, \href
  {http://adsabs.harvard.edu/abs/2015ApJ...801...84G} {801, 84}

\bibitem[\protect\citeauthoryear{{Harris}}{{Harris}}{1962}]{Harris1962}
{Harris} E.~G.,  1962, {Nuovo Cim.}, \href
  {http://adsabs.harvard.edu/abs/2016arXiv160207703H} {23, 115}

\bibitem[\protect\citeauthoryear{{Hennebelle} \& {Fromang}}{{Hennebelle} \&
  {Fromang}}{2008}]{Hennebelle2008b}
{Hennebelle} P.,  {Fromang} S.,  2008, \mn@doi [\aap]
  {10.1051/0004-6361:20078309}, \href
  {http://adsabs.harvard.edu/abs/2008A%26A...477....9H} {477, 9}

\bibitem[\protect\citeauthoryear{{Hennebelle} \& {Teyssier}}{{Hennebelle} \&
  {Teyssier}}{2008}]{Hennebelle2008}
{Hennebelle} P.,  {Teyssier} R.,  2008, \mn@doi [\aap]
  {10.1051/0004-6361:20078310}, \href
  {http://adsabs.harvard.edu/abs/2008A%26A...477...25H} {477, 25}

\bibitem[\protect\citeauthoryear{{Hennebelle}, {Commer{\c c}on}, {Joos},
  {Klessen}, {Krumholz}, {Tan}  \& {Teyssier}}{{Hennebelle}
  et~al.}{2011}]{Hennebelle2011}
{Hennebelle} P.,  {Commer{\c c}on} B.,  {Joos} M.,  {Klessen} R.~S.,
  {Krumholz} M.,  {Tan} J.~C.,   {Teyssier} R.,  2011, \mn@doi [\aap]
  {10.1051/0004-6361/201016052}, \href
  {http://adsabs.harvard.edu/abs/2011A%26A...528A..72H} {528, A72}

\bibitem[\protect\citeauthoryear{Henson \& Yang}{Henson \&
  Yang}{2002}]{Henson2002}
Henson V.~E.,  Yang U.~M.,  2002, \mn@doi [Applied Numerical Mathematics]
  {http://dx.doi.org/10.1016/S0168-9274(01)00115-5}, 41, 155

\bibitem[\protect\citeauthoryear{{Hopkins}}{{Hopkins}}{2017}]{Hopkins2017}
{Hopkins} P.~F.,  2017, \mn@doi [\mnras] {10.1093/mnras/stw3306}, \href
  {http://adsabs.harvard.edu/abs/2017MNRAS.466.3387H} {466, 3387}

\bibitem[\protect\citeauthoryear{{Hopkins} \& {Raives}}{{Hopkins} \&
  {Raives}}{2016}]{Hopkins2016b}
{Hopkins} P.~F.,  {Raives} M.~J.,  2016, \mn@doi [\mnras]
  {10.1093/mnras/stv2180}, \href
  {http://adsabs.harvard.edu/abs/2016MNRAS.455...51H} {455, 51}

\bibitem[\protect\citeauthoryear{{Hull} et~al.,}{{Hull}
  et~al.}{2017}]{Hull2017}
{Hull} C.~L.~H.,  et~al., 2017, \mn@doi [\apjl] {10.3847/2041-8213/aa71b7},
  \href {http://adsabs.harvard.edu/abs/2017ApJ...842L...9H} {842, L9}

\bibitem[\protect\citeauthoryear{{Hunter}}{{Hunter}}{2007}]{Matplotlib}
{Hunter} J.~D.,  2007, \mn@doi [Computing In Science \& Engineering]
  {10.1109/MCSE.2007.55}, 9, 90

\bibitem[\protect\citeauthoryear{{Iffrig} \& {Hennebelle}}{{Iffrig} \&
  {Hennebelle}}{2017}]{Iffrig2017}
{Iffrig} O.,  {Hennebelle} P.,  2017, \mn@doi [\aap]
  {10.1051/0004-6361/201630290}, \href
  {http://adsabs.harvard.edu/abs/2017A%26A...604A..70I} {604, A70}

\bibitem[\protect\citeauthoryear{{Kannan}, {Springel}, {Pakmor}, {Marinacci}
  \& {Vogelsberger}}{{Kannan} et~al.}{2016}]{Kannan2016}
{Kannan} R.,  {Springel} V.,  {Pakmor} R.,  {Marinacci} F.,   {Vogelsberger}
  M.,  2016, \mn@doi [\mnras] {10.1093/mnras/stw294}, \href
  {http://adsabs.harvard.edu/abs/2016MNRAS.458..410K} {458, 410}

\bibitem[\protect\citeauthoryear{{Kannan}, {Vogelsberger}, {Pfrommer},
  {Weinberger}, {Springel}, {Hernquist}, {Puchwein}  \& {Pakmor}}{{Kannan}
  et~al.}{2017}]{Kannan2017}
{Kannan} R.,  {Vogelsberger} M.,  {Pfrommer} C.,  {Weinberger} R.,  {Springel}
  V.,  {Hernquist} L.,  {Puchwein} E.,   {Pakmor} R.,  2017, \mn@doi [\apjl]
  {10.3847/2041-8213/aa624b}, \href
  {http://adsabs.harvard.edu/abs/2017ApJ...837L..18K} {837, L18}

\bibitem[\protect\citeauthoryear{{Klimchuk}}{{Klimchuk}}{2006}]{Klimchuk2006}
{Klimchuk} J.~A.,  2006, \mn@doi [\solphys] {10.1007/s11207-006-0055-z}, \href
  {http://adsabs.harvard.edu/abs/2006SoPh..234...41K} {234, 41}

\bibitem[\protect\citeauthoryear{{Kotera} \& {Olinto}}{{Kotera} \&
  {Olinto}}{2011}]{Kotera2011}
{Kotera} K.,  {Olinto} A.~V.,  2011, \mn@doi [\araa]
  {10.1146/annurev-astro-081710-102620}, \href
  {http://adsabs.harvard.edu/abs/2011ARA%26A..49..119K} {49, 119}

\bibitem[\protect\citeauthoryear{{Krasnopolsky}, {Li}  \&
  {Shang}}{{Krasnopolsky} et~al.}{2010}]{Krasnopolsky2010}
{Krasnopolsky} R.,  {Li} Z.-Y.,   {Shang} H.,  2010, \mn@doi [\apj]
  {10.1088/0004-637X/716/2/1541}, \href
  {http://adsabs.harvard.edu/abs/2010ApJ...716.1541K} {716, 1541}

\bibitem[\protect\citeauthoryear{{Landi} \& {Bettarini}}{{Landi} \&
  {Bettarini}}{2012}]{Landi2012}
{Landi} S.,  {Bettarini} L.,  2012, \mn@doi [\ssr] {10.1007/s11214-011-9824-6},
  \href {http://adsabs.harvard.edu/abs/2012SSRv..172..253L} {172, 253}

\bibitem[\protect\citeauthoryear{{Landi}, {Londrillo}, {Velli}  \&
  {Bettarini}}{{Landi} et~al.}{2008}]{Landi2008}
{Landi} S.,  {Londrillo} P.,  {Velli} M.,   {Bettarini} L.,  2008, \mn@doi
  [Physics of Plasmas] {10.1063/1.2825006}, \href
  {http://adsabs.harvard.edu/abs/2008PhPl...15a2302L} {15, 012302}

\bibitem[\protect\citeauthoryear{{Lesur}, {Kunz}  \& {Fromang}}{{Lesur}
  et~al.}{2014}]{Lesur2014}
{Lesur} G.,  {Kunz} M.~W.,   {Fromang} S.,  2014, \mn@doi [\aap]
  {10.1051/0004-6361/201423660}, \href
  {http://adsabs.harvard.edu/abs/2014A%26A...566A..56L} {566, A56}

\bibitem[\protect\citeauthoryear{{Li}, {McKee}, {Klein}  \& {Fisher}}{{Li}
  et~al.}{2008}]{Li2008}
{Li} P.~S.,  {McKee} C.~F.,  {Klein} R.~I.,   {Fisher} R.~T.,  2008, \mn@doi
  [\apj] {10.1086/589874}, \href
  {http://adsabs.harvard.edu/abs/2008ApJ...684..380L} {684, 380}

\bibitem[\protect\citeauthoryear{{Li}, {Krasnopolsky}  \& {Shang}}{{Li}
  et~al.}{2011}]{Li2011}
{Li} Z.-Y.,  {Krasnopolsky} R.,   {Shang} H.,  2011, \mn@doi [\apj]
  {10.1088/0004-637X/738/2/180}, \href
  {http://adsabs.harvard.edu/abs/2011ApJ...738..180L} {738, 180}

\bibitem[\protect\citeauthoryear{{Mac Low}, {Norman}, {Konigl}  \&
  {Wardle}}{{Mac Low} et~al.}{1995}]{MacLow1995}
{Mac Low} M.-M.,  {Norman} M.~L.,  {Konigl} A.,   {Wardle} M.,  1995, \mn@doi
  [\apj] {10.1086/175477}, \href
  {http://adsabs.harvard.edu/abs/1995ApJ...442..726M} {442, 726}

\bibitem[\protect\citeauthoryear{{Machida}, {Inutsuka}  \&
  {Matsumoto}}{{Machida} et~al.}{2007}]{Machida2007}
{Machida} M.~N.,  {Inutsuka} S.-i.,   {Matsumoto} T.,  2007, \mn@doi [\apj]
  {10.1086/521779}, \href {http://adsabs.harvard.edu/abs/2007ApJ...670.1198M}
  {670, 1198}

\bibitem[\protect\citeauthoryear{{Marinacci} \& {Vogelsberger}}{{Marinacci} \&
  {Vogelsberger}}{2016}]{Marinacci2016}
{Marinacci} F.,  {Vogelsberger} M.,  2016, \mn@doi [\mnras]
  {10.1093/mnrasl/slv176}, \href
  {http://adsabs.harvard.edu/abs/2016MNRAS.456L..69M} {456, L69}

\bibitem[\protect\citeauthoryear{{Marinacci}, {Vogelsberger}, {Mocz}  \&
  {Pakmor}}{{Marinacci} et~al.}{2015}]{Marinacci2015}
{Marinacci} F.,  {Vogelsberger} M.,  {Mocz} P.,   {Pakmor} R.,  2015, \mn@doi
  [\mnras] {10.1093/mnras/stv1692}, \href
  {http://adsabs.harvard.edu/abs/2015MNRAS.453.3999M} {453, 3999}

\bibitem[\protect\citeauthoryear{{Marinacci} et~al.,}{{Marinacci}
  et~al.}{2017}]{Marinacci2017}
{Marinacci} F.,  et~al., 2017, preprint, \href
  {http://adsabs.harvard.edu/abs/2017arXiv170703396M} {} (\mn@eprint {arXiv}
  {1707.03396})

\bibitem[\protect\citeauthoryear{{Masson}, {Teyssier}, {Mulet-Marquis},
  {Hennebelle}  \& {Chabrier}}{{Masson} et~al.}{2012}]{Masson2012}
{Masson} J.,  {Teyssier} R.,  {Mulet-Marquis} C.,  {Hennebelle} P.,
  {Chabrier} G.,  2012, \mn@doi [\apjs] {10.1088/0067-0049/201/2/24}, \href
  {http://adsabs.harvard.edu/abs/2012ApJS..201...24M} {201, 24}

\bibitem[\protect\citeauthoryear{{Matsushita}, {Machida}, {Sakurai}  \&
  {Hosokawa}}{{Matsushita} et~al.}{2017}]{Matsushita2017}
{Matsushita} Y.,  {Machida} M.~N.,  {Sakurai} Y.,   {Hosokawa} T.,  2017,
  \mn@doi [\mnras] {10.1093/mnras/stx893}, \href
  {http://adsabs.harvard.edu/abs/2017MNRAS.470.1026M} {470, 1026}

\bibitem[\protect\citeauthoryear{{Mestel} \& {Spitzer}}{{Mestel} \&
  {Spitzer}}{1956}]{Mestel1956}
{Mestel} L.,  {Spitzer} Jr. L.,  1956, \mn@doi [\mnras]
  {10.1093/mnras/116.5.503}, \href
  {http://adsabs.harvard.edu/abs/1956MNRAS.116..503M} {116, 503}

\bibitem[\protect\citeauthoryear{{Mignone}, {Bodo}, {Massaglia}, {Matsakos},
  {Tesileanu}, {Zanni}  \& {Ferrari}}{{Mignone} et~al.}{2007}]{Mignone2007}
{Mignone} A.,  {Bodo} G.,  {Massaglia} S.,  {Matsakos} T.,  {Tesileanu} O.,
  {Zanni} C.,   {Ferrari} A.,  2007, \mn@doi [\apjs] {10.1086/513316}, \href
  {http://adsabs.harvard.edu/abs/2007ApJS..170..228M} {170, 228}

\bibitem[\protect\citeauthoryear{{Mignone}, {Zanni}, {Tzeferacos}, {van
  Straalen}, {Colella}  \& {Bodo}}{{Mignone} et~al.}{2012}]{Mignone2012}
{Mignone} A.,  {Zanni} C.,  {Tzeferacos} P.,  {van Straalen} B.,  {Colella} P.,
    {Bodo} G.,  2012, \mn@doi [\apjs] {10.1088/0067-0049/198/1/7}, \href
  {http://adsabs.harvard.edu/abs/2012ApJS..198....7M} {198, 7}

\bibitem[\protect\citeauthoryear{{Mocz}, {Vogelsberger}  \& {Hernquist}}{{Mocz}
  et~al.}{2014}]{Mocz2014}
{Mocz} P.,  {Vogelsberger} M.,   {Hernquist} L.,  2014, \mn@doi [\mnras]
  {10.1093/mnras/stu865}, \href
  {http://adsabs.harvard.edu/abs/2014MNRAS.442...43M} {442, 43}

\bibitem[\protect\citeauthoryear{{Mocz}, {Pakmor}, {Springel}, {Vogelsberger},
  {Marinacci}  \& {Hernquist}}{{Mocz} et~al.}{2016}]{Mocz2016}
{Mocz} P.,  {Pakmor} R.,  {Springel} V.,  {Vogelsberger} M.,  {Marinacci} F.,
  {Hernquist} L.,  2016, \mn@doi [\mnras] {10.1093/mnras/stw2004}, \href
  {http://adsabs.harvard.edu/abs/2016MNRAS.463..477M} {463, 477}

\bibitem[\protect\citeauthoryear{{Mocz}, {Burkhart}, {Hernquist}, {McKee}  \&
  {Springel}}{{Mocz} et~al.}{2017}]{Mocz2017}
{Mocz} P.,  {Burkhart} B.,  {Hernquist} L.,  {McKee} C.~F.,   {Springel} V.,
  2017, \mn@doi [\apj] {10.3847/1538-4357/aa6475}, \href
  {http://adsabs.harvard.edu/abs/2017ApJ...838...40M} {838, 40}

\bibitem[\protect\citeauthoryear{{Mouschovias}}{{Mouschovias}}{1976a}]{Mouschovias1976a}
{Mouschovias} T.~C.,  1976a, \mn@doi [\apj] {10.1086/154436}, \href
  {http://adsabs.harvard.edu/abs/1976ApJ...206..753M} {206, 753}

\bibitem[\protect\citeauthoryear{{Mouschovias}}{{Mouschovias}}{1976b}]{Mouschovias1976b}
{Mouschovias} T.~C.,  1976b, \mn@doi [\apj] {10.1086/154478}, \href
  {http://adsabs.harvard.edu/abs/1976ApJ...207..141M} {207, 141}

\bibitem[\protect\citeauthoryear{{Nakano}, {Nishi}  \& {Umebayashi}}{{Nakano}
  et~al.}{2002}]{Nakano2002}
{Nakano} T.,  {Nishi} R.,   {Umebayashi} T.,  2002, \mn@doi [\apj]
  {10.1086/340587}, \href {http://adsabs.harvard.edu/abs/2002ApJ...573..199N}
  {573, 199}

\bibitem[\protect\citeauthoryear{{Ntormousi}, {Hennebelle}, {Andr{\'e}}  \&
  {Masson}}{{Ntormousi} et~al.}{2016}]{Ntormousi2016}
{Ntormousi} E.,  {Hennebelle} P.,  {Andr{\'e}} P.,   {Masson} J.,  2016,
  \mn@doi [\aap] {10.1051/0004-6361/201527400}, \href
  {http://adsabs.harvard.edu/abs/2016A%26A...589A..24N} {589, A24}

\bibitem[\protect\citeauthoryear{{Pakmor} \& {Springel}}{{Pakmor} \&
  {Springel}}{2013}]{Pakmor2013}
{Pakmor} R.,  {Springel} V.,  2013, \mn@doi [\mnras] {10.1093/mnras/stt428},
  \href {http://adsabs.harvard.edu/abs/2013MNRAS.432..176P} {432, 176}

\bibitem[\protect\citeauthoryear{{Pakmor}, {Bauer}  \& {Springel}}{{Pakmor}
  et~al.}{2011}]{Pakmor2011}
{Pakmor} R.,  {Bauer} A.,   {Springel} V.,  2011, \mn@doi [\mnras]
  {10.1111/j.1365-2966.2011.19591.x}, \href
  {http://adsabs.harvard.edu/abs/2011MNRAS.418.1392P} {418, 1392}

\bibitem[\protect\citeauthoryear{{Pakmor}, {Marinacci}  \& {Springel}}{{Pakmor}
  et~al.}{2014}]{Pakmor2014}
{Pakmor} R.,  {Marinacci} F.,   {Springel} V.,  2014, \apjl, \href
  {http://adsabs.harvard.edu/abs/2013arXiv1312.2620P} {783, L20}

\bibitem[\protect\citeauthoryear{{Pakmor}, {Springel}, {Bauer}, {Mocz},
  {Munoz}, {Ohlmann}, {Schaal}  \& {Zhu}}{{Pakmor} et~al.}{2016}]{Pakmor2016}
{Pakmor} R.,  {Springel} V.,  {Bauer} A.,  {Mocz} P.,  {Munoz} D.~J.,
  {Ohlmann} S.~T.,  {Schaal} K.,   {Zhu} C.,  2016, \mn@doi [\mnras]
  {10.1093/mnras/stv2380}, \href
  {http://adsabs.harvard.edu/abs/2016MNRAS.455.1134P} {455, 1134}

\bibitem[\protect\citeauthoryear{{Pakmor} et~al.,}{{Pakmor}
  et~al.}{2017}]{Pakmor2017}
{Pakmor} R.,  et~al., 2017, \mn@doi [\mnras] {10.1093/mnras/stx1074}, \href
  {http://adsabs.harvard.edu/abs/2017MNRAS.469.3185P} {469, 3185}

\bibitem[\protect\citeauthoryear{{Parker}}{{Parker}}{1983}]{Parker1983}
{Parker} E.~N.,  1983, \mn@doi [\apj] {10.1086/160637}, \href
  {http://adsabs.harvard.edu/abs/1983ApJ...264..642P} {264, 642}

\bibitem[\protect\citeauthoryear{{Powell}, {Roe}, {Linde}, {Gombosi}  \& {De
  Zeeuw}}{{Powell} et~al.}{1999}]{Powell1999}
{Powell} K.~G.,  {Roe} P.~L.,  {Linde} T.~J.,  {Gombosi} T.~I.,   {De Zeeuw}
  D.~L.,  1999, \mn@doi [Journal of Computational Physics]
  {10.1006/jcph.1999.6299}, \href
  {http://adsabs.harvard.edu/abs/1999JCoPh.154..284P} {154, 284}

\bibitem[\protect\citeauthoryear{{Rembiasz}, {Obergaulinger},
  {Cerd{\'a}-Dur{\'a}n}, {Aloy}  \& {M{\"u}ller}}{{Rembiasz}
  et~al.}{2017}]{Rembiasz2017}
{Rembiasz} T.,  {Obergaulinger} M.,  {Cerd{\'a}-Dur{\'a}n} P.,  {Aloy}
  M.-{\'A}.,   {M{\"u}ller} E.,  2017, \mn@doi [\apjs]
  {10.3847/1538-4365/aa6254}, \href
  {http://adsabs.harvard.edu/abs/2017ApJS..230...18R} {230, 18}

\bibitem[\protect\citeauthoryear{Saad \& Schultz}{Saad \&
  Schultz}{1986}]{Saad1986}
Saad Y.,  Schultz M.~H.,  1986, \mn@doi [SIAM Journal on Scientific and
  Statistical Computing] {10.1137/0907058}, 7, 856

\bibitem[\protect\citeauthoryear{{Santos-Lima}, {de Gouveia Dal Pino}  \&
  {Lazarian}}{{Santos-Lima} et~al.}{2012}]{Santos-Lima2012}
{Santos-Lima} R.,  {de Gouveia Dal Pino} E.~M.,   {Lazarian} A.,  2012, \mn@doi
  [\apj] {10.1088/0004-637X/747/1/21}, \href
  {http://adsabs.harvard.edu/abs/2012ApJ...747...21S} {747, 21}

\bibitem[\protect\citeauthoryear{{Seaton}, {Bartz}  \& {Darnel}}{{Seaton}
  et~al.}{2017}]{Seaton2017}
{Seaton} D.~B.,  {Bartz} A.~E.,   {Darnel} J.~M.,  2017, \mn@doi [\apj]
  {10.3847/1538-4357/835/2/139}, \href
  {http://adsabs.harvard.edu/abs/2017ApJ...835..139S} {835, 139}

\bibitem[\protect\citeauthoryear{{Seifried}, {Pudritz}, {Banerjee}, {Duffin}
  \& {Klessen}}{{Seifried} et~al.}{2012}]{Seifried2012}
{Seifried} D.,  {Pudritz} R.~E.,  {Banerjee} R.,  {Duffin} D.,   {Klessen}
  R.~S.,  2012, \mn@doi [\mnras] {10.1111/j.1365-2966.2012.20610.x}, \href
  {http://adsabs.harvard.edu/abs/2012MNRAS.422..347S} {422, 347}

\bibitem[\protect\citeauthoryear{{Shu}, {Adams}  \& {Lizano}}{{Shu}
  et~al.}{1987}]{Shu1987}
{Shu} F.~H.,  {Adams} F.~C.,   {Lizano} S.,  1987, \mn@doi [\araa]
  {10.1146/annurev.aa.25.090187.000323}, \href
  {http://adsabs.harvard.edu/abs/1987ARA%26A..25...23S} {25, 23}

\bibitem[\protect\citeauthoryear{{Springel}}{{Springel}}{2010}]{Arepo}
{Springel} V.,  2010, \mn@doi [\mnras] {10.1111/j.1365-2966.2009.15715.x},
  \href {http://adsabs.harvard.edu/abs/2010MNRAS.401..791S} {401, 791}

\bibitem[\protect\citeauthoryear{{Stone}, {Gardiner}, {Teuben}, {Hawley}  \&
  {Simon}}{{Stone} et~al.}{2008}]{Stone2008}
{Stone} J.~M.,  {Gardiner} T.~A.,  {Teuben} P.,  {Hawley} J.~F.,   {Simon}
  J.~B.,  2008, \mn@doi [\apjs] {10.1086/588755}, \href
  {http://adsabs.harvard.edu/abs/2008ApJS..178..137S} {178, 137}

\bibitem[\protect\citeauthoryear{{Tilley} \& {Balsara}}{{Tilley} \&
  {Balsara}}{2011}]{Tilley2011}
{Tilley} D.~A.,  {Balsara} D.~S.,  2011, \mn@doi [\mnras]
  {10.1111/j.1365-2966.2011.18982.x}, \href
  {http://adsabs.harvard.edu/abs/2011MNRAS.415.3681T} {415, 3681}

\bibitem[\protect\citeauthoryear{{Vall\'ee}}{{Vall\'ee}}{1998}]{Vallee1998}
{Vall\'ee} J.~P.,  1998, \fcp, \href
  {http://adsabs.harvard.edu/abs/1998FCPh...19..319V} {19, 319}

\bibitem[\protect\citeauthoryear{{Zhu}, {Liu}, {Alexander}  \& {McAteer}}{{Zhu}
  et~al.}{2016}]{Zhu2016}
{Zhu} C.,  {Liu} R.,  {Alexander} D.,   {McAteer} R.~T.~J.,  2016, \mn@doi
  [\apjl] {10.3847/2041-8205/821/2/L29}, \href
  {http://adsabs.harvard.edu/abs/2016ApJ...821L..29Z} {821, L29}

\bibitem[\protect\citeauthoryear{{de Avillez} \& {Breitschwerdt}}{{de Avillez}
  \& {Breitschwerdt}}{2005}]{deAvillez2005}
{de Avillez} M.~A.,  {Breitschwerdt} D.,  2005, \mn@doi [\aap]
  {10.1051/0004-6361:20042146}, \href
  {http://adsabs.harvard.edu/abs/2005A%26A...436..585D} {436, 585}

\makeatother
\end{thebibliography}

\label{lastpage}

\end{document}